\documentclass[english,11pt, a4paper]{article}
\usepackage[T1]{fontenc}
\usepackage[latin9]{inputenc}
\usepackage{jheppub}
\usepackage{color}
\usepackage{dsfont}
\usepackage{amsmath}
\usepackage{amssymb}
\usepackage{graphicx}
\usepackage{cancel}
\usepackage[normalem]{ulem}
\usepackage{tikz-feynman}
\usepackage{soul}
\usepackage{booktabs}
\usepackage{tabularx}
\usepackage{array}

\soulregister\ref7
\soulregister\eqref7
\soulregister\pageref7
\soulregister\cite7
\soulregister\citep7

\makeatletter
\usepackage{slashed}
\usepackage[dvipsnames]{xcolor}
\allowdisplaybreaks[2]
\usepackage{braket}
\usepackage{footnote}
\usepackage{babel}

\title{\boldmath Heavy quarkonium spin transport in a hot medium}

\author[a,b,c]{Nora Brambilla,}
\author[d,a]{Shuo Fang,}
\author[a]{Tom Magorsch,}
\author[a]{Panayiotis Panayiotou}

\affiliation[a]{Technical University of Munich, TUM School of Natural Sciences, Physics Department, James-Franck-Strasse 1, 85748 Garching, Germany}
\affiliation[b]{Technical University of Munich, Institute for Advanced Study, Lichtenbergstrasse 2 a, 85748 Garching, Germany}
\affiliation[c]{Technical University of Munich, Munich Data Science Institute, Walther-von-Dyck-Strasse 10, 85748 Garching, Germany}
\affiliation[d]{Department of Modern Physics, University of Science and Technology of China, Anhui 230026, China}

\emailAdd{nora.brambilla@tum.de}
\emailAdd{shuo.fang@tum.de}
\emailAdd{tom.magorsch@tum.de}
\emailAdd{panayiotis.panayiotou@tum.de}

\preprint{TUM-EFT 222/26}

\abstract{We investigate the in-medium spin transport of heavy quarkonium using an open-quantum-system approach based on potential nonrelativistic QCD. We express the heavy quarkonium dilepton decay angular distribution in terms of late-time quarkonium spin-dependent distribution functions. Working in the quantum-optical regime, where the inverse quarkonium size is much larger than the binding energy and the binding energy is comparable to or larger than the medium temperature $T$, we derive a spin-dependent quantum master equation incorporating both dissociation and regeneration, followed by the corresponding projected semiclassical Boltzmann equations. At leading order in the coupling, we retain the recoil corrections of relative order $T/M$ that contribute to the leading spin-dependent collision kernel. Here $M$ denotes the heavy-quark mass. 
At zeroth order in fluid gradients, we show that the relative motion between quarkonium and the medium supplies the leading anisotropy, while spin-chromomagnetic and spin-orbit interactions both contribute to the leading nonzero spin-alignment kernel.
As an exploratory phenomenological application, we also estimate the medium-induced polarization parameter $\lambda_{\theta}$ of bottomonium $\Upsilon(1\text{S})$ using a simplified rate equation with background density matrices obtained from the Lindblad equation using QTraj. Within this simplified treatment, the resulting medium-induced polarization is found to be very small while a naive rescaling for charmonium can qualitatively describe the polarization of charmonium.}

\makeatother

\begin{document}
\maketitle
\flushbottom

\section{Introduction}\label{sec:Introduction}
Spin polarization provides an important probe of various features of
Quantum Chromodynamics (QCD), for example, the hadron structure
\cite{Aidala:2012mv}. Pertinent to this work, spin observables reveal
novel properties of the quark-gluon plasma (QGP), a deconfined phase
of QCD matter produced in relativistic heavy ion collisions. Specifically, the global
polarization of $\Lambda$ hyperons demonstrates that QGP possesses
a large vorticity of order $10^{21}\,\mathrm{s}^{-1}$ \cite{STAR:2017ckg,Becattini:2020ngo}, 
while local polarization reveals the interplay between spin and hydrodynamical
gradients  \cite{STAR:2019erd,Fu:2021pok,Becattini:2021iol,Yi:2021ryh}. Recently,
relativistic heavy-ion experiments at the RHIC and the LHC have observed significant spin alignment of light or heavy vector mesons along the normal vector of the event plane
\cite{STAR:2022fan,ALICE:2019aid,ALICE:2023jad,ALICE:2025cdf}.
 The former can be explained qualitatively by effective models focusing on
the idea that coalescence of constituent quarks induces spin correlations. This indicates the existence of strong-field correlations \cite{Sheng:2019kmk,Sheng:2022wsy}.
However, the mechanism for heavy-flavored mesons such as $J/\psi$ and $D^{*+}$
is still under debate \cite{Muller:2021hpe,Kumar:2023ghs,Yang:2024ejk,Chen:2025mrf,Yan:2025tlx,Sheng:2024kgg}. 

Heavy quarkonium is an important tool in understanding
strongly interacting hot matter \cite{Rothkopf:2019ipj}.
Heavy quarkonium suppression is viewed as one of the formation signals of a QGP \cite{Matsui:1986dk}, and through the nuclear modification factor ($R_{AA}$), it provides insights into QCD forces within the QGP.  
Besides Debye screening, heavy quarkonium
experiences dissociation during the medium
evolution. This is due to an imaginary part in the heavy quark potential \cite{Laine:2006ns,Brambilla:2008cx,Beraudo:2007ky,Escobedo:2008sy,Brambilla:2010vq}, which  endows heavy quarkonia with temperature-dependent widths. These are shown to dominate screening in a realistic QGP environment.
Dissociation has two origins: gluodissociation via dipole transition
\cite{Brambilla:2008cx,Brambilla:2011sg} and Landau damping via parton
scattering \cite{Laine:2006ns,Brambilla:2013dpa}. Moreover, another  phenomenologically important mechanism is regeneration, where heavy quarkonium
is formed from either correlated or  uncorrelated heavy quarks \cite{Braun-Munzinger:2000csl,Thews:2000rj}.
A consistent framework for heavy quarkonium transport with the  inclusion
of these effects, although very challenging (see also \cite{Andronic:2024oxz}
for a review), is made possible with the framework of open quantum system (OQS) \cite{Akamatsu:2011se,Akamatsu:2014qsa,Brambilla:2016wgg,Brambilla:2017zei,Katz:2015qja,Blaizot:2015hya,Blaizot:2017ypk,Blaizot:2018oev,Yao:2018nmy,Yao:2020xzw,Yao:2020eqy,Sharma:2019xum}
(see \cite{Akamatsu:2020ypb,Yao:2021lus} for recent reviews). Leveraging QCD effective field theories (EFT), we treat the heavy quark-antiquark
pair as a subsystem and the QGP as an environment.
A nonrelativistic particle in vacuum, with relative velocity $v\ll1$, follows the hierarchy of scales 
\begin{equation}
M\gg Mv\gg Mv^2\gtrsim \Lambda_{\rm QCD} \label{eq:energy-hierarchy-1}
\end{equation}
where $M$ is the heavy quark mass, $Mv$ is the typical momentum transfer and $Mv^2$ is the binding energy. The dynamics of a heavy quark-antiquark pair following the scale hierarchy in eq.~\eqref{eq:energy-hierarchy-1}
are described by a weakly-coupled potential nonrelativistic QCD (pNRQCD) i.e. $Mv\gg \Lambda_{\rm QCD}$, once the heavy
quark mass and the soft momentum transfer, $p\sim Mv$, are integrated out \cite{Pineda:1997bj,Brambilla:1999xf,Brambilla:2004jw}. The presence of the QGP introduces additional thermal scales, such as the local  temperature $T$. Including the center-of-mass motion, we focus on the kinetic regime in which the quarkonium center-of-mass kinetic energy satisfies $\boldsymbol{P}^2/4M\sim T$.
In a typical QGP environment, the heavy-quarkonium binding energy is comparable to $T$ which is justified for low-lying bottomonium. Therefore, in the presence of these thermal scales, the hierarchy of scales reads  
\begin{equation}
M\gg Mv\gtrsim\left|\boldsymbol{P}\right|\sim\sqrt{MT}\gg Mv^{2}\gtrsim T.\label{eq:energy-hierarchy-2}
\end{equation}
The moving quarkonium interacts with the QGP via dipole interactions that are small. Neglecting recoil corrections, heavy quarkonium dynamics can be factorized from the medium evolution
\cite{Brambilla:2016wgg,Brambilla:2017zei}, and the information from the QGP 
is encoded in the heavy quarkonium transport coefficients, which
can be extracted from lattice data \cite{Brambilla:2019tpt,Larsen:2019bwy} (see also the perturbative calculations in \citep{New1}).
A Lindblad equation based on this setup is derived up to NLO in an $E/T$ expansion
under the hierarchy $Mv\gg T\gg Mv^{2}$ in \cite{Brambilla:2022ynh}, is simulated with a quantum
trajectory algorithm \cite{Omar:2021kra} and can describe $R_{AA}$ as a function of centrality
and transverse momentum $p_{{\rm T}}$ remarkably well \cite{Brambilla:2022ynh,Brambilla:2023hkw,Brambilla:2024tqg}. On the other hand, the hierarchy with center-of-mass momentum was discussed in context of dark matter dynamics~\cite{Biondini:2024aan}, but, to our knowledge, it has not previously been incorporated systematically into an open-system description of quarkonium transport. 

Using EFTs and OQS, the heavy quarkonium spin density matrix elements can be formulated systematically in field theory. Such matrix elements, as we will show, are related to quarkonium decay through the experimentally clean dilepton channel. Moreover, most of the previous works on in-medium spin
transport relied on either the assumption of a weakly-coupled QGP \cite{Hidaka:2022dmn}
or on phenomenological models \cite{Sheng:2022ffb,Kumar:2023ghs}. However, the phenomenologically relevant temperatures observed indicate a strongly-coupled QGP for the majority of its lifetime. To this end, a
theory-based QCD spin transport equation incorporating strongly-coupled QGP
information is worth pursuing. Within heavy quarkonium transport,
both the heavy quarkonium diffusion coefficient $\kappa$ and its dispersive counterpart, $\gamma$, are related to a gluonic correlator
\cite{Brambilla:2017zei,Brambilla:2019tpt}.
As shown in \cite{Yang:2024ejk}, in the case of spin transport, the chromomagnetic-magnetic correlator is the leading contribution. However, as suggested in \cite{Chen:2025mrf}, the relative motion
between heavy quarkonium and the medium also generates polarization which reveals the coupling between
flowing medium and heavy quarkonium motion. As a step toward this goal, we develop the recoil-dependent formulation at leading order in the coupling.

In this work, we develop a leading-coupling, recoil-dependent open-quantum-system framework for heavy-quarkonium spin transport in the quantum-optical regime. We first express the experimentally accessible polarization parameter $\lambda_{\theta}$ in terms of pNRQCD correlation functions.
We then derive a spin-dependent quantum master equation and projected semiclassical Boltzmann equations including color-singlet dissociation, color-octet regeneration, and the recoil corrections of relative order $T/M$ relevant to the leading spin-dependent collision kernel. At zeroth order in fluid gradients, spin-orbit interactions contribute alongside spin-chromomagnetic interactions at the leading nonzero order in spin alignment. In the regeneration kernels, we retain the off-diagonal momentum-space octet density matrix and the associated coherence effects.
We end by providing an exploratory estimate for the medium-induced polarization of lowest-lying bottomonium in Pb-Pb collisions at $\sqrt{s_{\rm NN}}=5.02\,\rm TeV$ using background singlet and octet density matrices evolved with QTraj, followed by a $J/\psi$ explanatory fitting. 

This paper is organized as follows: In section \ref{sec:Heavy-quarkonium-spin}
we show how starting from heavy-quarkonium decays, the angular distribution of dileptons can be expressed in terms of  spin density matrix
elements with heavy quark current correlators. Following, in section \ref{subsec:Potential-NRQCD-with},
we discuss the spin-dependent interactions that contribute to the
traceless pNRQCD correlators and reduce the 
 heavy quarkonium spin density matrix elements in terms of density operator and distribution function.
 In section \ref{sec:Evolution-of-heavy-quarkonium}, we derive a spin-dependent quantum master equation from real-time field theory, retaining the recoil corrections of relative order $T/M$ relevant to the leading spin-dependent collision kernel, and obtain the corresponding Boltzmann equations in both the quarkonium and medium rest frames.
We numerically estimate the solution of simplified spin Boltzmann equation
and $\Upsilon(1\text{S})$ polarization $q_{\rm T}$ dependence and make physical
explanations by comparing anisotropic dissociation and number damping rates in section \ref{sec:Numerical-estimation}. We include also a set of complementary appendices following the conclusions in section \ref{sec:Conclusions}. We work in the mostly-minus metric signature. 

\section{Heavy quarkonium spin polarization from dilepton decay distribution}\label{sec:Heavy-quarkonium-spin}
Our starting point is the thermal decay rate of heavy quarkonium in the  dilepton channel,
e.g. $\Upsilon \to l^{+}l^{-}$. The Lorentz-invariant decay rate is defined
as the probability of $\Upsilon $ with momentum $q^{\mu}=k_{1}^{\mu}+k_{2}^{\mu}$
to decay into a lepton pair $l^{+}l^{-}$ with momentum $k_{1}^{\mu}$
and $k_{2}^{\mu}$. After integrating over $k_2$ and $|\boldsymbol{k}_1|$, the angular distribution of the dilepton in the quarkonium rest frame is \citep{Speranza:2018osi}
\begin{eqnarray}
\frac{{\rm d}\mathcal{R}}{{\rm d}^{4}X{\rm d}^{4}q{\rm d}\Omega} & = & \frac{-Q^{2}e^{4}}{2(2\pi)^{6}q^{4}}\sqrt{1-\frac{4m_{l}^{2}}{q^{2}}}L^{\mu\nu}(q,\hat{\boldsymbol{k}}_{1})J_{\mu\nu}(q,X).
\end{eqnarray}
Here  $Q$ is the electric charge of the heavy (anti)quark and $m_l$ is the lepton mass. The leptonic tensor $L^{\mu\nu}$ is
defined as
\begin{eqnarray}
L^{\mu\nu}(q,\hat{\boldsymbol{k}}_{1}) & = & 2k_1^{\mu}k_1^{\nu}-q^{\mu}k_{1}^{\nu}-q^{\nu}k_{1}^{\mu}+\eta^{\mu\nu}q\cdot k_{1},
\end{eqnarray}
where  $\hat{\boldsymbol{k}}_{1}$ represents
the azimuthal and polar angle part of $\boldsymbol{k}_{1}$ with respect to a specific axis. The heavy-quark current correlation function is 
\begin{eqnarray}
J_{\mu\nu}\left(q,X=\frac{x_1+x_2}{2}\right) & = & \int_{x_{1}-x_{2}}e^{-iq\cdot(x_{1}-x_{2})}{\rm Tr}\left\{\hat{\rho}(t_{0})j_{\mu}(x_{1})j_{\nu}(x_{2})\right\},\label{eq:Heavy_quark_current_correlator}
\end{eqnarray}
where the trace is over spin and color indices; $j_{\mu}(x)$ is the heavy quark current, and
$\hat{\rho}(t_{0})$ is the system density operator at some initial time $t_{0}$. Throughout this work, we use the following shorthand notations for integral measures,
\begin{eqnarray}
\int_{x}=\int{\rm d}^{4}x & ,\; & \int_{\boldsymbol{x}}=\int{\rm d}^{3}\boldsymbol{x},\;\int_{p}=\int\frac{{\rm d}^{4}p}{(2\pi)^{4}},\;\int_{\boldsymbol{p}}=\int\frac{{\rm d}^{3}\boldsymbol{p}}{(2\pi)^{3}}.
\end{eqnarray}
We further choose a specific quantization axis for $k_{1}^{\mu}$
by parameterizing
\begin{eqnarray}
k_{1}^{\mu} & = & (\sqrt{\boldsymbol{k}_{1}^{2}+m_{l}^{2}},|\boldsymbol{k}_{1}|\sin\theta\cos\phi,|\boldsymbol{k}_{1}|\sin\theta\sin\phi,|\boldsymbol{k}_{1}|\cos\theta),
\end{eqnarray}
with $|\boldsymbol{k}_{1}|=\sqrt{\frac{q^{2}}{4}-m_{l}^{2}}$, then
$(\theta,\phi)$ can be defined as the polar and azimuthal angle with
respect to the quantization $z$-axis in the decay distribution $\displaystyle\frac{{\rm d}\mathcal{R}(\theta,\phi)}{{\rm d}^{4}x{\rm d}^{4}q{\rm d}\Omega}$.
In the helicity frame, $z$-direction is chosen as the 3-momentum
direction of $\boldsymbol{q}$ in the laboratory frame such
that $\displaystyle \cos\theta=\frac{\boldsymbol{k}_{1}\cdot\boldsymbol{q}}{|\boldsymbol{k_1}||\boldsymbol{q}|}$.
The definition of polarization angles $(\theta,\phi)$ varies in different polarization frames as discussed in \cite{Baym:2017qxy}.
 In the dilepton center of mass frame, the
polarization vectors read
\begin{eqnarray}
\varepsilon_{\lambda}^{\mu} & = & (0,\boldsymbol{\varepsilon}_{\lambda}),\;\lambda=0,\pm1,
\label{eq:Polarization_vectors}
\end{eqnarray}
where $|\boldsymbol{\varepsilon}_{\lambda}|^{2}=1$, $\boldsymbol{\varepsilon}_{\lambda}\cdot\boldsymbol{\varepsilon}_{\lambda^{\prime}}^{*}=\delta_{\lambda\lambda^{\prime}}$
and the transverse condition $\varepsilon_{\lambda}\cdot q=0$ is satisfied. We can decompose the leptonic momentum
$k_{1}$ with respect to the polarization vectors in the dilepton rest-frame
where $q^{\mu}=(M_{Q\bar{Q}},\boldsymbol{0})$,
\begin{eqnarray}
k_{1}^{\mu} & = & \frac{k_{1}\cdot q}{q^{2}}q^{\mu}+\sum_{\lambda}(-k_{1}\cdot\varepsilon_{\lambda}^{*})\varepsilon_{\lambda}^{\mu},\label{eq:k1-decomposition}
\end{eqnarray}
 and integrate over the azimuthal angle to obtain 
 \begin{eqnarray}
\frac{{\rm d}\mathcal{R}}{{\rm d}^{4}X{\rm d}^{4}q{\rm d}\cos\theta} & = & \frac{Q^{2}\alpha_{e}^{2}{\rm Tr}\rho_{\lambda\lambda^{\prime}}}{2(2\pi)^{3}q^{2}}\left[1+\rho_{00}^{\text{N}}(q)\right]\left[1+\lambda_{\theta}\cos^{2}\theta\right],\label{eq:Decay_distribution_massless_l}
\end{eqnarray}
where the polarization parameter, $\lambda_{\theta}$, and the normalized
density matrix elements are defined as 
\begin{eqnarray}
\lambda_{\theta} & = & \frac{1-3\rho_{00}^{{\rm N}}(q)}{1+\rho_{00}^{{\rm N}}(q)},\;\rho_{\lambda\lambda^{\prime}}^{{\rm N}}=\frac{\rho_{\lambda\lambda^{\prime}}}{{\rm Tr}\rho_{\lambda\lambda^{\prime}}},\label{eq:Normalized-SDM}
\end{eqnarray}
and where the spin density matrix elements
of the heavy quarkonium read 
\begin{eqnarray}
\rho_{\lambda\lambda^{\prime}}(q,x) & = & \varepsilon_{\lambda}^{\mu}(q)\varepsilon_{\lambda^{\prime}}^{*\rho}(q)J_{\mu\rho}(q,x),\label{eq:Density_matrix}\\
J_{\mu\nu}(q,x) & = & \sum_{\lambda,\lambda^{\prime}}\varepsilon_{\mu}^{*\lambda}(q)\varepsilon_{\nu}^{\lambda^{\prime}}(q)\rho_{\lambda\lambda^{\prime}}(q,x).
\end{eqnarray}
Here we work in the $m_l=0$ limit and the decay distribution in eq.~\eqref{eq:Decay_distribution_massless_l} has the same form in the linear and circular polarization bases for a fixed quantization axis. A discussion in the circular basis of $\{\varepsilon_{\lambda}^{\mu}\}$
involving complex $\varepsilon_{\pm}^{\mu}$ is listed in Appendix \ref{sec:Dilepton-decay-distribution}.
In the derivation of eq.~\eqref{eq:Decay_distribution_massless_l}, we have used that the heavy vector current $j^{\mu}$ is conserved such that only the transverse projection of $J_{\mu\nu}$ contributes,
i.e. $\int_{x}q^{\mu}J_{\mu\rho}(q,x)=0$ where the $\partial_x$ term in the phase-space Ward identity gives zero due to the divergence theorem,
and $\rho_{\mu}^{\mu}(q)=-{\rm Tr}\rho_{\lambda\lambda^{\prime}}(q)$.
Note that the polarization vectors $\varepsilon_{\lambda}^{\mu}(q)$ endow $\rho_{\lambda\lambda^{\prime}}$ with polarization frame dependence while in a given polarization basis, $\rho_{\lambda\lambda^{\prime}}$ are defined covariantly. $\varepsilon_{\lambda}^{\mu}$ can still have $\hat{\boldsymbol{q}}$ dependence in the quarkonium rest
frame (e.g. in the helicity frame). 

Our aim is to relate the polarization parameter, $\lambda_{\theta}$, with correlators in pNRQCD. 
As in \cite{Brambilla:2017zei}, $j_{0}(x)$
is related to the annihilation and creation of heavy quark (and/or antiquark) and thus does not contribute
in our interested kinematic regime.
Therefore we focus on $J_{ij}(q,X)$
in the \textit{quarkonium rest frame}.  Inserting the heavy electromagnetic current in such frame 
\begin{eqnarray}
j_{i}(x) & = & e^{2iMt}\psi^{\dagger}\boldsymbol{\sigma}_{i}\chi+e^{-2iMt}\chi^{\dagger}\boldsymbol{\sigma}_{i}\psi
\end{eqnarray} into the heavy quark current correlator (\ref{eq:Heavy_quark_current_correlator})
\begin{eqnarray}
J_{ij}(q,X) & = & \int_{x_{1}-x_{2}}e^{i(2Mv-q)\cdot(x_{1}-x_{2})}\left\langle \psi^{\dagger}(x_{1})\boldsymbol{\sigma}_{i}\chi(x_{1})\chi^{\dagger}(x_{2})\boldsymbol{\sigma}_{j}\psi(x_{2})\right\rangle ,
\end{eqnarray}where the Pauli spinors $\psi^{\dagger}$ creates a heavy quark field while $\chi^{\dagger}$
annihilates a heavy antiquark field, so $\psi^{\dagger}\boldsymbol{\sigma}\chi$ and 
$\chi^{\dagger}\boldsymbol{\sigma}\psi$ correspond to creating and annihilating
a heavy quark-antiquark pair. The expectation value is understood as $\langle\hat{O}\rangle={\rm Tr}[\hat{\rho}(t_{0})\hat{O}]$. The density operator $\hat{\rho}(t_{0})$
at initial time $t_{0}$ is determined by initial conditions, and the subsequent time evolution is carried by $\hat{O}$ in the Heisenberg picture and is represented through contour evolution operators \cite{Blaizot:2001nr}.
If we consider only the diagonal components of the general heavy quark current correlator $J_{ij}=\frac{1}{3}\delta_{ij}J_{kk}+\delta J_{ij}$,
then 
$\rho_{\lambda\lambda^{\prime}}\propto\varepsilon_{\lambda}^{i}\varepsilon_{\lambda^{\prime}}^{*i}=\delta_{\lambda\lambda^{\prime}}$ leads to $\lambda_\theta=0$. 
This implies no polarization will be produced. Therefore, polarization depends only on the \textit{traceless anisotropic} elements of the spin indices which is the focus of this work. 

For concreteness, we denote the
deviation of the spin density matrix from the isotropic unpolarized part 
\begin{eqnarray}
\delta\rho_{\lambda\lambda^{\prime}}(q) & = & \rho_{\lambda\lambda^{\prime}}(q)-\frac{1}{3}\delta_{\lambda\lambda^{\prime}}{\rm Tr}\rho_{\lambda\lambda^{\prime}}(q)=\Delta_{\lambda\lambda^{\prime}}^{ij}J_{ij}(q),\label{eq:Delta_SDME}
\end{eqnarray}
where we have introduced
the projector 
$\displaystyle \Delta_{\lambda\lambda^{\prime}}^{ij}=\varepsilon_{\lambda}^{i}\varepsilon_{\lambda^{\prime}}^{*j}-\frac{1}{d-1}\delta_{\lambda\lambda^{\prime}}\delta^{ij}$. 
As discussed in section~\ref{sec:Introduction}, accounting for the recoil effects we focus on the following hierarchy 
\begin{equation}
M\gg Mv\sim1/r\gtrsim\left|\boldsymbol{P}\right|\sim\sqrt{MT}\gg Mv^{2}\gtrsim T,\Lambda_{{\rm QCD}},\label{eq:Energy-Hierarchy}
\end{equation}
where the heavy quark mass is much larger than the soft scale, the
typical relative momentum in the heavy quark-antiquark pair $Mv\sim r^{-1}$,
and it is further much larger than the binding energy $E\sim Mv^{2}$
and medium temperature $T$. We assume the thermal medium is in local
equilibrium and the  heavy quark-antiquark pair is kicked by gluons
with typical energy of $T$. Therefore, the change of center-of-mass kinetic energy of the pair follows $\Delta\left(\frac{\boldsymbol{P}^{2}}{4M}\right)\sim T$. Different from charm quarks \citep{PHENIX:2010xji,ALICE:2014qvj,STAR:2017kkh}, bottom quarks and bottomonia, in general, are not thermalized in the QGP produced by current high-energy experiments. Though there is no direct measurement of elliptic flow of $B^0$ meson, the $v_2$ of electrons from bottom hadron semileptonic decays indicates bottom quarks are only partially thermalized at small $p_{\rm T}$ \cite{ALICE:2020hdw}; and the collective flow of non-prompt $D^0$ also indicates the non-zero $v_2$ of $b$ quarks at small $p_{\rm T}$ \citep{CMS:2022vfn}. Nevertheless, the hierarchy in eq.~\eqref{eq:Energy-Hierarchy} does not necessarily require a thermalized quarkonium distribution, and we restrict the center-of-mass momentum to the thermal scale, $|\boldsymbol{P}|\sim\sqrt{MT}$ in this work. In general, $|\boldsymbol{P}|$ can be larger than $Mv$, but in such case, the potentials of singlet/octet fields get modified by integrating out the new scale $|\boldsymbol{P}|$ and are still unknown. We thereby adopt this center-of-mass momentum scaling for simplicity.
In OQS, certain approximations can be taken based on hierarchical relations of the characteristic timescales that define the total system. These are: the typical environment correlation time scale $\tau_{{\rm E}}\sim T^{-1}$
governed by the QGP evolution, the system intrinsic scale $\tau_{{\rm S}}\sim E^{-1}$
and the system relaxation scale $\tau_{{\rm R}}\sim\Gamma_{Q\bar{Q}}^{-1}$
which is the inverse of the typical in-medium quarkonium damping rate.
$\Gamma_{Q\overline{Q}}$ can be estimated from the imaginary part of the
finite-temperature quarkonium self-energy in perturbative calculations in a weakly-coupled plasma and is of $\alpha_{{\rm s}}^{2}r^{2}T^{3}$
for $E\ll T$ \cite{Brambilla:2008cx,Brambilla:2010vq} and of $\alpha_{{\rm s}}r^{2}E^{3}$
for $T\lesssim E$ \cite{Brambilla:2011sg}. Therefore under our chosen
hierarchy of scales, eq.~\eqref{eq:Energy-Hierarchy}, it follows that,
\begin{equation}
\tau_{{\rm R}}\gg\tau_{{\rm E}},\;\tau_{{\rm S}}.\label{eq:Quantum_optical_regime}
\end{equation}
Therefore in the language of OQS, we are in  the \textit{quantum optical regime} \cite{Akamatsu:2020ypb}. We note that the condition 
$\tau_{{\rm R}}\gg\tau_{{\rm E}}$ guarantees the applicability of
the Markovian approximation, while the Born approximation assumes the backward influence of system
evolution on environment can be neglected, and $\tau_{{\rm R}}\gg\tau_{{\rm S}}$
allows us to utilize the \textit{rotating-wave approximation}, see section~\ref{sec:Evolution-of-heavy-quarkonium}.
We remark that the quantum optical hierarchy works well for the lowest-lying
heavy quarkonia, e.g. $\Upsilon$(1S). 
For higher excitations the hierarchy $Mv\gg T\gg Mv^{2}$ may be more appropriate.

Because of the hierarchy $M\gg Mv\gg Mv^{2}$, the heavy (anti)quark
field can be integrated out and the interpolating heavy quark-antiquark
pair with ultrasoft energy is left whose dynamics is described by
the (weakly-coupled) potential nonrelativistic QCD (pNRQCD) effective field theory.
Projecting the bilinear heavy quark field to an interpolating quark-antiquark
field gives
\begin{eqnarray}
 \Psi_{m,\alpha;n,\beta}(\boldsymbol{x}_{1},&\boldsymbol{x}_{2},& t)  \sim  \psi_{m,\alpha}(\boldsymbol{x}_{1},t)\chi^{\dagger}_{n,\beta}(\boldsymbol{x}_{2},t)\nonumber \\
 & = & \frac{1}{N_{c}}\delta_{mn}\left(\psi{\chi}^{\dagger}\right)_{\alpha\beta}(\boldsymbol{R},\boldsymbol{r},t)+\frac{1}{T_{{\rm F}}}T_{mn}^{a}\left(\psi T^{a}\chi^{\dagger}\right)_{\alpha\beta}(\boldsymbol{R},\boldsymbol{r},t)
\end{eqnarray}
where $\{m,n\}$ are the color indices and $\{\alpha,\beta\}$ are
the spin indices and we define $\boldsymbol{r}=\boldsymbol{x}_1-\boldsymbol{x}_2$ and $\boldsymbol{R}=\frac{\boldsymbol{x}_1+\boldsymbol{x}_2}{2}$. 
In the expression of $J_{ij}(q)$, only the color-singlet
components $S_{\alpha\beta}\sim\left(\psi{\chi}^{\dagger}\right)_{\alpha\beta}$
enter: 
\begin{eqnarray}
 &  & J_{ij}(q,X)\nonumber \\
 & = & \int_{x_{1}-x_{2}}e^{i(2Mv-q)\cdot(x_{1}-x_{2})}\left\langle \boldsymbol{\sigma}_{i,\alpha\beta}\left(\psi^{\dagger}\chi\right)_{\alpha\beta}(\boldsymbol{x}_1,\boldsymbol{0},t_{1})\left({\chi}^{\dagger}\psi\right)_{\gamma\delta}(\boldsymbol{x}_{2},\boldsymbol{0},t_{2})\boldsymbol{\sigma}_{j,\gamma\delta}\right\rangle \nonumber \\
 & = & \int_{x_{1}-x_{2}}e^{i(2Mv-q)\cdot(x_{1}-x_{2})}\left\langle {\rm tr}\left(\boldsymbol{\sigma}_{i}S^{\dagger}(\boldsymbol{x}_{1},\boldsymbol{0},t_{1})\right){\rm tr}\left(S(\boldsymbol{x}_{2},\boldsymbol{0},t_{2})\boldsymbol{\sigma}_{j}\right)\right\rangle .\label{eq:Singlet_corr_fun_1}
\end{eqnarray}
We proceed by discussing the Pauli-matrix components of the color-singlet field $S$. For $S$-wave quarkonia such as $\Upsilon(1S)$ and  $J/\psi$,
the total angular momentum state of the meson is a simple superposition
of spin of heavy quark and heavy antiquark. It reads
\begin{eqnarray}
S_{++} & : & |\!\!\uparrow\uparrow\rangle,\;S_{--}:|\!\!\downarrow\downarrow\rangle,\;S_{+-}:|\!\!\uparrow\downarrow\rangle,\;S_{-+}:|\!\!\downarrow\uparrow\rangle.
\end{eqnarray}
The color-singlet field $S$ itself is a $2\times2$ matrix in
spin space. We introduce a suitable basis by redefining $\chi^{\prime}=\chi(-i\sigma^{2})$
such that
\begin{eqnarray}
\psi\chi^{\dagger} & = & \psi(-i\sigma^{2}\chi^{\prime,\dagger}),
\end{eqnarray}
under which the Lagrangian of potential NRQCD remains same. Under this basis, we can decompose a matrix valued quark-antiquark field into 
\begin{eqnarray}
S_{\alpha\beta} & = & \frac{\delta_{\alpha\beta}}{\sqrt{2}}S_{\text{sin}}+\frac{(\boldsymbol{\sigma}^{i})_{\alpha\beta}}{\sqrt{2}}S^{i}=\frac{\delta_{\alpha\beta}S_{\text{sin}}}{\sqrt{2}}+\frac{(\boldsymbol{\sigma}^{i})_{\alpha\beta}}{\sqrt{2}}\sum_{\lambda=0,\pm1}\left(\boldsymbol{\epsilon}_{\lambda}^{i}S_{\lambda}\right),\label{eq:S_Eigenvector_basis_Expansion}
\end{eqnarray}
where 
\begin{eqnarray*}
S_{\text{sin}} & = & \frac{1}{\sqrt{2}}{\rm Tr}(S),\;S^{i}=\frac{1}{\sqrt{2}}{\rm Tr}(S\boldsymbol{\sigma}^{i}),
\end{eqnarray*}
and
\begin{eqnarray}
\boldsymbol{\epsilon}_{1}=\frac{1}{\sqrt{2}}(1,i,0),\; & \boldsymbol{\epsilon}_{-1}=\frac{1}{\sqrt{2}}(-1,i,0),\; & \boldsymbol{\epsilon}_{0}=(0,0,-1),
\end{eqnarray}
with $\sum_{\lambda}\boldsymbol{\epsilon}_{\lambda}^{*,i}\boldsymbol{\epsilon}_{\lambda}^{j}=\delta^{ij}$,
$\boldsymbol{\epsilon}_{\lambda}^{*}\cdot\boldsymbol{\epsilon}_{\lambda^{\prime}}=\delta_{\lambda\lambda^{\prime}}$ and the trace is understood over only the spin space.
In eq.~\eqref{eq:S_Eigenvector_basis_Expansion}, we find $S_{{\rm sin}}=(|\!\!\downarrow\uparrow\rangle-|\!\!\uparrow\downarrow\rangle)/\sqrt{2}$
is indeed the spin-singlet state, while $S_{\lambda=1}=|\!\!\uparrow\uparrow\rangle$,
$S_{\lambda=-1}=|\!\!\downarrow\downarrow\rangle$ and $S_{\lambda=0}=(|\!\!\downarrow\uparrow\rangle+|\!\!\uparrow\downarrow\rangle)/\sqrt{2}$
are the spin-triplet eigenstates\footnote{Note that we can also decompose $S_{\alpha\beta}$ using Pauli matrix basis $S_{\alpha\beta}=\delta_{\alpha\beta}S_{\boldsymbol{1}}/\sqrt{2}+(\boldsymbol{\sigma}^{i})_{\alpha\beta}S_{\boldsymbol{3}}^{i}/\sqrt{2}$.
However, these spin components are not the eigenstates of spin angular
momentum operator, for example, $S_{\boldsymbol{1}}=(|\!\!\uparrow\uparrow\rangle+|\!\!\downarrow\downarrow\rangle)/\sqrt{2}$.
}. We note $\boldsymbol{\epsilon}_{\lambda}$
should be distinguished from polarization vectors $\boldsymbol{\varepsilon}_{\lambda}$ in eq.~\eqref{eq:Polarization_vectors} defined under a general quantization
axis, which renders different definitions for the spin distribution
function from \cite{Yang:2024ejk}. The color-singlet-field correlation function in eq.~\eqref{eq:Singlet_corr_fun_1}
now reduces to 
\begin{eqnarray}
J_{ij}\left(q,\frac{R+R^{\prime}}{2}\right) & = & 2\int_{R-R^{\prime}}e^{i(2Mv-q)\cdot(R-R^{\prime})}\left\langle S_{i}^{\dagger}(\boldsymbol{R},\boldsymbol{0},t)S_{j}(\boldsymbol{R}^{\prime},\boldsymbol{0},t^{\prime})\right\rangle. \label{eq:Singlet_corr_fun_2}
\end{eqnarray}
We also use the notation $R=(t,\boldsymbol{R})$ in what follows. 

Finally, isolating the nonvanishing contribution to $\lambda_\theta$, we consider the operation of Pauli matrices on the interpolating
field. We define
\begin{eqnarray}
\left(\boldsymbol{\sigma}_{(1)}S\right)_{\alpha\beta} & = & \left(\boldsymbol{\sigma}\psi\right)_{\alpha}(-i\sigma^{2}\chi^{\prime,\dagger})_{\beta},
\end{eqnarray}
and 
\begin{eqnarray}
\left(\boldsymbol{\sigma}_{(2)}S\right)_{\alpha\beta} & = & \psi_{\alpha}\left(\boldsymbol{\sigma}(-i\sigma^{2}\chi^{\prime,\dagger})\right)_{\beta}=\psi_{\alpha}\left(-i\sigma^{2}\sigma^{2}\boldsymbol{\sigma}(\sigma^{2}\chi^{\prime,\dagger})\right)_{\beta}\nonumber \\
 & = & \psi_{\alpha}\left(-i\chi^{\prime,\dagger}\boldsymbol{\sigma}\sigma^{2}\right)_{\beta}=-\psi_{\alpha}\left(-i\sigma^{2}(\chi^{\prime,\dagger}\boldsymbol{\sigma})\right)_{\beta},
\end{eqnarray}
where $\boldsymbol{\sigma}_{(1)}$ only acts on the heavy quark field
while $\boldsymbol{\sigma}_{(2)}$ only acts on the heavy antiquark
fields and $\sigma_{2}\boldsymbol{\sigma}\sigma_{2}=-\boldsymbol{\sigma}^{T}$.
In a compact form we can write
\begin{eqnarray}
\boldsymbol{\sigma}_{(1)}S & = & \boldsymbol{\sigma}S,\;\boldsymbol{\sigma}_{(2)}S=-S\boldsymbol{\sigma}.\label{eq:Pauli_matrix_bilinear_field}
\end{eqnarray}
From the above, the bilinear terms we encounter in the pNRQCD Lagrangian, under their spin decomposition, read
\begin{align}
    &\mathrm{Tr}\left(S^\dagger AS\right)=S_{\mathrm{sin}}^{\dagger}A S_{\mathrm{sin}}+S^{i,\dagger}A S^{i}\nonumber, \\   \mathrm{Tr}\left(S^\dagger\frac{i\left(\sigma^{i}_{(1)}+\sigma^{i}_{(2)}\right)}{2} S\right)\!\!=&\epsilon^{ijk}S^{j,\dagger}S^k,\,\mathrm{Tr}\left(S^\dagger\frac{\left(\sigma^{i}_{(1)}-\sigma^{i}_{(2)}\right)}{2} S\right)\!\!=S^{i,\dagger} S_{\mathrm{sin}}+S_{\mathrm{sin}}^{\dagger}S^{i}.\label{eq: bilin2}
\end{align}
where $A$ represents operators that do not involve Pauli matrices. For non-vanishing spin polarization,  we focus on bilinear terms like in eq.~\eqref{eq: bilin2}.

\section{Spin-dependent interactions in pNRQCD\label{subsec:Potential-NRQCD-with}}
According to section \ref{sec:Heavy-quarkonium-spin}, we are interested in the traceless components of the singlet correlator $\Delta_{\lambda\lambda^{\prime}}^{ij}J_{ij}$ which generate
polarization of the heavy meson i.e $\lambda_\theta\neq0$. To this end, 
we need to incorporate interactions sensitive to the quarkonium spin
structure. 
Using eqs. \eqref{eq:S_Eigenvector_basis_Expansion} and \eqref{eq: bilin2}  we find that the leading order potential and electric dipole transition do not polarize quarkonium.
Possible sources are the spin-dependent potential among color-singlet/octet
fields, spin-dependent dipole transition like the spin-spin interaction
and spin-magnetic, spin-orbital coupling.
Moreover, relative motion between quarkonium and the thermal medium introduces a preferred direction and allows center-of-mass-dependent spin-orbit interactions to contribute to quarkonium spin alignment.

Before stating our chosen pNRQCD Lagrangian in the laboratory frame, we comment on our choice of spin-dependent interactions.
Substituting the spin decomposition for $S$-wave states
\begin{eqnarray}
{\rm S} & = & \frac{\mathds{1}_{c}}{\sqrt{N_{c}}}S=\frac{\mathds{1}_{c}}{\sqrt{2N_{c}}}\left(\mathds{1}_{s}S_{{\rm sin}}+\boldsymbol{\sigma}^{i}S^{i}\right),\\
{\rm O} & = & \frac{1}{\sqrt{T_{{\rm F}}}}O^{a}T^{a}=\frac{1}{\sqrt{2T_{{\rm F}}}}T^{a}\left(\mathds{1}_{s}O_{\text{sin}}^{a}+\boldsymbol{\sigma}^{i}O^{a,i}\right),
\end{eqnarray}
to the Lagrangian up to leading order (LO) in multipole expansion,
\begin{eqnarray}
\mathcal{L}_{{\rm pNRQCD}}^{(\text{LO})} & = & \int_{\boldsymbol{r}}{\rm Tr}\left[{\rm S}^{\dagger}(i\partial_{0}-h_{s}){\rm S}+{\rm O}^{\dagger}(iD_{0}-h_{o}){\rm O}\right]\nonumber \\
 & = & \int_{\boldsymbol{r}}\left[S_{{\rm sin}}^{\dagger}(i\partial_{0}-h_{s}^{\text{sin}})S_{{\rm sin}}+S^{\dagger,i}(i\partial_{0}\delta^{ij}-h_{s}^{{\rm tri},ij})S^{j}\right]\nonumber \\
 &  & +\int_{\boldsymbol{r}}\left[O_{\text{sin}}^{\dagger,a}(iD_{0}-h_{o})O_{\text{sin}}^{a}+O^{\dagger,a,i}(iD_{0}\delta^{ij}-h_{o}^{{\rm tri},ij})O^{a,j}\right]\label{eq:LO_pNRQCD}
\end{eqnarray}
where 
$D_{0}=\partial_{0}+igA_{0}$ with $A_0$ the adjoint gluon field. 
Note that we have not redefined the octet field following~\cite{Brambilla:2017zei}.
Here we explicitly
separate the Hamiltonian for the spin singlet and spin triplet fields, denoted
as $h_{s/o}$. The Hamiltonians contain spin-dependent interactions related to Pauli
matrix and they read
\begin{eqnarray}
h_{s/o}^{\text{sin}}(r,P) & = & \frac{\boldsymbol{p}^{2}}{M}-\frac{\boldsymbol{p}^{4}}{4M^{3}}+\frac{\boldsymbol{P}^{2}}{4M}+V_{s/o}^{\text{sin}},\label{eq:Spin-s-hamilton}\\
h_{s/o}^{{\rm tri},ij}(r,P) & = & \left(\frac{\boldsymbol{p}^{2}}{M}-\frac{\boldsymbol{p}^{4}}{4M^{3}}+\frac{\boldsymbol{P}^{2}}{4M}+V_{s/o,0}^{{\rm tri}}\right)\delta^{ij}+\delta V_{s/o}^{{\rm tri},ij}\equiv h_{s/o,0}^{\text{tri}}\delta^{ij}+\delta V_{s/o}^{{\rm tri},ij},\label{eq:Spin-t-hamilton}
\end{eqnarray}
where $\boldsymbol{p}=-i\boldsymbol{\nabla}_{\boldsymbol{r}}$ and
$\boldsymbol{P}=-i\boldsymbol{D}_{\boldsymbol{R}}$ are the relative-
and center-of-mass-momentum operators, and $M$ is the heavy quark
mass. $V_{s/o}^{{\rm sin}}$ and $V_{s/o}^{{\rm tri},ij}\equiv V_{s/o,0}^{{\rm tri}}\delta^{ij}+\delta V_{s/o}^{{\rm tri},ij}$
are the spin-dependent potentials for spin-singlet
and spin-triplet fields, respectively.  $V_{s/o,0}^{{\rm tri}}$ and $h_{s/o,0}^{{\rm tri}}$ stand for the diagonal component of the potential and Hamiltonian, which ultimately do not generate spin polarization. Up to $\mathcal{O}(M\alpha_{{\rm s}}^{3})$, we have $V_{s/o}^{\text{sin}}=V_{s/o,0}^{{\rm tri}}\equiv V_{s/o}$, and the leading-order potentials read
\begin{eqnarray}
V_{s}=-C_{{\rm F}}\frac{\alpha_{{\rm s}}(r)}{r} & \;, & V_{o}=\left(\frac{C_{{\rm A}}}{2}-C_{{\rm F}}\right)\frac{\alpha_{{\rm s}}(r)}{r}.
\end{eqnarray}
The spin-dependent corrections start from $\mathcal{O}\left(M\alpha_{{\rm s}}^{4}\right)$, possible contributions in the spin-dependent potential to traceless
correlators at $\mathcal{O}(M\alpha_{{\rm s}}^{4})$ are 
\begin{eqnarray}
\delta\mathcal{L}_{{\rm pNRQCD}}^{(\text{LO})} & = & \int_{\boldsymbol{r}}\left\{ \frac{V_{{\rm LS,CM}}^{(2)}(r)}{2M^{2}}\left[S_{{\rm sin}}^{\dagger}(\boldsymbol{r}\times\boldsymbol{P})^{i}S^{i}+{\rm H.c.}\right]-\frac{3V_{\boldsymbol{S}_{12}}^{(2)}(r)}{8M^{2}r^{2}}S^{\dagger,i}\boldsymbol{r}^{i}\boldsymbol{r}^{j}S^{j}\right\} \nonumber \\
 &  & \;+(\text{color-octet part}),\label{eq:Spin-dependent-potential}
\end{eqnarray}
where the Wilson coefficients $V_{{\rm LS,CM}}^{(2)}(r),V_{\boldsymbol{S}_{12}}^{(2)}(r)$
are of $\mathcal{O}(Er^{-2})$ \cite{Brambilla:1999xj,Brambilla:2004jw}.
The first term of eq.~\eqref{eq:Spin-dependent-potential} gives an effective potential for spin-triplet fields through a self-energy diagram, and contributes at $\mathcal{O}(1/M^4)$.

The chromoelectric dipole interaction terms of pNRQCD Lagrangian, at the first order in multipole expansion, are given by 
\begin{eqnarray}
\mathcal{L}_{{\rm pNRQCD}}^{(\text{NLO})} & = & \int_{\boldsymbol{r}}V_{{\rm A}}(r)\sqrt{\frac{T_{{\rm F}}}{N_{c}}}\left[S_{{\rm sin}}^{\dagger}\boldsymbol{r}\cdot g\boldsymbol{E}^{a}O_{{\rm sin}}^{a}+S^{\dagger,i}\boldsymbol{r}\cdot g\boldsymbol{E}^{a}O^{a,i}+{\rm H.c.}\right]\nonumber \\
 &  & +\int_{\boldsymbol{r}}\frac{V_{{\rm B}}(r)d^{abc}}{4T_{{\rm F}}}\left[O_{{\rm sin}}^{\dagger,a}\boldsymbol{r}\cdot g\boldsymbol{E}^{c}O_{{\rm sin}}^{b}+O^{\dagger,a,i}\boldsymbol{r}\cdot g\boldsymbol{E}^{c}O^{b,i}\right].\label{eq:NLO-pNRQCD-L}
\end{eqnarray}
Here and for what follows, the chromoelectric  and chromomagnetic fields are defined as $\boldsymbol{E}^{i}=F^{i0}$
and $\boldsymbol{B}^{i}=-\epsilon^{ijk}F^{jk}/2$ respectively. Such interactions contribute to $\mathcal{O}(\alpha_{{\rm s}}^{3})$
corrections to correlator $J^{ij}\propto\delta^{ij}$ up to leading
order. The leading-order contribution to $\Delta_{ij}^{\lambda\lambda^{\prime}}J^{ij}$
comes from the spin-dependent potential shown in eq.~\eqref{eq:Spin-dependent-potential}.
Since the electric-electric correlator $\langle\boldsymbol{E}\cdot\boldsymbol{E}\rangle\sim T^{4}$,
those color-singlet spin singlet/triplet vertex gives corrections
at least of $\mathcal{O}(\alpha_{{\rm s}}^{7})$ in the electric dipole
transition channel. As a comparison, the magnetic dipole transition
gives $\mathcal{O}(\alpha_{{\rm s}}^{5})$, therefore, the leading-order chromoelectric dipole interaction does not contribute up to the order we are interested in.

The next-to-leading order dipole transition consists of different
terms, but the spin-dependent vertices are the spin-magnetic coupling
and its minimal Lorentz invariant extension,
\begin{eqnarray}
\mathcal{L}_{\text{pNRQCD}}^{\text{spin}} & = & \int_{\boldsymbol{r}}{\rm Tr}{\rm S}^{\dagger}\left(\frac{c_{{\rm F}}V_{{\rm SO}b}^{(1,0)}}{2M}g\boldsymbol{B}(R)+\frac{c_{{\rm S}}V_{{\rm SO}a}^{(2,0)}[\boldsymbol{P}\times,g\boldsymbol{E}(R)]}{16M^{2}}\right)\cdot(\boldsymbol{\sigma}_{(1)}-\boldsymbol{\sigma}_{(2)}){\rm O}+{\rm H.c.}\nonumber \\
 & = & 2\sqrt{\frac{T_{{\rm F}}}{N_{c}}}\int_{\boldsymbol{r}}\left[S_{{\rm sin}}^{\dagger}\left(\frac{c_{{\rm F}}V_{{\rm SO}b}^{(1,0)}}{2M}g\boldsymbol{B}^{a}+c_{{\rm S}}V_{{\rm SO}a}^{(2,0)}\frac{[\boldsymbol{P}\times,g\boldsymbol{E}^{a}]}{16M^{2}}\right)^{i}O^{a,i}+{\rm H.c.}\right.\nonumber \\
 &  & \;\left.+S^{\dagger,i}\left(\frac{c_{{\rm F}}V_{{\rm SO}b}^{(1,0)}}{2M}g\boldsymbol{B}^{a}+c_{{\rm S}}V_{{\rm SO}a}^{(2,0)}\frac{[\boldsymbol{P}\times,g\boldsymbol{E}^{a}]}{16M^{2}}\right)^{i}O_{\text{sin}}^{a}+{\rm H.c.}\right],\label{eq:spin-dependent-pNRQCD-L}
\end{eqnarray}
where $[\boldsymbol{X}\times,\boldsymbol{Y}]=\boldsymbol{X}\times\boldsymbol{Y}-\boldsymbol{Y}\times\boldsymbol{X}$
and the above tree-level matching coefficients are consistent with
that constraint from Poincar\'e invariance \cite{Brambilla:2003nt,Berwein:2018fos}.
Here we keep leading recoil corrections to the leading spin-dependent
interactions but drop the terms at same order $\frac{V_{{\rm SO}c}^{(1,0)}}{2Mr^{2}}(\boldsymbol{r}\cdot\boldsymbol{B})\boldsymbol{r}\cdot(\boldsymbol{\sigma}^{(1)}-\boldsymbol{\sigma}^{(2)})$
as the matching coefficient $V_{{\rm SO}c}^{(1,0)}\sim\mathcal{O}(\alpha^{1})$
is beyond our interested order \cite{Brambilla:2003nt}. In NRQCD, $\frac{1}{M^{2}}[\boldsymbol{P}\times,g\boldsymbol{E}^{a}]$
originates from the spin-orbital coupling and is of same order as
spin-magnetic coupling $\frac{1}{M}\boldsymbol{\sigma}\cdot g\boldsymbol{B}$, so we also expect it to give contributions in pNRQCD.
In the end, we list the heavy quarkonium Feynman propagators with time-ordering $\mathcal{T}$ up to
leading order from eq.~\eqref{eq:LO_pNRQCD}: for the spin-singlet
and spin-triplet fields,
\begin{eqnarray}
\langle \mathcal{T}S_{\text{sin}}(R,\boldsymbol{r})S_{\text{sin}}^{\dagger}(R^{\prime},\boldsymbol{r}^{\prime})\rangle & = & \theta(t-t^{\prime})e^{-ih_{s}^{\text{sin}}(t-t^{\prime})}\delta^{(3)}(\boldsymbol{R}-\boldsymbol{R}^{\prime})\delta^{(3)}(\boldsymbol{r}-\boldsymbol{r}^{\prime})\\
 & = & \int_{p}e^{-ip\cdot(R-R^{\prime})}\frac{i}{p_{0}-h_{s}^{\text{sin}}(r;p)+i\epsilon} \delta^{(3)}(\boldsymbol{r}-\boldsymbol{r}^{\prime}),\label{eq:Spin-s-color-s_11GF}
\end{eqnarray}
and
\begin{eqnarray}
 &  & \langle \mathcal{T}S^{k}(R,\boldsymbol{r})S^{\dagger,j}(R^{\prime},\boldsymbol{r}^{\prime})\rangle\nonumber \\
 & = & \int_{P}e^{-iP\cdot(R-R^{\prime})}\frac{i}{P_{0}-h_{s,0}^{\text{tri}}(r;P)+i\epsilon}\left[\delta^{kj}+\mathcal{O}\left(\frac{1}{M^{2}}\right)\right] \delta^{(3)}(\boldsymbol{r}-\boldsymbol{r}^{\prime}),\label{eq:Spin-t-color-s_11GF}
\end{eqnarray}
where we keep only the vacuum part because the finite-$T$ contribution
is proportional to the density of heavy quarkonium and can be dropped
in the quarkonium transport equation in the dilute limit (linear order
in quarkonium density). Here we introduce the convention $\mathcal{T}(\bar{\mathcal{T}})$ for (anti-)time ordering operators.  Similarly for the color-octet part,
\begin{eqnarray}
\langle \mathcal{T}O_{\text{sin}}^{a}(R,\boldsymbol{r})O_{\text{sin}}^{\dagger,b}(R^{\prime},\boldsymbol{r}^{\prime})\rangle & = & \theta(t-t^{\prime})\left(\mathcal{T}e^{-i\int_{t^\prime}^t \text{d}t_1 h_o^{\text{sin}}(t_1,\boldsymbol{R},\boldsymbol{r})}\right)_{ab}\delta^{(3)}(\boldsymbol{r}-\boldsymbol{r}^{\prime})\delta^{(3)}(\boldsymbol{R}-\boldsymbol{R}^\prime),\nonumber\\\label{eq:Spin-s-color-o_11GF}\\
\langle \mathcal{T}O^{a,i}(R,\boldsymbol{r})O^{\dagger,b,j}(R^{\prime},\boldsymbol{r}^{\prime})\rangle & = & \delta^{ij}\theta(t-t^{\prime})\left(\mathcal{T}e^{-i\int_{t^\prime}^t \text{d}t_1 h_o^{\text{tri}}(t_1,\boldsymbol{R},\boldsymbol{r})}\right)_{ab}\delta^{(3)}(\boldsymbol{r}-\boldsymbol{r}^{\prime})\delta^{(3)}(\boldsymbol{R}-\boldsymbol{R}^\prime).\nonumber\\\label{eq:Spin-t-color-o_11GF}
\end{eqnarray}
For what follows, we remain at relative leading order in coupling and therefore the time-evolution kernels $\left(\mathcal{T}e^{-i\int_{t^\prime}^t \text{d}t_1 h_o^{\text{tri/sin}}(t_1,\boldsymbol{R},\boldsymbol{r})}\right)_{ab}$ reduces to $e^{-i(t-t^{\prime}) h_o^{\text{tri/sin}}(R,\boldsymbol{r})}\delta_{ab}$. It is equivalent to setting the Wilson line in the octet propagator to $U^{ab}(R,R^{\prime})=\delta^{ab}\delta^{(3)}(\boldsymbol{R}-\boldsymbol{R}')$. 
The spin-triplet correlator, eq.~\eqref{eq:Singlet_corr_fun_2}, is a
``12'' Wightman function.
In thermal equilibrium it is proportional to the quarkonium number density
according to Kubo-Martin-Schwinger relation. Shown in Appendix \ref{sec:Spin-dependent-density-matrix}, its non-equilibrium generalization is related to the density operator, 
\begin{eqnarray}
  \left\langle S_{i}^{\dagger}(R,\boldsymbol{r})S_{j}(R^{\prime},\boldsymbol{r}^{\prime})\right\rangle 
 & = & \langle\boldsymbol{R}^\prime,\boldsymbol{r}^{\prime}|\hat{\rho}_{s}^{ji}(t^{\prime};t)|\boldsymbol{R},\boldsymbol{r}\rangle,\label{eq:Singlet_12_correlator}
\end{eqnarray}
where $\hat{\rho}_s^{ji}$ is the spin-dependent spin-triplet density operator defined in eq.~\eqref{eq:Spin-dependent-DO}. It is the spin-dependent generalization of eq. (23) of \cite{Brambilla:2017zei} and is verified 
in Appendix \ref{sec:Spin-dependent-density-matrix}. Eq.~\eqref{eq:Singlet_12_correlator} implicitly assumes that we work with a single heavy quark-antiquark pair in the dilute limit. The tree-level ``12'' propagators whose dynamics is governed only by the system Hamiltonian, represented by $\langle...\rangle^{0}$, are, e.g. for $S_{i}$ field, 
\begin{eqnarray}
\langle S_{i}^{\dagger}(R,\boldsymbol{r})S_{j}(R^{\prime},\boldsymbol{r}^{\prime})\rangle^{\text{0}} & = & \langle\boldsymbol{R}^{\prime},\boldsymbol{r}^{\prime}|e^{-ih_{s}^{{\rm tri}}(t^{\prime}-t_{0})}\hat{\rho}_{s,0}^{ji}(t_0;t_0)e^{ih_{s}^{{\rm tri}}(t-t_{0})}|\boldsymbol{R},\boldsymbol{r}\rangle.\label{eq:Tree-level-12-propagator}
\end{eqnarray}
According to \cite{New1}, the ``12'' Wightman function for a spin-singlet color-octet
field correlator is not well-defined, instead, our master equations deal with correlators that transform as a color-singlet. However, we do not delve further into this, as we are interested in the leading-order relative contribution. Such diagonal color-octet density operator in the color space is
\begin{align}
    \mathrm{Tr}\left[O^{a \dagger}(R,\boldsymbol{r})O^a(R^{\prime},\boldsymbol{r}^{\prime})\hat{\rho}\right]=\langle\boldsymbol{R}^{\prime},\boldsymbol{r}^{\prime}|\hat\rho_{o}(t^{\prime};t)|\boldsymbol{R},\boldsymbol{r}\rangle,\label{eq:Tree-level-Octet-12-propagator}
\end{align}
Lastly, with eq.~\eqref{eq:Singlet_12_correlator}, we now express the spin density matrix elements of eq.~(\ref{eq:Normalized-SDM}) in terms of the full quarkonium density operator. The
Wigner-transformed color-singlet heavy-current correlator (\ref{eq:Singlet_corr_fun_2}) becomes
\begin{eqnarray}
 &  & \int_{\frac{R+R^{\prime}}{2}}J_{ij}\left(q,\frac{R+R^{\prime}}{2}\right)\nonumber \\
 & = & 2\int_{t_{0}}^{+\infty}{\rm d}t{\rm d}t^{\prime}e^{i(2M-q_{0})(t-t^{\prime})}\sum_{\lambda,\lambda^{\prime}}\boldsymbol{\epsilon}_{i}^{\lambda,*}\boldsymbol{\epsilon}_{j}^{\lambda^{\prime}}\sum_{n,n^{\prime}}\phi_{n,\lambda}^{*}(\boldsymbol{0})\phi_{n^{\prime},\lambda^{\prime}}(\boldsymbol{0})\langle n^{\prime},\lambda^{\prime},\boldsymbol{q}|\hat{\rho}_{s}^{{\rm tri}}(t^{\prime};t)|n,\lambda,\boldsymbol{q}\rangle,\nonumber \\
 & \approx & 2\sum_{\lambda,\lambda^{\prime}}\sum_{n,n^{\prime}}\boldsymbol{\epsilon}_{i}^{\lambda,*}\boldsymbol{\epsilon}_{j}^{\lambda^{\prime}}\tilde{\phi}_{n,\lambda}^{*}(\boldsymbol{0})\tilde{\phi}_{n^{\prime},\lambda^{\prime}}(\boldsymbol{0})\langle n^{\prime},\lambda^{\prime},\boldsymbol{q}|\hat{\rho}_{s}^{{\rm tri}}(t_{{\rm F}};t_{{\rm F}})|n,\lambda,\boldsymbol{q}\rangle,
\end{eqnarray}
where we exploit the fact that the evolution of quarkonium after freeze-out
dominates the whole stage of the evolution and substitute the integral
lower bound $t_{0}\to t_{{\rm F}}$. Here $\tilde{\phi}_{n,\lambda_{1}}(\boldsymbol{0})=\phi_{n,\lambda_{1}}(\boldsymbol{0})(2M-q_{0}+E_{n\lambda_{1}}^{\boldsymbol{q}*})^{-1}$
is the mass weighted color-singlet bound-state wavefunction at the origin. $|n,\lambda,\boldsymbol{q}\rangle$ is the bound-state eigenstate with discrete quantum number $\{n,\lambda\}$ and continuous center-of-mass momentum $\boldsymbol{q}$. The spin-triplet color-singlet density operator is related to the spin-dependent one as shown in Appendix \ref{sec:Spin-dependent-density-matrix}. For the quarkonia in a specific state
characterized by quantum number $n$ such as $\Upsilon(1S)$, the
normalized spin density matrix elements of eq.~(\ref{eq:Normalized-SDM}) reduce to
\begin{eqnarray}
 &  & \delta\rho_{\lambda\lambda^{\prime}}^{n,\rm N}(q)=\left(\int_{X}\delta\rho_{\lambda\lambda^{\prime}}^{n}(q,X)\right)\left/\left(\int_X\text{Tr}\rho_{\lambda\lambda^{\prime}}^{n}(q,X)\right)\right.\nonumber \\
 & \!\!\!= & \varepsilon_{\lambda}^{i}(q)\varepsilon_{\lambda^{\prime}}^{*j}(q)\frac{\underset{\lambda_{1},\lambda_{2}}{\sum}\boldsymbol{\epsilon}_{i}^{\lambda_{2},*}\boldsymbol{\epsilon}_{j}^{\lambda_{1}}\tilde{\phi}_{n,\lambda_{1}}(\boldsymbol{0})\tilde{\phi}_{n,\lambda_{2}}^{*}(\boldsymbol{0})\langle n,\lambda_{1},\boldsymbol{q}|\hat{\rho}_{s}^{{\rm tri}}(t_{{\rm F}};t_{{\rm F}})|n,\lambda_{2},\boldsymbol{q}\rangle}{\underset{\lambda_{3}}{\sum}|\tilde{\phi}_{n,\lambda_{3}}(\boldsymbol{0})|^{2}\langle n,\lambda_{3},\boldsymbol{q}|\hat{\rho}_{s}^{{\rm tri}}(t_{{\rm F}};t_{{\rm F}})|n,\lambda_{3},\boldsymbol{q}\rangle}-\frac{\delta_{\lambda\lambda^{\prime}}}{d-1}, \label{eq: SDME-DO-form}
\end{eqnarray}
where we use the definition in eq.~\eqref{eq:Density_matrix}. We emphasize that eq.~\eqref{eq: SDME-DO-form} is defined in a general frame but we will work in the quarkonium rest frame such that we can eliminate the momentum dependence in the polarization vectors $\varepsilon_{\lambda}^i$. Since only the bound-state density operator enters here, we will concentrate on bound-state evolution in the following.

\section{Evolution of heavy-quarkonium spin density matrix in a moving medium\protect\label{sec:Evolution-of-heavy-quarkonium}}

In this section, we derive the evolution equation of heavy-quarkonium
spin distribution function. We start from the integral form of the Kadanoff-Baym equation
up to leading spin-dependent multipole (inverse mass) expansion in
pNRQCD, we derive an equation of motion for the spin density operator
and we project it on bound states to obtain Boltzmann transport equations
in both quarkonium and medium rest frames afterwards. We focus on the zeroth-order contributions in the fluid-gradient expansion.

\subsection{Derivation of quantum master equation \protect\label{subsec:Dyson-Schwinger-equation}}

As shown in eqs.~\eqref{eq:Singlet_12_correlator} and \eqref{eq:Tree-level-Octet-12-propagator},
the ``12'' Wightman functions are proportional to the quarkonium density
operator, and we aim to derive its evolution equation from that of
$\langle S_{i}^{\dagger}(R,\boldsymbol{r})S_{j}(R^{\prime},\boldsymbol{r}^{\prime})\rangle$.
We note that in eq.~\eqref{eq:Delta_SDME}, only the correlation function
subtracted by $\delta_{ij}$ part  contributes to $\delta\rho_{\lambda\lambda^{\prime}}$,
so we focus on the following traceless propagator by subtracting terms proportional to $\delta_{ij}$. Due to the discussion in Section~\ref{subsec:Potential-NRQCD-with}, the electric
dipole transition contributes to $\lambda_{\theta}$ only with propagators
modified by spin-dependent potential and is suppressed compared with
magnetic dipole transition. Inserting the tree-level propagators and working in the dilute limit
in heavy quarkonium number, only 3 diagrams are important
 and then the leading-order traceless propagator
with minimal center-of-mass ($\boldsymbol{P}$-dependent) effects
is,
\begin{eqnarray}
    & & \delta\langle S_{i}^{\dagger}(R,\boldsymbol{r})S_{j}(R^{\prime},\boldsymbol{r}^{\prime})\rangle  =  \langle S_{i}^{\dagger}(R,\boldsymbol{r})S_{j}(R^{\prime},\boldsymbol{r}^{\prime})\rangle-\frac{\delta_{ji}}{d-1}\langle S_{l}^{\dagger}(R,\boldsymbol{r})S_{l}(R^{\prime},\boldsymbol{r}^{\prime})\rangle\nonumber\\
    & = & \delta\langle S_{i}^{\dagger}(R,\boldsymbol{r})S_{j}(R^{\prime},\boldsymbol{r}^{\prime})\rangle^{0}\nonumber\\
    &  & +\vcenter{\hbox{\begin{tikzpicture}
\begin{feynman}
\vertex (a);
\vertex at ($(a)+(2cm,0cm)$) (c1);
\vertex at ($(a)+(4.0cm,0cm)$) (d);
\tikzfeynmanset{every vertex={dot,minimum size=2mm}}
\vertex at ($(a)+(1.2cm,0cm)$) (b);
\vertex at ($(b)+(0.05cm,0.84cm)$) (m1) {\(q\)};
\vertex at ($(a)+(0cm,-0.4cm)$) (v1) {\(1\)};
\vertex at ($(a)+(0.6cm,0.24cm)$) (s3) {\(P\)};
\vertex at ($(b)+(0cm,-0.4cm)$) (v2) {\(1\)};
\tikzfeynmanset{every vertex={dot,minimum size=2mm}}
\vertex at ($(a)+(2.8cm,0cm)$) (c);
\tikzfeynmanset{every vertex={blob,minimum size=5mm}}
\vertex at ($(a)+(2cm,0.8cm)$) (c2);
\vertex at ($(c1)+(0.0cm,0.24cm)$) (s4) {\(P-q\)};
\vertex at ($(c)+(0cm,-0.4cm)$) (v3) {\(1\)};
\vertex at ($(d)+(0cm,-0.4cm)$) (v4) {\(2\)};
\diagram [layered layout]{
(a)--[insertion=0.5](b)--[double,insertion=1](c1)--[double](c)--[line width=2pt](d),
(b)--[gluon,quarter left](c2)--[gluon,quarter left](c),
};
\end{feynman}
\end{tikzpicture} 
}}
+
\vcenter{\hbox{
\begin{tikzpicture}
\begin{feynman}
\vertex (a);
\vertex at ($(a)+(2cm,0cm)$) (c1);
\vertex at ($(a)+(4.0cm,0cm)$) (d);
\tikzfeynmanset{every vertex={dot,minimum size=2mm}}
\vertex at ($(a)+(1.2cm,0cm)$) (b);
\vertex at ($(b)+(0.05cm,0.84cm)$) (m1) {\(q\)};
\vertex at ($(a)+(0cm,-0.4cm)$) (v1) {\(1\)};
\vertex at ($(a)+(0.6cm,0.24cm)$) (s3) {\(P\)};
\vertex at ($(b)+(0cm,-0.4cm)$) (v2) {\(1\)};
\tikzfeynmanset{every vertex={dot,minimum size=2mm}}
\vertex at ($(a)+(2.8cm,0cm)$) (c);
\tikzfeynmanset{every vertex={blob,minimum size=5mm}}
\vertex at ($(a)+(2cm,0.8cm)$) (c2);
\vertex at ($(c1)+(0.0cm,0.24cm)$) (s4) {\(P-q\)};
\vertex at ($(c)+(0cm,-0.4cm)$) (v3) {\(2\)};
\vertex at ($(d)+(0cm,-0.4cm)$) (v4) {\(2\)};
\diagram [layered layout]{
(a)--[insertion=0.5](b)--[double,line width=1pt](c1)--[double,line width=1pt](c)--[insertion=0.5](d),
(b)--[gluon,quarter left](c2)--[gluon,quarter left](c),
};
\end{feynman}
\end{tikzpicture}
}}
+
\vcenter{\hbox{
\begin{tikzpicture}
\begin{feynman}
\vertex (a);
\vertex at ($(a)+(2cm,0cm)$) (c1);
\vertex at ($(a)+(4.0cm,0cm)$) (d);
\tikzfeynmanset{every vertex={dot,minimum size=2mm}}
\vertex at ($(a)+(1.2cm,0cm)$) (b);
\vertex at ($(b)+(0.05cm,0.84cm)$) (m1) {\(q\)};
\vertex at ($(a)+(0cm,-0.4cm)$) (v1) {\(1\)};
\vertex at ($(a)+(0.6cm,0.24cm)$) (s3) {\(P\)};
\vertex at ($(b)+(0cm,-0.4cm)$) (v2) {\(2\)};
\tikzfeynmanset{every vertex={dot,minimum size=2mm}}
\vertex at ($(a)+(2.8cm,0cm)$) (c);
\tikzfeynmanset{every vertex={blob,minimum size=5mm}}
\vertex at ($(a)+(2cm,0.8cm)$) (c2);
\vertex at ($(c1)+(0.0cm,0.24cm)$) (s4) {\(P-q\)};
\vertex at ($(c)+(0cm,-0.4cm)$) (v3) {\(2\)};
\vertex at ($(d)+(0cm,-0.4cm)$) (v4) {\(2\)};
\diagram [layered layout]{
(a)--[line width=2pt](b)--[double,insertion=1](c1)--[double](c)--[insertion=0.5](d),
(b)--[gluon,quarter left](c2)--[gluon,quarter left](c),
};
\end{feynman}
\end{tikzpicture}
}}\nonumber
\end{eqnarray}
\begin{eqnarray}
\label{eq:Diagrammatic-rep-QME}
\end{eqnarray}
\begin{figure}
    \centering   
\begin{eqnarray*}
\vcenter{\hbox{\begin{tikzpicture}
\begin{feynman}
\vertex (a);
\vertex at ($(a)+(0cm,-0.2cm)$) (a1) {1};
\vertex at ($(a)+(1.2cm,0cm)$) (c);
\vertex at ($(c)+(0cm,-0.2cm)$) (c1) {2};
\diagram [layered layout]{
(a)--[line width=1.5pt](c),
};
\end{feynman}
\end{tikzpicture}
}}
&=& e^{-i h_{s}^{\rm tri} (t^{\prime}-t_0)} \hat{\rho}^{ji}_{s,0}(t_0,t_0)e^{i h_{s}^{\rm tri} (t-t_0)},\quad 
\vcenter{\hbox{\begin{tikzpicture}
\begin{feynman}
\vertex (a);
\vertex at ($(a)+(0cm,-0.2cm)$) (a1) {1};
\vertex at ($(a)+(1.2cm,0cm)$) (c);
\vertex at ($(c)+(0cm,-0.2cm)$) (c1) {2};
\diagram [layered layout]{
(a)--[double,line width=1.0pt](c),
};
\end{feynman}
\end{tikzpicture}
}}
= e^{-i h_{o}^{\rm sin} (t^{\prime}-t_0)} \hat{\rho}^{\rm sin}_{o,0}(t_0,t_0)e^{i h_{o}^{\rm sin} (t-t_0)}, \\    
\\
\vcenter{\hbox{\begin{tikzpicture}
\begin{feynman}
\vertex (a);
\vertex at ($(a)+(0cm,-0.2cm)$) (a1) {1};
\vertex at ($(a)+(1.2cm,0cm)$) (c);
\vertex at ($(c)+(0cm,-0.2cm)$) (c1) {1};
\diagram [layered layout]{
(a)--[insertion=0.5](c),
};
\end{feynman}
\end{tikzpicture}
}} 
&=& \theta(t-t^{\prime}) \delta^{(3)}(\boldsymbol{R}-\boldsymbol{R}^{\prime})\delta^{(3)}(\boldsymbol{r}-\boldsymbol{r}^{\prime}) e^{-ih_s^{\rm sin}(t-t^{\prime})},\\
\\
\vcenter{\hbox{\begin{tikzpicture}
\begin{feynman}
\vertex (a);
\vertex at ($(a)+(0cm,-0.2cm)$) (a1) {2};
\vertex at ($(a)+(1.2cm,0cm)$) (c);
\vertex at ($(c)+(0cm,-0.2cm)$) (c1) {2};
\diagram [layered layout]{
(a)--[insertion=0.5](c),
};
\end{feynman}
\end{tikzpicture}
}} 
&=& \theta(t^{\prime}-t) \delta^{(3)}(\boldsymbol{R}-\boldsymbol{R}^{\prime})\delta^{(3)}(\boldsymbol{r}-\boldsymbol{r}^{\prime}) e^{-ih_s^{\rm sin}(t-t^{\prime})},
\end{eqnarray*}

\begin{eqnarray*}
\begin{tikzpicture}
\begin{feynman}
\vertex (a);
\vertex at ($(a)+(2.0cm,0cm)$) (c);
\vertex at ($(a)+(1cm, 1cm)$) (d);
\tikzfeynmanset{every vertex={dot,minimum size=2mm}}
\vertex at ($(a)+(1cm,0cm)$) (b);
\diagram [layered layout]{
(a)--(b)--[double](c),
(b)--[gluon](d)
};
\end{feynman}
\end{tikzpicture}
&=&
\begin{tikzpicture}
\begin{feynman}
\vertex (a);
\vertex at ($(a)+(2.0cm,0cm)$) (c);
\vertex at ($(a)+(1cm, 1cm)$) (d);
\vertex at ($(a)+(1.5cm,0.5cm)$) (n) {\(\frac{gB^a_i}{2M}\)};
\tikzfeynmanset{every vertex={crossed dot,minimum size=2mm}}
\vertex at ($(a)+(1cm,0cm)$) (b);
\diagram [layered layout]{
(a)--(b)--[double](c),
(b)--[gluon](d)
};
\end{feynman}
\end{tikzpicture}
+
\begin{tikzpicture}
\begin{feynman}
\vertex (a);
\vertex at ($(a)+(2.0cm,0cm)$) (c);
\vertex at ($(a)+(1cm, 1cm)$) (d);
\vertex at ($(a)+(1.9cm,0.5cm)$) (n) {\(\frac{[P\times,gE^a]_i}{16M^2}\)};
\tikzfeynmanset{every vertex={crossed dot,minimum size=2mm,rotate=45}}
\vertex at ($(a)+(1cm,0cm)$) (b);
\diagram [layered layout]{
(a)--(b)--[double](c),
(b)--[gluon](d)
};
\end{feynman}
\end{tikzpicture}
\end{eqnarray*}
 \caption{
 Pictorial illustration of Feynman rules for propagators and vertices which enter the leading spin-dependent collision kernel up to relative order $T/M$ recoil corrections.
Double line: color-octet propagator; single line: color-singlet
propagator; spring line: gluon propagator; bold line: 12 propagator
expressed by the density operator. The cross insertion: propagator
with recoil corrections. The gluon propagator with blob insertion
denotes the full propagator at finite temperature. The black dotted
vertex includes both chromomagnetic and chromoelectric dipole transition.}
    \label{fig:Dipole-transition}
\end{figure}
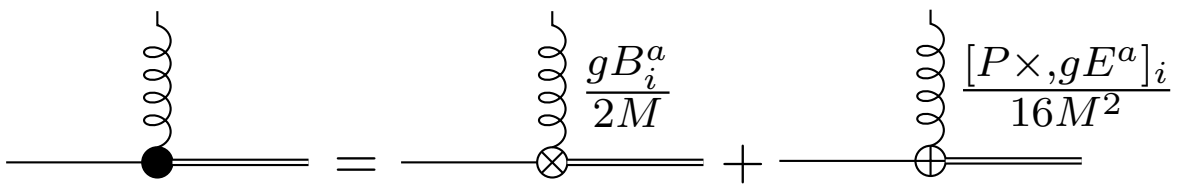

Here the tree-level Wightman ``12'' function $\langle...\rangle^0$ is defined in eq.~\eqref{eq:Tree-level-12-propagator}. In what follows, the collision terms should be understood as the projected part under $\Theta_{ijmn}=\delta_{im}\delta_{jn}-\frac{1}{d-1}\delta_{mn}\delta_{ij}$ in accordance with the l.h.s. of the quantum master equation~ \eqref{eq:Diagrammatic-rep-QME}. In deriving the operator-level equation \eqref{eq:Diagrammatic-rep-QME}, we integrate over $\boldsymbol{R},\boldsymbol{R}^{\prime},\boldsymbol{r},\boldsymbol{r}^{\prime}$
on both sides with multiplying $\delta^{(3)}(\boldsymbol{R}-\boldsymbol{R}^{\prime})\delta^{(3)}(\boldsymbol{r}-\boldsymbol{r}^{\prime})$,
use the identity
\begin{eqnarray}
\int_{\boldsymbol{R},\boldsymbol{r},\boldsymbol{R}^{\prime},\boldsymbol{r}^{\prime}}\delta^{(3)}(\boldsymbol{R}^{\prime}-\boldsymbol{R})\delta^{(3)}(\boldsymbol{r}^{\prime}-\boldsymbol{r})\langle\boldsymbol{R}^{\prime},\boldsymbol{r}^{\prime}|F(\boldsymbol{R},\boldsymbol{r})|\boldsymbol{R},\boldsymbol{r}\rangle & = & {\rm Tr}F(\hat{\boldsymbol{R}},\hat{\boldsymbol{r}}),
\end{eqnarray}
and extract the kernel on both sides. As discussed in Section \ref{sec:Heavy-quarkonium-spin}, we will focus on the leading non-zero contribution to quarkonium polarization in the weakly-coupled limit so the Wilson-line insertions are replaced by unit operators.

  With Feynman rules displayed in Figure \ref{fig:Dipole-transition}, each diagram in eq.~\eqref{eq:Diagrammatic-rep-QME} is represented by operator functionals:
\begin{eqnarray}
\vcenter{\hbox{
\begin{tikzpicture}
\begin{feynman}
\vertex (a);
\vertex at ($(a)+(2cm,0cm)$) (c1);
\vertex at ($(a)+(4.0cm,0cm)$) (d);
\tikzfeynmanset{every vertex={dot,minimum size=2mm}}
\vertex at ($(a)+(1.2cm,0cm)$) (b);
\vertex at ($(b)+(0.05cm,0.84cm)$) (m1) {\(q\)};
\vertex at ($(a)+(0cm,-0.4cm)$) (v1) {\(1\)};
\vertex at ($(a)+(0.6cm,0.24cm)$) (s3) {\(P\)};
\vertex at ($(b)+(0cm,-0.4cm)$) (v2) {\(1\)};
\tikzfeynmanset{every vertex={dot,minimum size=2mm}}
\vertex at ($(a)+(2.8cm,0cm)$) (c);
\tikzfeynmanset{every vertex={blob,minimum size=5mm}}
\vertex at ($(a)+(2cm,0.8cm)$) (c2);
\vertex at ($(c1)+(0.0cm,0.24cm)$) (s4) {\(P-q\)};
\vertex at ($(c)+(0cm,-0.4cm)$) (v3) {\(1\)};
\vertex at ($(d)+(0cm,-0.4cm)$) (v4) {\(2\)};
\diagram [layered layout]{
(a)--[insertion=0.5](b)--[double,insertion=1](c1)--[double](c)--[line width=2pt](d),
(b)--[gluon,quarter left](c2)--[gluon,quarter left](c),
};
\end{feynman}
\end{tikzpicture} 
}}
& = & \;-\int_{t_{0}}^{t}{\rm d}t_{1}e^{-ih_{s}^{{\rm tri}}(t-t_{1})}\hat{\Sigma}_{s,jm}(t_{1})e^{-ih_{s}^{{\rm tri}}(t_{1}-t_{0})}\hat{\rho}_{s,0}^{mi}(t_{0};t_{0})e^{ih_{s}^{{\rm tri}}(t-t_{0})},\nonumber\\ \\
\vcenter{\hbox{
\begin{tikzpicture}
\begin{feynman}
\vertex (a);
\vertex at ($(a)+(2cm,0cm)$) (c1);
\vertex at ($(a)+(4.0cm,0cm)$) (d);
\tikzfeynmanset{every vertex={dot,minimum size=2mm}}
\vertex at ($(a)+(1.2cm,0cm)$) (b);
\vertex at ($(b)+(0.05cm,0.84cm)$) (m1) {\(q\)};
\vertex at ($(a)+(0cm,-0.4cm)$) (v1) {\(1\)};
\vertex at ($(a)+(0.6cm,0.24cm)$) (s3) {\(P\)};
\vertex at ($(b)+(0cm,-0.4cm)$) (v2) {\(1\)};
\tikzfeynmanset{every vertex={dot,minimum size=2mm}}
\vertex at ($(a)+(2.8cm,0cm)$) (c);
\tikzfeynmanset{every vertex={blob,minimum size=5mm}}
\vertex at ($(a)+(2cm,0.8cm)$) (c2);
\vertex at ($(c1)+(0.0cm,0.24cm)$) (s4) {\(P-q\)};
\vertex at ($(c)+(0cm,-0.4cm)$) (v3) {\(2\)};
\vertex at ($(d)+(0cm,-0.4cm)$) (v4) {\(2\)};
\diagram [layered layout]{
(a)--[insertion=0.5](b)--[double,line width=1pt](c1)--[double,line width=1pt](c)--[insertion=0.5](d),
(b)--[gluon,quarter left](c2)--[gluon,quarter left](c),
};
\end{feynman}
\end{tikzpicture}
}}
& = &\;\int_{t_{0}}^{t}{\rm d}t_{1}e^{-ih_{s}^{{\rm tri}}(t-t_{1})}\hat{\Xi}_{s,ji}\left(\rho_{o,0}^{{\rm sin}}(t_{0};t_{0});t_{1}\right)e^{ih_{s}^{{\rm tri}}(t-t_{1})},\\\nonumber\\
\vcenter{\hbox{
\begin{tikzpicture}
\begin{feynman}
\vertex (a);
\vertex at ($(a)+(2cm,0cm)$) (c1);
\vertex at ($(a)+(4.0cm,0cm)$) (d);
\tikzfeynmanset{every vertex={dot,minimum size=2mm}}
\vertex at ($(a)+(1.2cm,0cm)$) (b);
\vertex at ($(b)+(0.05cm,0.84cm)$) (m1) {\(q\)};
\vertex at ($(a)+(0cm,-0.4cm)$) (v1) {\(1\)};
\vertex at ($(a)+(0.6cm,0.24cm)$) (s3) {\(P\)};
\vertex at ($(b)+(0cm,-0.4cm)$) (v2) {\(2\)};
\tikzfeynmanset{every vertex={dot,minimum size=2mm}}
\vertex at ($(a)+(2.8cm,0cm)$) (c);
\tikzfeynmanset{every vertex={blob,minimum size=5mm}}
\vertex at ($(a)+(2cm,0.8cm)$) (c2);
\vertex at ($(c1)+(0.0cm,0.24cm)$) (s4) {\(P-q\)};
\vertex at ($(c)+(0cm,-0.4cm)$) (v3) {\(2\)};
\vertex at ($(d)+(0cm,-0.4cm)$) (v4) {\(2\)};
\diagram [layered layout]{
(a)--[line width=2pt](b)--[double,insertion=1](c1)--[double](c)--[insertion=0.5](d),
(b)--[gluon,quarter left](c2)--[gluon,quarter left](c),
};
\end{feynman}
\end{tikzpicture}
}}
& = &\;-\int_{t_{0}}^{t}{\rm d}t_{1}e^{-ih_{s}^{{\rm tri}}(t-t_{0})}\hat{\rho}_{s,0}^{jk}(t_{0};t_{0})e^{ih_{s}^{{\rm tri}}(t_{1}-t_{0})}\hat{\Sigma}_{s,ki}^{\dagger}(t_{1})e^{-ih_{s}^{{\rm tri}}(t_{1}-t)},\nonumber\\
\end{eqnarray}
where the self-energy
and transition operators are defined as 
\begin{eqnarray}
\hat{\Sigma}_{s,jm}(t_{1}) & = & -i^{2}g^{2}\left(2\sqrt{\frac{T_{{\rm F}}}{N_{c}}}\right)^{2}\int_{t_{0}}^{t_{1}}{\rm d}t_{2}\theta(t_{1}-t_{2})\nonumber \\
 &  & \;\times\left\langle \mathcal{T}\left(\frac{1}{2M}\boldsymbol{B}^{a}(\hat{\boldsymbol{R}},t_{1})+\frac{[\hat{\boldsymbol{P}}\times,\boldsymbol{E}^{a}(\hat{\boldsymbol{R}},t_{1})]}{16M^{2}}\right)_{j}\right.\nonumber \\
 &  & \;\left.e^{-ih_{o}^{{\rm sin}}(t_{1}-t_{2})}\left(\frac{1}{2M}\boldsymbol{B}^{a}(\hat{\boldsymbol{R}},t_{2})+\frac{[\hat{\boldsymbol{P}}\times,\boldsymbol{E}^{a}(\hat{\boldsymbol{R}},t_{2})]}{16M^{2}}\right)_{m}\right\rangle e^{ih_{s}^{{\rm tri}}(t_{1}-t_{2})},\label{eq:SE}
\end{eqnarray}
and 
\begin{eqnarray}
 &  & \hat{\Xi}_{s,ji}\left(\rho_{o,0}^{{\rm sin}}(t_{0};t_{0});t_{1}\right)\nonumber \\
 & = & -\frac{i^{2}g^{2}}{N_{c}^{2}-1}\left(2\sqrt{\frac{T_{{\rm F}}}{N_{c}}}\right)^{2}\int_{t_{0}}^{t_{1}}{\rm d}t_{2}\theta(t_{1}-t_{2})\nonumber \\
 &  & \quad\times\biggl\langle \bar{\mathcal{T}}\left(\frac{1}{2M}\boldsymbol{B}^{a}(\hat{\boldsymbol{R}},t_{1})+\frac{[\hat{\boldsymbol{P}}\times,\boldsymbol{E}^{a}(\hat{\boldsymbol{R}},t_{1})]}{16M^{2}}\right)_{j}e^{-ih_{o}^{{\rm sin}}(t_{1}-t_{0})}\hat{\rho}_{o,0}^{{\rm sin}}(t_{0};t_{0})\nonumber \\
 &  & \quad\times e^{ih_{o}^{{\rm sin}}(t_{2}-t_{0})}\left(\frac{1}{2M}\boldsymbol{B}^{a}(\hat{\boldsymbol{R}},t_{2})+\frac{[\hat{\boldsymbol{P}}\times,\boldsymbol{E}^{a}(\hat{\boldsymbol{R}},t_{2})]}{16M^{2}}\right)_{i}\biggr\rangle e^{ih_{s}^{{\rm tri}}(t_{1}-t_{2})}+{\rm H.c.}\label{eq:Transition-operator}
\end{eqnarray}
Here both $\hat{\boldsymbol{P}}$ and spatial arguments $\hat{\boldsymbol{R}}$ are operators in the Hilbert space.

Differentiation with time $t$ leads to 
\begin{eqnarray}
\frac{{\rm d}}{{\rm d}t}\delta\hat{\rho}_{s}^{ji}(t;t) & = & -i\left[h_{s}^{{\rm tri}},\delta\hat{\rho}_{s}^{ji}(t;t)\right]-\hat{\Sigma}_{s,jm}(t)e^{-ih_{s}^{{\rm tri}}(t-t_{0})}\hat{\rho}_{s}^{mi}(t_{0};t_{0})e^{ih_{s}^{{\rm tri}}(t-t_{0})}\nonumber \\
 &  & -e^{-ih_{s}^{{\rm tri}}(t-t_{0})}\hat{\rho}_{s}^{jk}(t_{0};t_{0})e^{ih_{s}^{{\rm tri}}(t-t_{0})}\hat{\Sigma}_{s,ki}^{\dagger}(t)+\hat{\Xi}_{s,ji}\left(\rho_{o,0}^{{\rm sin}}(t_{0};t_{0});t\right).\label{eq:QME-temp}
\end{eqnarray}
where $\delta\hat{\rho}^{ji}_{s}=\hat{\rho}^{ji}_{s}-\delta^{ji}\hat{\rho}^{kk}_{s}/(d-1)$. Within the Born-Markovian treatment, we replace 
\begin{eqnarray}
\hat{\rho}_{o,0}^{{\rm sin}}(t)=e^{-ih_{o}^{{\rm sin}}(t-t_{0})}\hat{\rho}_{o}^{\text{sin}}(t_{0};t_{0})e^{ih_{o}^{{\rm sin}}(t-t_{0})} & \to & \hat{\rho}_{o}^{{\rm sin}}(t),\label{eq:BM-approximation}
\end{eqnarray}
in the collision kernel up to the retained perturbative order. This is equivalent to neglecting the higher-order corrections from system-environment
interactions to $\hat{\rho}_{s/o}$. We arrive at the quantum master
equation, 
\begin{eqnarray}
\frac{{\rm d}}{{\rm d}t}\delta\hat{\rho}_{s}^{ji}(t) & = & -i\left[h_{s}^{{\rm tri}},\delta\hat{\rho}_{s}^{ji}(t)\right]-\hat{\Sigma}_{s,jm}(t)\hat{\rho}_{s}^{mi}(t)-\hat{\rho}_{s}^{jk}(t)\hat{\Sigma}_{s,ki}^{\dagger}(t)\nonumber \\
 &  & \quad+\hat{\Xi}_{s,ji}\left(\rho_{o}^{{\rm sin}}(t);t\right),\label{eq:QME}
\end{eqnarray}
where the transition operator as a functional of full density operator
$\hat{\rho}_{o}^{{\rm sin}}(t)$ is obtained by making the replacement in eq.~\eqref{eq:BM-approximation}. For simplicity we will denote $\rho(t)\equiv\rho(t;t)$ from now on. We remark that a simple jump operator in Ref.~\cite{Brambilla:2017zei} cannot be defined here because $\hat{h}_o^{\rm sin}$ does not commute with $\hat{\boldsymbol{R}}$ and therefore we cannot isolate the environment operator from subsystem Hamiltonian. We can nonetheless observe such factorization in the Boltzmann equation instead in the next subsection.

\subsection{Boltzmann equation of spin distribution function\protect\label{subsec:Boltzmann-equation-of}}

We believe the center-of-mass effects are easier to depict using a phase-space
formulation of the quantum master equation~\eqref{eq:QME}. This can be
achieved by sandwiching the operator equation with states characterized
by specific quantum numbers. Before that, we note the traceless
density operator $\delta\hat{\rho}_{s}^{ji}(t)$ is a small correction suppressed by $1/M^2$
to its trace part in $\hat{\rho}_{s}^{ji}(t)$ as illustrated in Appendix \ref{sec:Spin-dependent-density-matrix}, and we thus
approximate in the collision term 
\begin{eqnarray}
\hat{\rho}_{s}^{ji}(t) & \approx & \frac{1}{d-1}\delta^{ji}\hat{\rho}_{s}^{{\rm tri}}(t),\label{eq:small-polar-approx}
\end{eqnarray}
and eq.~\eqref{eq:QME} now reads
\begin{eqnarray}
\frac{{\rm d}}{{\rm d}t}\delta\hat{\rho}_{s}^{ji}(t)+i\left[h_{s}^{{\rm tri}},\delta\hat{\rho}_{s}^{ji}(t)\right] & = & -\frac{1}{d-1}\hat{\Sigma}_{s,ji}(t)\hat{\rho}_{s}^{{\rm tri}}(t)-\frac{1}{d-1}\hat{\rho}_{s}^{{\rm tri}}(t)\hat{\Sigma}_{s,ji}^{\dagger}(t)\nonumber \\
 &  & \;+\hat{\Xi}_{s,ji}\left(\rho_{o}^{{\rm sin}}(t);t\right).\label{eq:QME-2}
\end{eqnarray}
This means the dynamics of the spin-dependent density operator is determined
by the subsystem Hamiltonian and by the dynamics of the unpolarized density operator
$\{\hat{\rho}_{s}^{{\rm tri}},\hat{\rho}_{o}^{{\rm sin}}\}$. The
quantum master equations of the latter have already been derived in~\citep{Brambilla:2017zei}.
We remark that the evolution of $\hat{\rho}_{o}^{{\rm sin}}$ is the same
as $\hat{\rho}_{o}^{{\rm tri}}$ because their Hamiltonians are the same
in the leading order and of dipole interactions as in eqs.~\eqref{eq:LO_pNRQCD} and \eqref{eq:NLO-pNRQCD-L}. The only difference is their different initial conditions from hard
processes. Another observation from eq.~\eqref{eq:QME-2} is that neglecting all the fluid-gradient fields up to leading gradient expansion, the
leading magnetic-magnetic correlator alone cannot induce nonzero polarization
since 
\begin{eqnarray}
\Delta_{\lambda\lambda^{\prime}}^{ij}\hat{\Sigma}_{s,ji}\sim\Delta_{\lambda\lambda^{\prime}}^{ij}\langle B^{j}B^{i}\rangle & \propto & \delta^{ji}\Delta_{\lambda\lambda^{\prime}}^{ij}=0.
\end{eqnarray}
This implies that the comoving effects from the gluon distribution function in a
moving medium $f_{{\rm B}}(p)=1\left/\left[\exp(\beta u\cdot p)-1\right]\right.$ with $u^{\mu}$ the medium fluid velocity
and the quarkonium center-of-mass motion are actually the leading
contribution to the quarkonium polarization. 

As discussed in the last subsection, such quantum master equation is not suitable for practical use because of the $\hat{\boldsymbol{R}}$ dependence in the self-energy operators. On the other hand, the separation of energy scales in the physical system allows us to obtain a well-defined distribution function for the bound states. Therefore, we will implement a semiclassical approximation and work with the Boltzmann equation in this work. 

According to eq.~\eqref{eq:Quantum_optical_regime}, the rotating-wave
approximation can be applied: the evolution time $t\sim\tau_{{\rm R}}\gg\tau_{{\rm S}}\sim(E_{n}-E_{n^{\prime}})^{-1}$
so the phase-factor $e^{i(E_{n}-E_{n^{\prime}})t}$
oscillates fast and gives zero on average unless $n^{\prime}=n$, 
\begin{eqnarray}
{}_{s}\langle n\lambda;\boldsymbol{P}^\prime|\hat{\rho}_{s}^{\text{tri}}(t)|n^{\prime}\lambda^{\prime};\boldsymbol{P}\rangle_{s} & \approx & \delta_{nn^{\prime}} {}_{s}\langle n\lambda;\boldsymbol{P}^\prime|\hat{\rho}_{s}^{\text{tri}}(t)|n\lambda^{\prime};\boldsymbol{P}\rangle_{s} ,\label{eq:RW-approx-fs}
\end{eqnarray}
where eq.~\eqref{eq:RW-approx-fs} is equivalent to working in the diagonal limit of the bound-state density operator $\hat{\rho}_s^{\rm{tri}}$. On the other hand, it is inappropriate to perform uniform  secular averaging of all relative-momentum coherence for the scattering states, because the intrinsic oscillation rate $\tau_{\rm{S}}^{-1}\sim(E_{\boldsymbol{p}}-E_{\boldsymbol{p}^\prime})$ can be arbitrarily small due to its gapless excitation. Therefore, as opposed to Refs.~\cite{Yao:2020eqy,Akamatsu:2020ypb,Hammou:2026ecw}, we will not perform $\boldsymbol{r}$ expansion for the scattering-state distribution function as will be shown later. In other words, we consider the coherence effects encoded in off-diagonal momentum-space octet density operator $\left\langle\boldsymbol{p}_1|\hat{\rho}_o|\boldsymbol{p}_2\right\rangle$ in this work.

We now define the following real bound-state spin-averaged
and spin-dependent distribution functions,
\begin{eqnarray}
f_{n\lambda}^{s}(t,\boldsymbol{x},\boldsymbol{P}_{{\rm cm}}) & = & \int_{\boldsymbol{P}_{{\rm rel}}}e^{i\boldsymbol{P}_{{\rm rel}}\cdot\boldsymbol{x}}{}_{s}\langle n\lambda;\boldsymbol{P}_{{\rm cm}}+\frac{\boldsymbol{P}_{{\rm rel}}}{2}|\hat{\rho}_{s}^{{\rm tri}}(t)|n\lambda;\boldsymbol{P}_{{\rm cm}}-\frac{\boldsymbol{P}_{{\rm rel}}}{2}\rangle_{s},\label{eq:Qkn-df}\\
\delta f_{ji,n\lambda}^{s}(t,\boldsymbol{x},\boldsymbol{P}_{{\rm cm}}) & = & \int_{\boldsymbol{P}_{{\rm rel}}}e^{i\boldsymbol{P}_{{\rm rel}}\cdot\boldsymbol{x}}{}_{s}\langle n\lambda;\boldsymbol{P}_{{\rm cm}}+\frac{\boldsymbol{P}_{{\rm rel}}}{2}|\delta\hat{\rho}_{s}^{ji}(t)|n\lambda;\boldsymbol{P}_{{\rm cm}}-\frac{\boldsymbol{P}_{{\rm rel}}}{2}\rangle_{s},\label{eq:Qkn-sdf-2}
\end{eqnarray}
and the scattering-state spin-singlet distribution function,
\begin{eqnarray}
f_{{\rm sin}}^{o}(t,\boldsymbol{x},\boldsymbol{P}_{{\rm cm}};\boldsymbol{r},\boldsymbol{p}_{{\rm cm}}) & = & \int_{\boldsymbol{P}_{{\rm rel}}}\int_{\boldsymbol{p}_{{\rm rel}}}e^{i\boldsymbol{P}_{{\rm rel}}\cdot\boldsymbol{x}}e^{i\boldsymbol{p}_{{\rm rel}}\cdot\boldsymbol{r}}{}_{o}\langle\boldsymbol{p}_{{\rm cm}}+\frac{\boldsymbol{p}_{{\rm rel}}}{2};\boldsymbol{P}_{{\rm cm}}+\frac{\boldsymbol{P}_{{\rm rel}}}{2}|\hat{\rho}_{o}^{{\rm sin}}(t)\nonumber \\
 &  & \quad|\boldsymbol{p}_{{\rm cm}}-\frac{\boldsymbol{p}_{{\rm rel}}}{2};\boldsymbol{P}_{{\rm cm}}-\frac{\boldsymbol{P}_{{\rm rel}}}{2}\rangle_{o}.\label{eq:Scattering-state-distribution}
\end{eqnarray}
Eq.~\eqref{eq:Qkn-sdf-2} can be reduced to $f_{n\lambda}^{s}(t,\boldsymbol{x},\boldsymbol{P}_{{\rm cm}})$,
the overlap of $\hat{\rho}_{s}^{{\rm tri}}(t)$, with the definition in eq.~\eqref{eq:Spin-dependent-DO}, 
\begin{eqnarray}
\delta f_{ji,n\lambda}^{s}(t,\boldsymbol{x},\boldsymbol{P}_{{\rm cm}}) & = & \left(\boldsymbol{\epsilon}_{i}^{\lambda,*}\boldsymbol{\epsilon}_{j}^{\lambda}-\frac{\delta^{ji}}{d-1}\right)f_{n\lambda}^{s}(t,\boldsymbol{x},\boldsymbol{P}_{{\rm cm}}).
\end{eqnarray}
The bound-state rotating-wave approximation, eq.~\eqref{eq:RW-approx-fs}, does not limit the off-diagonal matrix elements in spin magnetic quantum number $\lambda$ in general. But to make use of the spin distribution function~\eqref{eq:Qkn-df} and its evolution equation, we neglect the off-diagonal matrix elements in eq.~\eqref{eq: SDME-DO-form} by setting $\lambda_1=\lambda_2$ similar to Ref.~\citep{Yang:2024ejk}. Together with the conventional rotating-wave approximation~\eqref{eq:RW-approx-fs},  eq.~\eqref{eq: SDME-DO-form} can now be rewritten as 
\begin{eqnarray}
\delta\rho_{\lambda\lambda^{\prime}}^{n,\text{N}}(q) & = & \varepsilon_{\lambda}^{i}(q)\varepsilon_{\lambda^{\prime}}^{*j}(q)\frac{\int_{\boldsymbol{R}_{1}}\underset{\lambda_{1}}{\sum}|\tilde{\phi}_{n,\lambda_{1}}(\boldsymbol{0})|^{2}\delta f_{ji,n\lambda_{1}}^{s}(t_{{\rm F}},\boldsymbol{R}_{1},\boldsymbol{q})}{\int_{\boldsymbol{R}_{2}}\underset{\lambda_{2}}{\sum}|\tilde{\phi}_{n,\lambda_{2}}(\boldsymbol{0})|^{2}f_{n\lambda_{2}}^{s}(t_{{\rm F}},\boldsymbol{R}_{2},\boldsymbol{q})}.\label{eq:SDME-sdf}
\end{eqnarray}
We emphasize that we do $not$ impose $\delta_{\lambda\lambda^{\prime}}$ as a result of eq.~\eqref{eq:RW-approx-fs} when deducing the Boltzmann equation.

The derivation of the Boltzmann equation also involves the projection
of operators like $e^{-ih_{s}^{{\rm tri}}(t-t_{1})}$ on the eigenstates,
e.g.
\begin{eqnarray}
h_{s}^{{\rm tri}}|n\lambda;\boldsymbol{P}\rangle_{s}=\left(E_{n}^{s}+\frac{\boldsymbol{P}^{2}}{4M}\right)|n\lambda;\boldsymbol{P}\rangle_{s} & ,\quad & h_{o}^{{\rm sin}}|\boldsymbol{p};\boldsymbol{P}\rangle_{o}=\left(\frac{\boldsymbol{p}^{2}}{M}+\frac{\boldsymbol{P}^{2}}{4M}\right)|\boldsymbol{p};\boldsymbol{P}\rangle_{o}.
\end{eqnarray}
The field operators like $\boldsymbol{B}(\hat{\boldsymbol{R}},t)$
in eqs.~(\ref{eq:SE})-(\ref{eq:Transition-operator}) are functionals
of Hilbert-space operators $\hat{\boldsymbol{R}}$, and to convert
them to c-number correlators, we need to insert the following complete
basis
\begin{eqnarray}
1 & = & \int_{\boldsymbol{R},\boldsymbol{r}}|\boldsymbol{R},\boldsymbol{r}\rangle\langle\boldsymbol{R},\boldsymbol{r}|.\label{eq:Unit-operator-r}
\end{eqnarray}
The following completeness relation will also be useful, 
\begin{eqnarray}
1 & = & \int_{\boldsymbol{P}}\left(\sum_{n\lambda}|n\lambda;\boldsymbol{P}\rangle_{s/o}{\,}_{s/o}\langle n\lambda;\boldsymbol{P}|+\int_{\boldsymbol{p}}\sum_{\lambda}|\boldsymbol{p}\lambda;\boldsymbol{P}\rangle_{s/o}{\,}_{s/o}\langle\boldsymbol{p}\lambda;\boldsymbol{P}|\right).\label{eq:Unit-operator-3}
\end{eqnarray}
We define the bound-state and scattering-state wave functions up to
leading order for later usage, 
\begin{eqnarray}
\langle\boldsymbol{R},\boldsymbol{r}|n\lambda;\boldsymbol{P}\rangle_{s} & = & e^{i\boldsymbol{P}\cdot\boldsymbol{R}}\phi_{n\lambda}(\boldsymbol{r})|\lambda\rangle_{s},\label{eq:Singlet-WF}\\
\langle\boldsymbol{R},\boldsymbol{r}|\boldsymbol{p};\boldsymbol{P}\rangle_{o} & = & e^{i\boldsymbol{P}\cdot\boldsymbol{R}}\Phi_{\boldsymbol{p}}(\boldsymbol{r}).\label{eq:Octet-WF}
\end{eqnarray}

We project equation~(\ref{eq:QME-2}) on bound states $\langle n\lambda;\boldsymbol{P}^{\prime}|$
and $|n\lambda;\boldsymbol{P}\rangle$ and perform a Wigner transform
by multiplying with $e^{i(\boldsymbol{P}^{\prime}-\boldsymbol{P})\cdot\boldsymbol{x}}$
and integrating over $\boldsymbol{P}^{\prime}-\boldsymbol{P}$. 
With $\boldsymbol{P}^{\prime}-\boldsymbol{P}\to-i\boldsymbol{\nabla}_{\boldsymbol{x}}$, we obtain
\begin{eqnarray}
 &  & \int_{\boldsymbol{P}^{\prime}-\boldsymbol{P}}e^{i(\boldsymbol{P}^{\prime}-\boldsymbol{P})\cdot\boldsymbol{x}}\langle n,\lambda;\boldsymbol{P}^{\prime}|i\left[h_{s}^{{\rm tri}},\delta\hat{\rho}_{s}^{ji}(t)\right]|n,\lambda;\boldsymbol{P}\rangle\nonumber \\
 & = & \int_{\boldsymbol{P}^{\prime}-\boldsymbol{P}}e^{i(\boldsymbol{P}^{\prime}-\boldsymbol{P})\cdot\boldsymbol{x}}i\frac{\boldsymbol{P}^{\prime,2}-\boldsymbol{P}^{2}}{4M}\langle n,\lambda;\boldsymbol{P}^{\prime}|\delta\hat{\rho}_{s}^{ji}(t)|n,\lambda;\boldsymbol{P}\rangle\nonumber \\
 & = & \frac{\boldsymbol{P}_{{\rm cm}}\cdot\boldsymbol{\nabla}_{\boldsymbol{x}}}{2M}\delta f_{ji,n\lambda}^{s}(t,\boldsymbol{x},\boldsymbol{P}_{{\rm cm}}),
\end{eqnarray}
and find that the left-hand side of eq.~\eqref{eq:QME-2} now becomes, 
\begin{equation}
\left(\frac{{\rm d}}{{\rm d}t}+\frac{\boldsymbol{P}_{{\rm cm}}\cdot\boldsymbol{\nabla}_{\boldsymbol{x}}}{2M}\right)\delta f_{ji,n\lambda}^{s}(t,\boldsymbol{x},\boldsymbol{P}_{{\rm cm}}).\label{eq:BE-lhs}
\end{equation} 
We are only interested in the leading-order corrections on both sides of the Boltzmann equation. The higher-order potentials in eq.~\eqref{eq:Spin-dependent-potential}, including spin-independent potentials beyond the Coulomb term and spin-dependent potential, only contribute the streaming part of the Boltzmann equations. Because we are only interested in the leading corrections on both sides of the Boltzmann equation, the first term of eq.~\eqref{eq:Spin-dependent-potential} gives an $\mathcal{O}(M^{-4})$ effective potential for spin-triplet fields through a self-energy diagram; thus it is  suppressed compared with leading recoil corrections $\boldsymbol{P}^2/M^2$ and can be neglected. The second term of eq.~\eqref{eq:Spin-dependent-potential} is the higher-order potential compared to the leading-order $P$-independent, the Coulomb potential,  and its first-order traceless diagonal matrix element vanishes by angular-momentum selection rule e.g. for the $S$-wave states. So it is not important after projection on states $\langle n\lambda;\boldsymbol{P}|$ and $| n\lambda;\boldsymbol{P}^{\prime}\rangle$ in the streaming part of the Boltzmann equation.

The derivation of the right-handed side is listed in Appendix \ref{Sec:Calculation-of-collision-kernel}. One critical step is the implementation of the rotating-wave approximation \eqref{eq:RW-approx-fs}. Keeping only the matrix elements of $\hat{\rho}_s^{\rm tri}$ diagonal in bound-state energy labels  manifests this secular approximation. 
Together with the small-polarization approximation~\eqref{eq:small-polar-approx}, only $f^{s}_{n\lambda}$ and general $f^{o}_{\rm sin}$ enter on the right-hand side. The resulting
Boltzmann equation now reads
\begin{eqnarray}
& &\left(\frac{{\rm d}}{{\rm d}t}+\frac{\boldsymbol{P}_{{\rm cm}}\cdot\boldsymbol{\nabla}_{\boldsymbol{x}}}{2M}\right)\left[\Delta_{\lambda_1\lambda_2}^{ji}\delta f_{ji,n\lambda}^{s}(t,\boldsymbol{x},\boldsymbol{P}_{{\rm cm}})\right]\nonumber\\ & = & \Delta_{\lambda_1\lambda_2}^{ji}\left(-\mathcal{C}_{ji,n\lambda}^{-,\rm mrf}[f_{n\lambda}^{s}](t,\boldsymbol{x},\boldsymbol{P}_{{\rm cm}})+\mathcal{C}_{ji,n\lambda}^{+, \rm mrf}[f_{{\rm sin}}^{o}](t,\boldsymbol{x},\boldsymbol{P}_{{\rm cm}}).\right)\label{eq:SBE-mrf}
\end{eqnarray}
Throughout the following discussions, we project the spin kinetic equations onto the selected spin-density matrix elements. Here the dissociation and regeneration parts of the collision kernel
are defined as 
\begin{eqnarray}
 &  & \Delta_{\lambda_1\lambda_2}^{ji}\mathcal{C}_{ji,n\lambda}^{-,\rm mrf}[f_{n\lambda}^{s}](t,\boldsymbol{x},\boldsymbol{P}_{{\rm cm}})\nonumber \\
 & = & \Delta_{\lambda_1\lambda_2}^{ji}\frac{g^{2}T_{{\rm F}}}{M^{2}(d-1)N_{c}}2{\rm Re}\int_{\boldsymbol{p}}\left|_{o}\langle\boldsymbol{p}|n\rangle_{s}\right|^{2}\nonumber \\
 &  & \quad\times\int_{q}\frac{i}{\Delta E_{n,p}^{so}+\frac{2\boldsymbol{P}_{{\rm cm}}\cdot\boldsymbol{q}-\boldsymbol{q}^{2}}{4M}-q_{0}+i\epsilon}\exp\left(-\frac{i}{2}\overleftarrow{\boldsymbol{\nabla}}_{\boldsymbol{P}_{{\rm cm}}}\cdot\boldsymbol{\nabla}_{\boldsymbol{x}}\right)\nonumber \\
 &  & \quad\times\left[\tilde{\mathcal{G}}_{11}^{ji}(q,P_{{\rm cm}})+\left(\epsilon^{iqp}\tilde{G}_{BE,11}^{jq}(q)+\epsilon^{jqp}\tilde{G}_{EB,11}^{qi}(q)\right)\frac{i\boldsymbol{\nabla}_{\boldsymbol{x}}^{p}}{8M}\right]f_{n\lambda}^{s}(t,\boldsymbol{x},\boldsymbol{P}_{{\rm cm}})\label{eq:CK-disso-1}
\end{eqnarray}
and 
\begin{eqnarray}
 &  & \Delta_{\lambda_1\lambda_2}^{ji}\mathcal{C}_{ji,n\lambda}^{+,\rm mrf}[f_{{\rm sin}}^{o}](t,\boldsymbol{x},\boldsymbol{P}_{{\rm cm}})\nonumber \\
 & = & +\Delta_{\lambda_1\lambda_2}^{ji}\frac{g^{2}T_{{\rm F}}}{M^{2}(N_{c}^{2}-1)N_{c}}2{\rm Re}\int_{q}\int_{\boldsymbol{p}_{1},\boldsymbol{p}_{2}}\left[_{s}\langle n|\boldsymbol{p}_{1}\rangle_{o}\right]\left[_{o}\langle\boldsymbol{p}_{2}|n\rangle_{s}\right]\nonumber \\
 &  & \quad\times\frac{i}{\Delta E_{n,p_{2}}^{so}+\frac{2\boldsymbol{P}_{{\rm cm}}\cdot\boldsymbol{q}-\boldsymbol{q}^{2}}{4M}-q_{0}+i\epsilon}\exp\left(\frac{i}{2}\overleftarrow{\boldsymbol{\nabla}}_{\boldsymbol{P}_{{\rm cm}}}\cdot\boldsymbol{\nabla}_{\boldsymbol{x}}\right)\nonumber \\
 &  & \quad\times\left[\tilde{\mathcal{G}}_{22}^{ji}(q,P_{{\rm cm}})+\left(\epsilon^{imn}\tilde{G}_{BE,22}^{jn}(q)-\epsilon^{jmn}\tilde{G}_{EB,22}^{ni}(q)\right)\frac{i\boldsymbol{\nabla}_{\boldsymbol{x}}^{m}}{8M}\right]\nonumber \\
 &  & \quad\times\int_{\boldsymbol{r}}e^{-i(\boldsymbol{p}_{1}-\boldsymbol{p}_{2})\cdot\boldsymbol{r}}f_{{\rm sin}}^{o}(t,\boldsymbol{x},\boldsymbol{P}_{{\rm cm}}-\boldsymbol{q};\boldsymbol{r},\frac{\boldsymbol{p}_{1}+\boldsymbol{p}_{2}}{2}).\label{eq:CK-regen-1}
\end{eqnarray}
The argument $\boldsymbol{P}_{{\rm cm}}$ in eq.~\eqref{eq:SBE-mrf}
is the heavy-quarkonium center-of-mass momentum in the medium-rest frame.
We introduce the energy difference $\Delta E_{n,p}^{so}=E_{n}^{s}-E_{p}^{o}=E_{n}^{s}-\frac{\boldsymbol{p}^{2}}{M}$.
We plug in the completeness relations eqs.(\ref{eq:Unit-operator-r})
and (\ref{eq:Unit-operator-3}) in the derivation, 
and the transition amplitude
$\langle\boldsymbol{p}(\lambda^{\prime})|\hat{\rho}(t)|n(\lambda)\rangle$
vanishes according to the rotating wave approximation. The wavefunction
overlap $_{s}\langle n|\boldsymbol{p}\rangle_{o}$ effectively
accounts for the effects for the multiple soft gluon exchange between
heavy quark and antiquark. We also take $t_{0}\to-\infty$ when performing
time integral.

The momentum-space gauge-field correlator in the medium-rest frame is defined with a tilde as
\begin{eqnarray}
\widetilde{G}_{BB,11}^{ji}\left(q,\frac{x+y}{2}\right) & = & \int_{x-y}e^{iq\cdot(x-y)}\langle \mathcal{T}\boldsymbol{B}_{{\rm rf}}^{a,j}(x)\boldsymbol{B}_{{\rm rf}}^{a,i}(y)\rangle.\label{eq:p-space-GF-correlator}
\end{eqnarray}
We are interested in the zeroth order in the fluid-gradient expansion in this work, and thus
the medium will be considered as being in local equilibrium at $\left\{u_{\mu}(X),T(X)\right\}$ during
the quarkonium evolution and derivatives of  $X=\frac{x+y}{2}$ of order $\mathcal{O}(\rm Kn^1)$ are neglected. Similarly, the $G_{22}$ represents the correlator with anti-time ordering and $G_{12},G_{21}$ represents the Wightman correlators. The subscript ``rf'' stands for ``medium rest frame''. The chromo magnetic-magnetic correlator
with recoil correction is
\begin{eqnarray}
\mathcal{G}_{11}^{ji}(q,P) & = & G_{BB,11}^{ji}(q)+\epsilon^{irs}\frac{q^{s}-2P^{s}}{8M}G_{BE,11}^{jr}(q)+\epsilon^{jpq}\frac{q^{q}-2P^{q}}{8M}G_{EB,11}^{pi}(q)\nonumber \\
 &  & \;+\epsilon^{irs}\epsilon^{jpq}\frac{q^{q}-2P^{q}}{8M}\frac{q^{s}-2P^{s}}{8M}G_{EE,11}^{pr}(q).\label{eq:Gauge-field-correlator}
\end{eqnarray}
We keep the relative $\mathcal{O}(\frac{T}{M})$ corrections in the
correlators in eqs.~(\ref{eq:CK-disso-1}) and (\ref{eq:CK-regen-1})
compared to the leading $G_{BB}^{ji}$. We assume the spatial derivative $\boldsymbol{P}_{{\rm rel}}=i\boldsymbol{\nabla}_{\boldsymbol{x}}\sim q$
is of the magnitude of the typical energy of hot gluons $\mathcal{O}(T)$
if $e^{iq_0(t-t_{0})}\sim\mathcal{O}(1)$, and we
thus kept only the linear $\frac{\boldsymbol{\nabla}_{\boldsymbol{x}}}{M}$
terms in the correlators. We comment that the spatial gradient on quarkonium distribution function $\nabla_{\boldsymbol{x}}$ is different from the hydrodynamic gradient $\partial_{X}$: the latter comes from the medium scattering length and arranged by Knudsen number expansion, and the scale of $\nabla_{\boldsymbol{x}}$ in this work is determined by  dynamical 
terms in the correlators. We focus only on the effects from the zeroth order in the fluid-gradient expansion, i.e. $\mathcal{O}(\rm Kn^0)$ and therefore effects from fluid vorticity or thermal shear are neglected. Here we summarize the energy scales of different derivative operators and momentum appearing in the Boltzmann equation in Table \ref{tab:energy-scales}.

\begin{table}[htbp]
    \centering    \renewcommand{\arraystretch}{1.4}
    \begin{tabularx}{\textwidth}{
        @{} |c |X| %>{\raggedright\arraybackslash}X @{}
    }
        \hline
        Physical quantities/operators   & Energy scaling  \\
        \hline
        \hline
         Quarkonium center-of-mass momentum kick $\boldsymbol{P}_{\rm rel}$        & $\sim T$ \\

        Quarkonium  center-of-mass momentum $\boldsymbol{P}_{\rm cm}\sim \left(\boldsymbol{\nabla}_{\boldsymbol{P}_{\rm cm}}\right)^{-1}$          & $\sim \sqrt{MT}$\\

         Spatial gradient of quarkonium distribution function $\boldsymbol{\nabla_x}$           &$\sim T$\\     

         Typical hot-gluon momentum $q$   & $\sim T$\\

         Dipolar transition energy gap $\Delta E^{so}_{n,p}$ & $\sim Mv^2$\\

         Scattering-state relative momentum $\frac{\boldsymbol{p}_1+\boldsymbol{p}_2}{2}$ and $\boldsymbol{p}_1-\boldsymbol{p}_2$ & $\sim Mv$\\

         Fluid-field gradient $\partial_{X}$ & $\sim \text{Kn} \;T$\\
\hline
        \hline
    \end{tabularx}
    \caption{Comparison of typical scales of operators and quantities in the heavy-quarkonium spin Boltzmann equation. Energy scales are expressed in terms of the relative velocity $v$, heavy quark mass $M$, the hot-medium temperature $T$ and hydrodynamical Knudsen number $\text{Kn}=\frac{l_{\rm hydro}}{L_{\rm hydro}}$. The hierarchy of in-medium heavy quarkonium energy scales is listed in eq.~\eqref{eq:Energy-Hierarchy}.}
    \label{tab:energy-scales}
\end{table}

The collision term only depends on the quarkonium number distribution
function whose evolution acts as a ``background''. The first term
$-\mathcal{C}_{ji,n\lambda}^{-,\rm mrf}[f_{n\lambda}^{s}]$ from the self-energy
projection is basically a rate equation describing the damping of
color-singlet quarkonium due to gluon exchange, part of which was
investigated in Ref.~\citep{Chen:2025mrf} (only influence of magnetic-field
correlator was calculated), and the second term $\mathcal{C}_{ji,n\lambda}^{+,\rm mrf}[f_{\text{sin}}^{o}]$
from transition operator projection describes regeneration contribution
from in-medium color-octet scattering state. Similar to the quarkonium
number transport \citep{Yao:2018nmy,Yao:2020eqy}, the spin collision
term here is also of factorized form: the QGP background is described by the gauge-field correlators,
the system-environment coupling is weak, and the separated subsystem dynamics is described by off-equilibrium
distributions. The wavefunction
overlap can be calculated perturbatively by solving the Schr\"{o}dinger
equation. Here we present the leading-order treatment of gluon-field correlators and a non-perturbative expression with chromofields dressed by Wilson lines with recoil corrections is beyond the scope of this work (also see Wilson-line discussions in Ref.~\cite{Yang:2024ejk} in the recoilless case). Interestingly, the Moyal-star $\star=\exp\left(\frac{i}{2}\overleftarrow{\boldsymbol{\nabla}}_{\boldsymbol{P}_{{\rm cm}}}\cdot\boldsymbol{\nabla}_{\boldsymbol{x}}-\frac{i}{2}\overleftarrow{\boldsymbol{\nabla}}_{\boldsymbol{x}}\cdot\boldsymbol{\nabla}_{\boldsymbol{P}_{{\rm cm}}}\right)$
structure that automatically emerges indicates a \textit{semiclassical
expansion} was implicitly performed in deriving the spin transport
equation. Roughly speaking, the recoil color-singlet propagator
$\frac{i}{q_{0}-\Delta E+i\epsilon}$ and the remaining parts in eqs.(\ref{eq:CK-disso-1})
and (\ref{eq:CK-regen-1}) can be rearranged as $S\star\Sigma$ with
$S$ and $\Sigma$ being quarkonium propagator and self energy. So
the $\boldsymbol{x}$ gradient contributing at relative order $\frac{T}{M}$
gives leading quantum corrections~\citep{Blaizot:2001nr} (also see
Ref.~\citep{Fang:2023bbw} for the role of Moyal product in spin transport). 

To make a comparison with the previous work in Ref.~\cite{Yang:2024ejk},
we only keep the classical parts of eqs.~(\ref{eq:CK-disso-1}) and
(\ref{eq:CK-regen-1}) by dropping all $\boldsymbol{\nabla}_{\boldsymbol{x}}$
terms on the r.h.s. Writing the gluon propagators in terms of ``12''
and ``21'' real-time components, the Boltzmann equation (\ref{eq:SBE-mrf})
now reads 
\begin{eqnarray}
 &  & \left(\frac{{\rm d}}{{\rm d}t}+\frac{\boldsymbol{P}_{{\rm cm}}}{2M}\cdot\boldsymbol{\nabla}_{\boldsymbol{R}}\right)\left[\Delta_{\lambda_1\lambda_2}^{ji}\delta f_{ji,n\lambda}^{s}(t,\boldsymbol{R},\boldsymbol{P}_{{\rm cm}})\right]\nonumber \\
 & = & \Delta_{\lambda_1\lambda_2}^{ji}\frac{g^{2}T_{{\rm F}}}{M^{2}N_{c}}{\rm Re}\int_{q}\int_{\boldsymbol{p}_{1},\boldsymbol{p}_{2}}\left[_{s}\langle n|\boldsymbol{p}_{1}\rangle_{o}\right]\left[_{o}\langle\boldsymbol{p}_{2}|n\rangle_{s}\right]2\pi\delta\left(q_{0}-\frac{\boldsymbol{P}_{{\rm cm}}\cdot\boldsymbol{q}}{2M}+\frac{\boldsymbol{q}^{2}}{4M}-\Delta E_{n,\boldsymbol{p}_{1}}^{so}\right)\nonumber \\
 &  & \;\times\left\{ \widetilde{\mathcal{G}}_{12}^{ji}(q,P_{{\rm cm}})\int_{\boldsymbol{r}}e^{-i(\boldsymbol{p}_{1}-\boldsymbol{p}_{2})\cdot\boldsymbol{r}}\frac{f_{{\rm sin}}^{o}(t,\boldsymbol{R},\boldsymbol{P}_{{\rm cm}}-\boldsymbol{q};\boldsymbol{r},\frac{\boldsymbol{p}_{1}+\boldsymbol{p}_{2}}{2})}{N_{c}^{2}-1}\right.\nonumber \\
 &  & \quad\left.-(2\pi)^3\delta^{(3)}(\boldsymbol{p}_{1}-\boldsymbol{p}_{2})\widetilde{\mathcal{G}}_{21}^{ji}(q,P_{{\rm cm}})\frac{f_{n\lambda}^{s}(t,\boldsymbol{R},\boldsymbol{P}_{{\rm cm}})}{d-1}\right\} .\label{eq:Classical-SBE}
\end{eqnarray}
where we used fact that the momentum correlators $\widetilde{\mathcal{G}}_{21}^{ji},\widetilde{\mathcal{G}}_{12}^{ji}$
are real and ${\rm Re}\frac{i}{x+i\epsilon}=\pi\delta(x)$. We find
the energy-conservation law during dissociation/recombination processes
are automatically satisfied. 
When subtracting the non-polarization part, eq.~\eqref{eq:Classical-SBE}
is consistent with the collision terms eqs. (4.20), (4.21) of  Ref.~\cite{Yang:2024ejk} where different definitions of distribution
functions are used. Moreover, our transport equation retains the recoil corrections of leading relative order $\frac{T}{M}$ that contribute to the leading spin-dependent collision term. Such recoil corrections to the gluon-field
correlators and gluon energies are
also included; our general equation (\ref{eq:SBE-mrf}) also
includes the leading quantum corrections to the collision kernel due
to Moyal product. Lastly, an intriguing property of our Boltzmann equation is that the relative-momentum coherence is kept explicitly, in contrast to the $\boldsymbol{r}$-expansion employed in Refs.~\citep{Yao:2020eqy,Akamatsu:2020ypb} (see also the recent work~\cite{Hammou:2026ecw}). Such an expansion requires a separation between the characteristic scale of $\Delta\boldsymbol{r}=\boldsymbol{r}-\boldsymbol{r}_0$ and the $\boldsymbol{r}$-spatial variation scale of the distribution functions, whereas invoking the semiclassical approximation alone
does not establish this scale separation. Moreover, as we will see in section~\ref{sec:Numerical-estimation}, the contributions from off-diagonal density matrix elements in $\boldsymbol{p}$ space are not negligible compared to those from the diagonal parts.

\subsection{Boltzmann equation in the quarkonium rest frame\protect\label{subsec:Boltzmann-equation-in}}

The physical observables are measured and calculated in the quarkonium-rest
frame, so it is also advantageous to derive the transport equation
in such frame. However, it is insufficient to simply set $\boldsymbol{P}_{{\rm cm}}=0$
in the spin Boltzmann equation (\ref{eq:SBE-mrf}), as the gluonic correlators $\mathcal{G}_{\alpha\beta}^{ji}$
now also contain information of relative motion between quarkonium and
medium.  In this
subsection, we perform a Lorentz transformation to gluon fields to include such relative-motion effects. Expressing the gluonic correlators in the medium-rest frame is also motivated by prospective extensions involving nonperturbative input from lattice simulations. A similar strategy was used in previous works
investigating $\phi$ meson spin alignment in a thermal medium \cite{Sheng:2022ffb}.

We project the quantum master equation (\ref{eq:QME}) on the states
$\langle n\lambda;\frac{\boldsymbol{P}_{{\rm rel}}}{2}|...|n\lambda;-\frac{\boldsymbol{P}_{{\rm rel}}}{2}\rangle$, Wigner transform it on both sides
and similarly obtain
\begin{eqnarray}
\frac{{\rm d}}{{\rm d}t}\left(\Delta_{\lambda_1\lambda_2}^{ji}\delta f_{ji,n\lambda}^{s}(t,\boldsymbol{x})\right) & = & \Delta_{\lambda_1\lambda_2}^{ji}\left(-\mathcal{C}_{ji,n\lambda}^{-,{\rm qrf}}[f_{n\lambda}^{s}](t,\boldsymbol{x})+\mathcal{C}_{ji,n\lambda}^{+,{\rm qrf}}[f_{{\rm sin}}^{o}](t,\boldsymbol{x})\right)\label{eq:SBE-qrf}
\end{eqnarray}
with 
\begin{eqnarray}
 &  & \Delta_{\lambda_1\lambda_2}^{ji}\mathcal{C}_{ji,n\lambda}^{-,{\rm qrf}}[f_{n\lambda}^{s}](t,\boldsymbol{x})\nonumber \\
 & = & \Delta_{\lambda_1\lambda_2}^{ji}\frac{g^{2}T_{{\rm F}}}{M^{2}(d-1)N_{c}}2{\rm Re}\int_{\boldsymbol{p}_{2}}\left|_{o}\langle\boldsymbol{p}_{2}|n\rangle_{s}\right|^{2}\int_{q}\frac{i}{\Delta E_{n,p_{2}}^{so}-\frac{\boldsymbol{q}^{2}}{4M}-q_{0}+i\epsilon}\left(1+\frac{i}{2}\overleftarrow{\boldsymbol{\nabla}}_{\boldsymbol{q}}\cdot\boldsymbol{\nabla}_{\boldsymbol{x}}\right)\nonumber \\
 &  & \;\times\left[\mathcal{G}_{11}^{ji}(q)+\left(\epsilon^{imn}G_{BE,11}^{jn}(q)+\epsilon^{jmn}G_{EB,11}^{ni}(q)\right)\frac{-i\boldsymbol{\nabla}_{\boldsymbol{x}}^{m}}{8M}\right]f_{n\lambda}^{s}(t,\boldsymbol{x})\label{eq:CK-disso-2}
\end{eqnarray}
and 
\begin{eqnarray}
 &  & \Delta_{\lambda_1\lambda_2}^{ji}\mathcal{C}_{ji,n\lambda}^{+,{\rm qrf}}[f_{{\rm sin}}^{o}](t,\boldsymbol{x})\nonumber \\
 & = & \Delta_{\lambda_1\lambda_2}^{ji}\frac{g^{2}T_{{\rm F}}}{M^{2}(N_{c}^{2}-1)N_{c}}2{\rm Re}\int_{q}\int_{\boldsymbol{p}_{1},\boldsymbol{p}_{2}}\left[_{s}\langle n|\boldsymbol{p}_{1}\rangle_{o}\right]\left[_{o}\langle\boldsymbol{p}_{2}|n\rangle_{s}\right]\frac{i}{\Delta E_{n,p_{2}}^{so}-\frac{\boldsymbol{q}^{2}}{4M}-q_{0}+i\epsilon}\nonumber \\
 &  & \;\times\left(1-\frac{i}{2}\overleftarrow{\boldsymbol{\nabla}}_{\boldsymbol{q}}\cdot\boldsymbol{\nabla}_{\boldsymbol{x}}\right)\left[\mathcal{G}_{22}^{ji}(q)+\left(\epsilon^{imn}G_{BE,22}^{jn}(q)-\epsilon^{jmn}G_{EB,22}^{ni}(q)\right)\frac{i\boldsymbol{\nabla}_{\boldsymbol{x}}^{m}}{8M}\right]\nonumber \\
 &  & \;\times\int_{\boldsymbol{r}}e^{-i(\boldsymbol{p}_{1}-\boldsymbol{p}_{2})\cdot\boldsymbol{r}}f_{{\rm sin}}^{o}(t,\boldsymbol{x},-\boldsymbol{q};\boldsymbol{r},\frac{\boldsymbol{p}_{1}+\boldsymbol{p}_{2}}{2})\label{eq:CK-regen-2}
\end{eqnarray}
where we defined 
\begin{eqnarray}
\delta f_{ji,n\lambda}^{s}(t,\boldsymbol{x})=\delta f_{ji,n\lambda}^{s}(t,\boldsymbol{x},\boldsymbol{P}_{{\rm cm}}=\boldsymbol{0}), & \quad & \mathcal{G}_{\alpha\beta}^{ji}(q)=\mathcal{G}_{\alpha\beta}^{ji}(q,P=0)
\end{eqnarray}
All the $\boldsymbol{P}_{{\rm cm}}$ are set to zero, and the momentum
gradient $\boldsymbol{\nabla}_{\boldsymbol{P}_{{\rm cm}}}$ now becomes
$\boldsymbol{\nabla}_{\boldsymbol{q}}$ as 
\begin{eqnarray}
e^{-i\frac{\boldsymbol{q}^{2}}{4M}(t-t_{2})}\boldsymbol{P}_{{\rm rel}}\cdot\frac{-i\boldsymbol{q}}{4M}(t-t_{2})e^{i\boldsymbol{P}_{{\rm rel}}\cdot\boldsymbol{x}} & = & e^{i(-\frac{\boldsymbol{q}^{2}}{4M})(t-t_{2})}\frac{-i}{2}\overleftarrow{\boldsymbol{\nabla}}_{\boldsymbol{q}}\cdot\overrightarrow{\boldsymbol{\nabla}}_{\boldsymbol{x}}e^{i\boldsymbol{P}_{{\rm rel}}\cdot\boldsymbol{x}}.
\end{eqnarray} 
Nevertheless, the rest-frame equation is still 
coupled with the evolution equation for $f^o_{\rm sin}$ in the medium-rest frame. Though the exponent in $\exp\left[{P}_{{\rm rel}}\cdot\frac{\boldsymbol{q}}{4M}(t-t_{2})\right]$ itself is not a small quantity, this exponential characterizes possible quantum corrections to the collision kernel because an $\hbar$ is implicitly accompanied, and we keep only the leading quantum corrections by expanding it  up to its next-to-leading order, namely, $1+\frac{i}{2}\overleftarrow{\boldsymbol{\nabla}}_{\boldsymbol{q}}\cdot\boldsymbol{\nabla}_{\boldsymbol{x}}$.

All the gluon field correlators in eqs.(\ref{eq:CK-disso-2}) and (\ref{eq:CK-regen-2}) are evaluated in the quarkonium
rest frame. The chromoelectric and chromomagnetic fields are related to their medium rest-frame counterparts via Lorentz transformations,
\begin{eqnarray}
\boldsymbol{E}^{a} & = & \gamma_{v}(\boldsymbol{E}_{{\rm rf}}^{a}-\boldsymbol{v}\times\boldsymbol{B}_{{\rm rf}}^{a})-\frac{\gamma_{v}^{2}}{\gamma_{v}+1}\boldsymbol{v}(\boldsymbol{v}\cdot\boldsymbol{E}_{{\rm rf}}^{a}),\\
\boldsymbol{B}^{a} & = & \gamma_{v}(\boldsymbol{B}_{{\rm rf}}^{a}+\boldsymbol{v}\times\boldsymbol{E}_{{\rm rf}}^{a})-\frac{\gamma_{v}^{2}}{\gamma_{v}+1}\boldsymbol{v}(\boldsymbol{v}\cdot\boldsymbol{B}_{{\rm rf}}^{a}),
\end{eqnarray}
where $\boldsymbol{v}$
is the velocity of the medium in the quarkonium rest frame and $\gamma_{v}=(1-\boldsymbol{v}^{2})^{-\frac{1}{2}}$. Within the energy hierarchy~\eqref{eq:Energy-Hierarchy}, $\boldsymbol{v}\sim\frac{\boldsymbol{P}_{\rm cm}}{2M}\sim\sqrt{\frac{T}{M}}$. The spacetime argument $x$ transforms to $x^{\prime}=\Lambda^{-1}(v)x$, so in the Fourier space, $q^{\mu}$ transforms to $q^{\prime}=\Lambda^{-1}(v)q$ as the conjugate to $x^\prime$.
 Since
we focus on corrections up to $\mathcal{O}\left(\frac{T}{M}\right)$, we will
only keep up to $\mathcal{O}(\boldsymbol{v}^{2})$.
We can then express e.g. $G_{BB,11}^{ji}$ in terms of $\tilde{G}$,
\begin{eqnarray}
G_{BB,11}^{ji} & = & (1+\boldsymbol{v}^2)\widetilde{G}_{BB,11}^{ji}+\boldsymbol{v}^{m}\left(\epsilon^{imn}\widetilde{G}_{BE,11}^{jn}+\epsilon^{jmn}\widetilde{G}_{EB,11}^{ni}\right)+\epsilon^{jmn}\epsilon^{ipq}\boldsymbol{v}^{m}\boldsymbol{v}^{p}\widetilde{G}_{EE,11}^{nq}\nonumber \\
 &  & \;-\frac{1}{2}\boldsymbol{v}^{k}\left(\boldsymbol{v}^{i}\widetilde{G}_{BB,11}^{jk}+\boldsymbol{v}^{j}\widetilde{G}_{BB,11}^{ki}\right)+\mathcal{O}(\boldsymbol{v}^{3}),\label{eq:moving-medium-BB-correlator}\\
G_{BE,11}^{jr} & = & (1+\boldsymbol{v}^2)\widetilde{G}_{BE,11}^{jr}+\epsilon^{jpq}\boldsymbol{v}^{p}\widetilde{G}_{EE,11}^{qr}-\epsilon^{rmn}\boldsymbol{v}^{m}\widetilde{G}_{BB,11}^{jn}-\epsilon^{jpq}\epsilon^{rmn}\boldsymbol{v}^{p}\boldsymbol{v}^{m}\widetilde{G}_{EB,11}^{qn}\nonumber \\
 &  & \;-\frac{1}{2}\boldsymbol{v}^{p}\left(\boldsymbol{v}^{r}\widetilde{G}_{BE,11}^{jp}+\boldsymbol{v}^{j}\widetilde{G}_{BE,11}^{pr}\right)+\mathcal{O}(\boldsymbol{v}^{3}),\\
G_{EE,11}^{in} & = & (1+\boldsymbol{v}^2)\widetilde{G}_{EE,11}^{in}-\boldsymbol{v}^{p}\left(\epsilon^{npq}\widetilde{G}_{EB,11}^{iq}+\epsilon^{ipq}\widetilde{G}_{BE,11}^{qn}\right)+\epsilon^{nrs}\epsilon^{ipq}\boldsymbol{v}^{p}\boldsymbol{v}^{r}\widetilde{G}_{BB,11}^{qs}\nonumber \\
 &  & \;-\frac{1}{2}\boldsymbol{v}^{p}\left(\boldsymbol{v}^{i}\widetilde{G}_{EE,11}^{pn}+\boldsymbol{v}^{n}\widetilde{G}_{EE,11}^{ip}\right)+\mathcal{O}(\boldsymbol{v}^{3}).
\end{eqnarray}
Here the momentum argument in these $\tilde{G}$ correlators is the transformed momentum $q^{\prime}=\Lambda^{-1} q$, i.e. up to relative $\mathcal{O}(\boldsymbol{v}^2)$,  $q^0=\left(1+\frac{\boldsymbol{v}^2}{2}\right)q^{\prime,0}+\boldsymbol{v}\cdot\boldsymbol{q}^{\prime}$ and $\boldsymbol{q}=\boldsymbol{q}^{\prime}+\boldsymbol{v}q_0^{\prime}+\frac{1}{2}\boldsymbol{v}(\boldsymbol{v}\cdot\boldsymbol{q}^{\prime})$.

Finally, we remark the relative velocity $\boldsymbol{v}$
is the addition of the fluid velocity $u^{\mu}=\gamma_u(1,\boldsymbol{u})$ describing medium evolution
and the quarkonium velocity $U^{\mu}=\frac{E_{\boldsymbol{q}}}{M_{Q\bar{Q}}}(1,\frac{\boldsymbol{q}}{E_{\boldsymbol{q}}})$ with $E_{\boldsymbol{q}}=\sqrt{M_{Q\bar{Q}}^2+\boldsymbol{q}^2}$ measured
in high-energy experiments, both of which are defined in the laboratory
frame, 
$\boldsymbol{v}=\boldsymbol{u}\oplus\boldsymbol{U}$, namely, 
\begin{eqnarray}
\boldsymbol{v}(\boldsymbol{q},\boldsymbol{u}) & = & \frac{M_{Q\bar{Q}}\boldsymbol{u}-\boldsymbol{q}+(E_{\boldsymbol{q}}-M_{Q\bar Q})\frac{1}{\boldsymbol{q}^{2}}\boldsymbol{q}(\boldsymbol{q}\cdot\boldsymbol{u})}{E_{\boldsymbol{q}}-\boldsymbol{q}\cdot\boldsymbol{u}}\label{eq:velocity-addition}
\end{eqnarray}
where the explicit form of $\boldsymbol{u}$ is obtained from hydrodynamical
simulations. We emphasize that the small relative velocity expansion over $\boldsymbol{v}$ is justified only when eq.~\eqref{eq:velocity-addition} is much smaller than 1, and practical implementation of fluid velocity and quarkonium momentum should guarantee this condition.

\subsection{Perturbative simplification of spin Boltzmann equations
\protect\label{subsec:Simplification-of-spin}}

For the phenomenological study, we work in the quarkonium rest frame because its polarization is measured within this frame in experiments. The chromoelectric and chromomagnetic correlators defined in this frame are related to those in the medium-rest frame by a Lorentz transformation. In this subsection, we calculate these correlators in the medium-rest frame, retaining the leading contributions that generate nonzero spin polarization. The simplification of the spin Boltzmann equation in the medium-rest frame and the perturbative calculation of collision kernel are listed in Appendix \ref{sec:Perturbative-collision-term}. We now simplify the spin collision terms in eqs.~\eqref{eq:CK-disso-2} and \eqref{eq:CK-regen-2}
by substituting the following perturbative gauge-field correlators
up to leading order in a weakly-coupled plasma where $T\gg m_{{\rm D}}$:
\begin{eqnarray}
\widetilde{G}_{BB,11}^{ji}(q) & = & (N_{c}^{2}-1)\epsilon^{jkl}\epsilon^{imn}q^{k}q^{m}D_{{\rm f}}^{ln}(q),\label{eq:mrf-BB-correlator}\\
\widetilde{G}_{BE,11}^{ip}(q) & = & \widetilde{G}_{EB,11}^{pi}(q)=(N_{c}^{2}-1)\epsilon^{imn}q^{m}q^{0}D_{{\rm f}}^{np}(q),\label{eq:mrf-BE-correlator}\\
\widetilde{G}_{EE,11}^{pr}(q) & = & (N_{c}^{2}-1)\left(q^{p}q^{r}D_{{\rm f}}^{00}(q)+(q^0)^2D_{{\rm f}}^{pr}(q)\right),\label{eq:mrf-EE-correlator}
\end{eqnarray}
where the leading-order gluon propagators in Coulomb gauge are 
\begin{eqnarray}
D_{{\rm f}}^{ij}(q)=\left(\delta^{ij}-\frac{\boldsymbol{q}^{i}\boldsymbol{q}^{j}}{\boldsymbol{q}^{2}}\right)\left[\frac{i}{q^{2}+i\epsilon}+2\pi\delta(q^{2})n_{{\rm B}}^{<}(|\boldsymbol{q}|)\right] & ,\; & D_{{\rm f}}^{00}(q)=\frac{i}{\boldsymbol{q}^{2}}.
\end{eqnarray}
Here $n_{{\rm B}}^{<}(|\boldsymbol{q}|)=\left[\exp(|\boldsymbol{q}|/T)-1\right]^{-1}$
is the Bose distribution function and the lesser symbol ``<'' comes
from the ``12'' Wightman correlator and we similarly introduce $n_{{\rm B}}^{>}=1+n_{{\rm B}}^{<}$. In what follows, we change the momentum argument $q$ in the collision integral eqs.~(\ref{eq:CK-disso-2}) and (\ref{eq:CK-regen-2}) to $q^{\prime}=\Lambda^{-1}(v) q$ so the gauge-field correlators are evaluated in the medium-rest frame. The heavy-quarkonium (anti)time-ordered propagator with dummy integral variable $q$ changes to 
\begin{equation}
\frac{\pm i}{q_{0}+\frac{1}{2}\boldsymbol{v}^{2}q_{0}+\boldsymbol{v}\cdot\boldsymbol{q}+\frac{\boldsymbol{q}^{2}}{4M}-\Delta E_{n,\boldsymbol{p}}^{so}\pm i\epsilon}\label{eq:Quarkonium-propagator-mrf}
\end{equation}
and the momentum derivative up to relative $\mathcal{O}(\boldsymbol{v}^2)$ changes to $\boldsymbol{\nabla_q}\to\boldsymbol{\nabla_q}+\frac{\boldsymbol{v}}{2}\boldsymbol{v}\cdot\boldsymbol{\nabla_q}$.

We only focus on the relative order $\frac{T}{M}$ contribution in
the collision terms. Plugging in the correlators in eq.~(\ref{eq:Gauge-field-correlator})
in the quarkonium rest frame, the dissociation term proportional
to $f_{n\lambda}^{s}$ up to relative $\mathcal{O}(\frac{T}{M})$
reads 
\begin{eqnarray}
& &\Delta_{\lambda_1\lambda_2}^{ji}\mathcal{C}_{ji,n\lambda}^{-,\rm qrf}[f_{n\lambda}^{s}] \nonumber\\& = & \Delta_{\lambda_1\lambda_2}^{ji}\frac{g^{2}T_{{\rm F}}}{M^{2}N_{c}(d-1)}\int_{\boldsymbol{p}}\left|_{s}\langle n|\boldsymbol{p}\rangle_{o}\right|^{2}2\text{Re}\int_{q}\frac{-i}{q_{0}+\boldsymbol{v\cdot q} +\frac{1}{2}\boldsymbol{v}^2 q_0 +\frac{\boldsymbol{q}^{2}}{4M}-\Delta E_{n,\boldsymbol{p}}^{so}-i\epsilon}\nonumber \\
 &  & \;\times\left\{ \left[G_{BB,11}^{ji}(q)+\left(\epsilon^{irs}\tilde{G}_{BE,11}^{jr}(q)+\epsilon^{jrs}\tilde{G}_{BE,11}^{ir}(q)\right)\frac{q^{s}-i\partial_{x}^{s}}{8M}\right]f_{n\lambda}^{s}(t,\boldsymbol{x})\right.\nonumber \\
 &  & \left.+\frac{i}{2}\overleftarrow{\boldsymbol{\nabla}}_{\boldsymbol{q}}\left(\tilde{G}_{BB,11}^{ji}+
\boldsymbol{v}^{m}\left[\epsilon^{imn}\widetilde{G}_{BE,11}^{jn}+\epsilon^{jmn}\widetilde{G}_{EB,11}^{ni}\right] \right)\cdot\boldsymbol{\nabla}_{\boldsymbol{x}}f_{n\lambda}^{s}(t,\boldsymbol{x})\right\}. \label{eq:pert-qrf-dis}
\end{eqnarray}
The tensor projections 
\begin{equation}
q^{i},\,q^{i}q^{j}q^{k}\to0,\;q^{i}q^{j}\to\frac{1}{3}\boldsymbol{q}^{2}\delta^{ij},\;q^{i}q^{j}q^{k}q^{l}\to\frac{\boldsymbol{q}^{4}}{15}(\delta^{ij}\delta^{kl}+\delta^{ik}\delta^{jl}+\delta^{il}\delta^{jk}),\label{eq:Tensor-projections}
\end{equation}
help simplify the dissociation term.  Note that  $\tilde{G}_{BB,11}^{ji}(q)$ and $\tilde{G}_{EB/BE,11}^{ji}(q)$ are even and odd
functions of $\boldsymbol{q}$ in the perturbative regime, respectively. We can show that, up to relative $\mathcal{O}(\boldsymbol{v}^2)$, the second term of eq.~\eqref{eq:pert-qrf-dis} from the Moyal-type derivative product vanishes due to the spatial reflection symmetry of the momentum-integral measure. Therefore
\begin{eqnarray}
 &  & \Delta_{\lambda_1\lambda_2}^{ji}\mathcal{C}_{ji,n\lambda}^{-,\text{qrf}}[f_{n\lambda}^{s}]\nonumber \\
 & = & f_{n\lambda}^{s}(t,\boldsymbol{x})\Delta_{\lambda_1\lambda_2}^{ji}\boldsymbol{v}^{i}\boldsymbol{v}^{j}\frac{g^{2}T_{{\rm F}}(N_{c}^{2}-1)}{(d-1)M^{2}N_{c}}\int_{\boldsymbol{p}}\left|_{s}\langle n|\boldsymbol{p}\rangle_{o}\right|^{2}\nonumber \\
 &  & \;2{\rm Re}\int_{q}\left[\frac{i}{q^{2}+i\epsilon}+2\pi\delta(q^{2})n_{{\rm B}}^{<}(|\boldsymbol{q}|)\right]\Biggl\{\frac{-i}{q_{0}-\Delta E_{n,\boldsymbol{p}}^{so}-i\epsilon}\frac{-2}{3}(\boldsymbol{q}^{2}+(q^{0})^{2})\nonumber \\
 &  & \;\;+\frac{-i}{\left(q_{0}-\Delta E_{n,\boldsymbol{p}}^{so}-i\epsilon\right)^{2}}\frac{2}{3}\boldsymbol{q}^{2}q^{0}+\frac{i}{\left(q_{0}-\Delta E_{n,\boldsymbol{p}}^{so}-i\epsilon\right)^{3}}\frac{2\boldsymbol{q}^{4}}{15}\Biggr\}.
\end{eqnarray}
Here the heavy-quarkonium propagator is expanded to the relative order $\frac{T}{M}$,   
\begin{eqnarray}
 &  & \frac{-i}{q_{0}+\frac{1}{2}\boldsymbol{v}^{2}q_{0}+\boldsymbol{v}\cdot\boldsymbol{q}+\frac{\boldsymbol{q}^{2}}{4M}-\Delta E_{n,\boldsymbol{p}}^{so}-i\epsilon}\nonumber \\
 & = & \frac{-i}{q_{0}-\Delta E_{n,\boldsymbol{p}}^{so}-i\epsilon}\Biggl[1-\frac{\boldsymbol{v}\cdot\boldsymbol{q}}{q_{0}-\Delta E_{n,\boldsymbol{p}}^{so}-i\epsilon}-\frac{\frac{1}{2}\boldsymbol{v}^{2}q_{0}+\frac{\boldsymbol{q}^{2}}{4M}}{q_{0}-\Delta E_{n,\boldsymbol{p}}^{so}-i\epsilon}\nonumber \\
 &  & \qquad+\frac{\left(\boldsymbol{v}\cdot\boldsymbol{q}\right)^{2}}{\left(q_{0}-\Delta E_{n,\boldsymbol{p}}^{so}-i\epsilon\right)^{2}}+\mathcal{O}\left(\left(\frac{T}{M}\right)^{\frac{3}{2}}\right)\Biggr].
\end{eqnarray}

The recombination term proportional to $f_{{\rm sin}}^{o}$ is more intricate
as $f_{{\rm sin}}^{o}$ also has an integral-momentum $\boldsymbol{q}$
dependence. We can exploit the energy hierarchy (\ref{eq:Energy-Hierarchy}) that gives $\boldsymbol{q}\sim T\ll\boldsymbol{P}_{{\rm cm}}\sim\sqrt{MT}$,
and perform the \textit{momentum gradient expansion}, 
\begin{eqnarray}
f_{{\rm sin}}^{o}(t,\boldsymbol{x},-\boldsymbol{q}^{\prime};\boldsymbol{p}) & = & \left[1-(\boldsymbol{q}+\boldsymbol{v} q_0)\cdot\boldsymbol{\nabla}_{\boldsymbol{P}_{{\rm cm}}}+\frac{1}{2!}(\boldsymbol{q}\cdot\boldsymbol{\nabla}_{\boldsymbol{P}_{{\rm cm}}})^{2}+...\right]f_{{\rm sin}}^{o}(t,\boldsymbol{x},\boldsymbol{0};\boldsymbol{p})
\end{eqnarray}
with $\boldsymbol{q}\cdot\boldsymbol{\nabla}_{\boldsymbol{P}_{{\rm cm}}}\sim\sqrt{\frac{T}{M}}$.
With angular projections (\ref{eq:Tensor-projections})
\begin{eqnarray}
& &\Delta_{\lambda_1\lambda_2}^{ij}\mathcal{C}_{ji,n\lambda}^{+,\rm qrf}[f_{{\rm sin}}^{o}]\nonumber\\
& = & \Delta_{\lambda_1\lambda_2}^{ij}\frac{g^{2}T_{{\rm F}}}{M^{2}N_{c}}2{\rm Re}\int_{\boldsymbol{p}_{1},\boldsymbol{p}_{2}}\left[_{s}\langle n|\boldsymbol{p}_{1}\rangle_{o}\right]\left[_{o}\langle\boldsymbol{p}_{2}|n\rangle_{s}\right]\int_{q}\left[\frac{i}{q^{2}+i\epsilon}+2\pi\delta(q^{2})n_{{\rm B}}^{<}(|\boldsymbol{q}|)\right]\nonumber\\
 &  & \;\times\Bigg\{\frac{i}{q_{0}-\Delta E_{n,\boldsymbol{p}_1}^{so}+i\epsilon}\left[-\frac{2}{3}(\boldsymbol{q}^{2}+(q^0)^2)\boldsymbol{v}^{i}\boldsymbol{v}^{j}+\frac{\boldsymbol{q}^{2}q^{0}}{3}\left(\boldsymbol{v}^{j}\boldsymbol{\nabla}_{\boldsymbol{P}_{{\rm cm}}}^{i}+\boldsymbol{v}^{i}\boldsymbol{\nabla}_{\boldsymbol{P}_{{\rm cm}}}^{j}\right)-\frac{\boldsymbol{q}^{4}}{15}\boldsymbol{\nabla}_{\boldsymbol{P}_{{\rm cm}}}^{i}\boldsymbol{\nabla}_{\boldsymbol{P}_{{\rm cm}}}^{j}\right]\nonumber \\
  &  & \quad+\frac{i}{\left(q_{0}-\Delta E_{n,\boldsymbol{p}_1}^{so}+i\epsilon\right)^{2}}\left[\frac{2}{3}\boldsymbol{v}^{i}\boldsymbol{v}^{j}\boldsymbol{q}^{2}q^{0}-\frac{\boldsymbol{q}^{4}}{15}\left(\boldsymbol{v}^{i}\boldsymbol{\nabla}_{\boldsymbol{P}_{{\rm cm}}}^{j}+\boldsymbol{v}^{j}\boldsymbol{\nabla}_{\boldsymbol{P}_{{\rm cm}}}^{i}\right)\right]\nonumber\\
 &  & \quad+\frac{-i}{\left(q_{0}-\Delta E_{n,\boldsymbol{p}_1}^{so}+i\epsilon\right)^{3}}\frac{2\boldsymbol{q}^{4}}{15}\boldsymbol{v}^{i}\boldsymbol{v}^{j}\Biggr\}\int_{\boldsymbol{r}}e^{-i(\boldsymbol{p}_{1}-\boldsymbol{p}_{2})\cdot\boldsymbol{r}}f_{{\rm sin}}^{o}(t,\boldsymbol{x},\boldsymbol{0};\boldsymbol{r},\frac{\boldsymbol{p}_{1}+\boldsymbol{p}_{2}}{2}),
\end{eqnarray}
where we dropped the dimensionless integrals over $\boldsymbol{q}$.

Since the scattering-state energy is always positive whereas the
bound-state energy is negative, we can impose the kinematic constraint
$\Delta E_{n,\boldsymbol{p}}^{so}<0$. Integrating over $q$, we now get
\begin{eqnarray}
\Delta_{\lambda_1\lambda_2}^{ji}\mathcal{C}_{ji,n\lambda}^{-,{\rm qrf}}[f_{n\lambda}^{s}] & = & -\frac{f_{n\lambda}^{s}(t,\boldsymbol{R})}{d-1}\Delta_{\lambda_1\lambda_2}^{ji}\boldsymbol{v}^{i}\boldsymbol{v}^{j}\Gamma_{n}^{s,\text{qrf}}\label{eq:qrf-pertur-disso}
\end{eqnarray}
and
\begin{eqnarray}
& &\Delta_{\lambda_1\lambda_2}^{ji}\mathcal{C}_{ji,n\lambda}^{+,{\rm qrf}}[f_{{\rm sin}}^{o}] \nonumber\\& = & \frac{1}{10}\Delta_{\lambda_1\lambda_2}^{ji}{\rm Re}\int_{\boldsymbol{p}_{1},\boldsymbol{p}_{2}}\left(1+n_{{\rm B}}^{<}(-\Delta E_{n,\boldsymbol{p}_{1}}^{so})\right)W_{n,\boldsymbol{p}_{1},\boldsymbol{p}_{2}}^{5}\nonumber \\
 &  & \;\Biggl\{\boldsymbol{v}^{i}\boldsymbol{v}^{j}\beta^{2}n_{{\rm B}}^{>}(\Delta E_{n,\boldsymbol{p}_{1}}^{so})\left(1+2n_{{\rm B}}^{<}(\Delta E_{n,\boldsymbol{p}_{1}}^{so})\right)-\beta n_{{\rm B}}^{>}(\Delta E_{n,\boldsymbol{p}_{1}}^{so})\left(\boldsymbol{v}^{i}\boldsymbol{\nabla}_{\boldsymbol{P}_{{\rm cm}}}^{j}+\boldsymbol{v}^{j}\boldsymbol{\nabla}_{\boldsymbol{P}_{{\rm cm}}}^{i}\right)\nonumber \\
 &  & \qquad+\boldsymbol{\nabla}_{\boldsymbol{P}_{{\rm cm}}}^{i}\boldsymbol{\nabla}_{\boldsymbol{P}_{{\rm cm}}}^{j}\Biggr\}\int_{\boldsymbol{r}}e^{-i(\boldsymbol{p}_{1}-\boldsymbol{p}_{2})\cdot\boldsymbol{r}}\frac{f_{{\rm sin}}^{o}(t,\boldsymbol{x},\boldsymbol{0};\boldsymbol{r},\frac{\boldsymbol{p}_{1}+\boldsymbol{p}_{2}}{2})}{N_{c}^{2}-1}\label{eq:qrf-pertur-reg}
\end{eqnarray}
where the vacuum part is calculated using residue theorem by first
integrating over $q_{0}$ (also see Ref.~\cite{Biondini:2023zcz}), and
finite-temperature part gives $\delta(q_{0}-\Delta E_{n,\boldsymbol{p}}^{so})$. We also keep only the imaginary part of the quarkonium self-energy to impose proper kinematic constraints in accordance with the optical theorem. Now it manifests that with positive gluon energy $\omega_{pn}=-\Delta E^{so}_{n,\boldsymbol{p}}$, the dissociation and regeneration parts are proportional to $n^<({\omega_{pn}})$ and $1+n^<({\omega_{pn}})$, satisfying detailed balance.
Here we introduce the scattering matrix elements 
\begin{eqnarray}
W_{n,\boldsymbol{p}_{1},\boldsymbol{p}_{2}}^{k} & = & \frac{4\alpha_{{\rm s}}C_{{\rm F}}}{3M^{2}}\left[_{s}\langle n|\boldsymbol{p}_{1}\rangle_{o}\right]\left[_{o}\langle\boldsymbol{p}_{2}|n\rangle_{s}\right](\Delta E_{n,\boldsymbol{p}_{1}}^{so})^{k}.\label{eq:Scattering-ME}
\end{eqnarray}
The anisotropic dissociation rate governing spin-alignment generation is defined as
\begin{eqnarray}
\Gamma_{n}^{s,\text{qrf}} & = & \int_{\boldsymbol{p}}\frac{1}{10}\beta^{2}W_{n\lambda,\boldsymbol{p},\boldsymbol{p}}^{5}n_{{\rm B}}^{<}(\Delta E_{n,\boldsymbol{p}}^{so})n_{{\rm B}}^{>}(\Delta E_{n,\boldsymbol{p}}^{so})\left(1+2n_{{\rm B}}^{<}(\Delta E_{n,\boldsymbol{p}}^{so})\right).\label{eq:Spin-dissociation-rate}
\end{eqnarray}
We remark that to get the traceless anisotropic part of $\rho_{\lambda\lambda^\prime}$,  the $i,j$ components of the collision term should be understood as the components that exist after contracting with the projector $\Delta_{\lambda\lambda^\prime}^{ij}$. The above perturbative collision kernel does not involve
information of $\boldsymbol{x}$-evolution, so the solution of $\delta f_{ji,n\lambda}^{s}(t,\boldsymbol{x})$
is simply the time integral over $f_{n\lambda}^{s}(t,\boldsymbol{x})$
and $f_{{\rm sin}}^{o}(t,\boldsymbol{x},\boldsymbol{P}_{{\rm cm}}=\boldsymbol{0};\boldsymbol{p})$ with collision-term weights.
If we assume the distribution function is locally isotropic around $\boldsymbol{P}_{\rm cm}=0$, its first-order momentum derivative vanishes and second-order momentum derivative is proportional to $\delta^{ij}$ which can be dropped after the traceless projection,  we may then drop the $\boldsymbol{P}_{{\rm cm}}$-derivative
terms in the recombination collision kernel, 
\begin{eqnarray}
&  & \Delta_{\lambda_1\lambda_2}^{ji}\frac{{\rm d}}{{\rm d}t}\delta f_{ji,n\lambda}^{s}(t,\boldsymbol{R})\nonumber \\
 & = & \Delta_{\lambda_1\lambda_2}^{ji}\boldsymbol{v}^{i}\boldsymbol{v}^{j}\int_{\boldsymbol{p}_{1},\boldsymbol{p}_{2}}W_{n,\boldsymbol{p}_{1},\boldsymbol{p}_{2}}^{5}\frac{1}{10}\beta^{2}n_{{\rm B}}^{<}(\Delta E_{n,\boldsymbol{p}_{1}}^{so})n_{{\rm B}}^{>}(\Delta E_{n,\boldsymbol{p}_{1}}^{so})\left(1+2n_{{\rm B}}^{<}(\Delta E_{n,\boldsymbol{p}}^{so})\right)\nonumber \\
 &  & \qquad\times\left[\frac{f_{n\lambda}^{s}(t,\boldsymbol{R})}{d-1}\delta_{\boldsymbol{p}_{1}\boldsymbol{p}_{2}}-\int_{\boldsymbol{r}}e^{-i(\boldsymbol{p}_{1}-\boldsymbol{p}_{2})\cdot\boldsymbol{r}}\frac{f_{{\rm sin}}^{o}(t,\boldsymbol{x},\boldsymbol{0};\boldsymbol{r},\frac{\boldsymbol{p}_{1}+\boldsymbol{p}_{2}}{2})}{N_{c}^{2}-1}\right],\label{eq:Simplified-SBE}
\end{eqnarray}
where $\delta_{\boldsymbol{p}_{1}\boldsymbol{p}_{2}}=(2\pi)^{3}\delta^{(3)}(\boldsymbol{p}_{1}-\boldsymbol{p}_{2})$. Note that $f_{n\lambda}^{s}(t,\boldsymbol{R})$ and $f_{{\rm sin}}^{o}(t,\boldsymbol{R},\boldsymbol{P}_{{\rm cm}}=\boldsymbol{0};\boldsymbol{p})$
at each time slice can be solved from Lindblad-type master equation
without knowing the $\{\boldsymbol{R},\boldsymbol{P}_{{\rm cm}}\}$
information. We will use the solution of eq.~\eqref{eq:Simplified-SBE}
to estimate the magnitude of spin polarization of heavy quarkonium
in the next section. 

Finally, the above perturbative calculation indicates the $\boldsymbol{R}$-evolution
automatically vanishes, and in such case eq.~\eqref{eq:SBE-qrf} can be rewritten as a Boltzmann equation with 
a classical collision term in the quarkonium rest frame. Rewriting the quarkonium-rest-frame propagator
$\mathcal{G}_{11}$ and quarkonium propagator in terms of Wightman
functions, eq.~\eqref{eq:SBE-qrf} reduces to
\begin{eqnarray}
 &  & \Delta_{\lambda\lambda^\prime}^{ij}\frac{{\rm d}}{{\rm d}t}\delta f_{ji,n\lambda}^{s}(t,\boldsymbol{R})\nonumber \\
 & = & \Delta_{\lambda\lambda^\prime}^{ij}\frac{g^{2}T_{{\rm F}}}{M^{2}N_{c}}{\rm Re}\int_{\boldsymbol{p}_{1},\boldsymbol{p}_{2}}\left[_{s}\langle n|\boldsymbol{p}_{1}\rangle_{o}\right]\left[_{o}\langle\boldsymbol{p}_{2}|n\rangle_{s}\right]\int_{q}2\pi\delta\left(q_{0}+\frac{\boldsymbol{q}^{2}}{4M}-\Delta E_{n,\boldsymbol{p}_{1}}^{so}\right)\nonumber \\
 &  & \;\times\left[\mathcal{G}_{12}^{ji}(q)\int_{\boldsymbol{r}}e^{-i(\boldsymbol{p}_{1}-\boldsymbol{p}_{2})\cdot\boldsymbol{r}}\frac{f_{\text{sin}}^{o}(t,\boldsymbol{R},-\boldsymbol{q};\boldsymbol{r},\frac{\boldsymbol{p}_{1}+\boldsymbol{p}_{2}}{2})}{N_{c}^{2}-1}-\mathcal{G}_{21}^{ji}(q)\delta_{\boldsymbol{p}_{1}\boldsymbol{p}_{2}}\frac{f_{n\lambda}^{s}(t,\boldsymbol{R})}{d-1}\right].\label{eq:CCK-qrf}
\end{eqnarray}

\section{Numerical estimation\protect\label{sec:Numerical-estimation}}

In this section, we solve the simplified spin Boltzmann equation in eq.~\eqref{eq:Simplified-SBE} with leading-order perturbative environment
correlator insertion. Despite ignoring the center-of-mass momentum dependence
in this equation, the solution of $\delta f_{ji,n\lambda}^{s}$ still
depends on the quarkonium momentum and the fluid velocity in the laboratory
frame through $\boldsymbol{v}^{i}(\boldsymbol{q},\boldsymbol{u})$.
From this, we are able to derive the transverse momentum $q_{{\rm T}}$
and rapidity $y$ dependence of $\lambda_{\theta}$ in the laboratory
frame with eq.~\eqref{eq:SDME-sdf}. We further show a qualitative phenomenological study of polarization parameter $\lambda_{\theta}$.

\subsection{Rate equation and observables\protect\label{subsec:Simplification-and-spin}}

We may further simplify eq.~\eqref{eq:Simplified-SBE} by dropping
the spacetime dependence of the fluid velocity during the evolution \footnote{Generally, the relative velocity $\boldsymbol{v}(\boldsymbol{q},\boldsymbol{u})$ varies in spacetime and should be included in the spacetime integration. For phenomenological estimates in this section, we treat $\boldsymbol{v}$ as a fixed parameter during the spin evolution and subsequently evaluate the distributions using the local fluid velocity at the observation time $t_{\rm F}$.  This approximates the evolving hydrodynamic background by a fixed relative velocity and neglects the dependence on its prior history.} and cast it to a rate equation, and then we find its solution in terms of overlaps of the density matrices,  
\begin{eqnarray}
 &  & \Delta_{\lambda_1\lambda_2}^{ji}\delta n_{ji,\text{1S}}^{s}(t_{{\rm F}};\boldsymbol{v})\nonumber \\
 & = & \Delta_{\lambda_1\lambda_2}^{ji}\delta n_{ji,\text{1S}}^{s}(t_{0};\boldsymbol{v})+\Delta_{\lambda_1\lambda_2}^{ji}\boldsymbol{v}^{i}\boldsymbol{v}^{j}\int_{t_{0}}^{t_{{\rm F}}}{\rm d}t\int_{\boldsymbol{p}_{1},\boldsymbol{p}_{2}}\left[\frac{n_{\text{1S}}^{s}(t,\boldsymbol{R})}{d-1}\delta_{\boldsymbol{p}_{1}\boldsymbol{p}_{2}}-\frac{_{o}\langle\boldsymbol{p}_{1}|\hat{\rho}_{o,l=0}^{\text{sin}}(t)|\boldsymbol{p}_{2}\rangle_{o}}{N_{c}^{2}-1}\right]\nonumber \\
 &  & \qquad\times W_{1{\rm S},\boldsymbol{p}_{1},\boldsymbol{p}_{2}}^{5}\frac{1}{10}\beta^{2}n_{{\rm B}}^{<}(\Delta E_{1{\rm S},\boldsymbol{p}_{1}}^{so})n_{{\rm B}}^{>}(\Delta E_{1{\rm S},\boldsymbol{p}_{1}}^{so})\left(1+2n_{{\rm B}}^{<}(\Delta E_{1{\rm S},\boldsymbol{p}}^{so})\right)
\label{eq:Int-form-SReq}
\end{eqnarray}
with $n(t)=\int{\rm d}^{3}\boldsymbol{R}\;f(t,\boldsymbol{R})$. The transition matrix element for the color-singlet quarkonium with quantum numbers $n$ is defined in eq.~\eqref{eq:Scattering-ME}.
The r.h.s. of the spin-rate equation \eqref{eq:Int-form-SReq} only depends
on the dynamics of the density operators of the spin-triplet color-singlet
$\hat{\rho}_{s}^{\text{tri}}(t)$ and the spin-singlet color-octet state
$\hat{\rho}_{o}^{\text{sin}}(t)$. Their overlap with the bound state $|n,\lambda\rangle$
and scattering state $|\boldsymbol{p}\rangle$ at each time slice is given respectively by
\begin{eqnarray}
n_{n\lambda}^{s}(t) & = & _{s}\langle n,\lambda|\hat{\rho}_{s}^{\text{tri}}(t)|n,\lambda\rangle_s,\label{eq:Singlet-overlap}
\end{eqnarray}
and
\begin{eqnarray}
\int_{\boldsymbol{r}}e^{-i(\boldsymbol{p}_{1}-\boldsymbol{p}_{2})\cdot\boldsymbol{r}}n_{\text{sin}}^{o}(t;\boldsymbol{r},\frac{\boldsymbol{p}_{1}+\boldsymbol{p}_{2}}{2}) & = & _{o}\langle\boldsymbol{p}_{1}|\hat{\rho}_{o,l=0}^{\text{sin}}(t)|\boldsymbol{p}_{2}\rangle_{o}.\label{eq:Octet-overlap}
\end{eqnarray}
Refs.~\cite{Brambilla:2016wgg,Brambilla:2017zei} derived a quantum master equation for the time evolution of the unpolarized
density operators $\hat{\rho}_{s/o}^{\text{tri}}$. Ref.~\cite{Brambilla:2022ynh} cast it into Lindblad form~\cite{Lindblad:1975ef,Gorini:1975nb}, which can be simulated efficiently with the open source QTraj~\cite{Omar:2021kra}. Previous phenomenological studies showed that QTraj can reproduce the bottomonium suppression pattern in different collision systems~\cite{Brambilla:2023hkw,Brambilla:2024tqg,Strickland:2023nfm,Brambilla:2026evh}. Since
the master equation for the spin-singlet state $\hat{\rho}_{s/o}^{\text{sin}}$
is the same as that for the spin-triplet counterpart up to leading multipole expansion, we simply use the same framework for $\hat{\rho}_{o}^{\text{sin}}$. Using QTraj, we obtain the full time evolution of the singlet and octet density matrices from which we can compute the necessary overlaps $_{s}\langle n,\lambda|\hat{\rho}_{s}^{\text{tri}}(t)|n,\lambda\rangle_s$ and $_{o}\langle\boldsymbol{p}_{1}|\hat{\rho}_{o,l=0}^{\text{sin}}(t)|\boldsymbol{p}_{2}\rangle_{o}$.

The initial condition $\delta n_{ji,1{\rm S}}^{s}(t_{0};\boldsymbol{v})$ of eq.~\eqref{eq:Int-form-SReq}
is usually tangled with various nuclear matter mechanisms, but as
pointed out in \cite{Cheung:2022nnq} with an improved color evaporation
model calculation, the final contribution from these cold nuclear
matter effects to the polarization tends to be zero at the kinematic region considered here with soft transverse momentum $q_{{\rm T}}<10$ GeV
in $\sqrt{s_{{\rm NN}}}=5.02$ TeV Pb-Pb collision for $J/\psi$.
We thus assume 
\begin{eqnarray}
\delta n_{ji,1{\rm S}}^{s}(t_{0};\boldsymbol{q},\boldsymbol{u}) & \approx & 0
\end{eqnarray}
 also holds for bottomonium. Nevertheless, as we will see, though the magnitude of $\delta n^s_{ji,\rm 1S}$ is small, it is the ratio of $\delta n^s_{ji,\rm 1S}$ and $n_{\rm 1S}(t_{\rm F})$ that contributes to $\delta\rho_{00}$ which can turn out to be non-negligible. This is important to the phenomenological study of $J/\psi$ polarization.

The quantum optical limit $T\lesssim E\ll r^{-1}$ works well for
the lowest-lying heavy quarkonium states, so we mainly focus on the
1S state in the following. We thus set the quantum number as $n=1,l=0$
for the color-singlet interpolating field, and according to the magnetic
selection rule, only the transition to $l=0$ color-octet state exists at leading order considered here:
namely the wavefunction overlap $_{s}\langle1{\rm S}|\boldsymbol{p}l\rangle_{o}\propto\int{\rm d}\Omega P_{l}(\hat{\boldsymbol{p}}\cdot\hat{\boldsymbol{r}})=4\pi\delta_{0l}$
after performing partial-wave expansion for the scattering-state wavefunction.

As the Lindblad equation derived in Ref.~\cite{Brambilla:2022ynh} is valid in the regime $T\gg Mv^2$, it assumes a different hierarchy than eq.~\eqref{eq:Int-form-SReq}, which was derived assuming~\eqref{eq:Energy-Hierarchy}. To be able to obtain a phenomenological estimate, we still use the QTraj framework to compute the necessary density matrix evolutions. A fully consistent treatment would require an efficient simulation scheme similar to QTraj in the quantum optical regime.

Obtaining the evolution of the bound-state particle
number $n_{1{\rm S}}^{s}(t)$ from the 1S overlap of $\hat{\rho}_{s}^{{\rm tri}}(t)$
is straightforward.
The scattering state overlap  $_{o}\langle\boldsymbol{p}_{1}|\hat{\rho}_{o,l=0}^{\text{sin}}(t)|\boldsymbol{p}_{2}\rangle_{o}$
is related to the $\boldsymbol{r}$-space overlap via inserting a complete basis, which is defined as 
\begin{eqnarray}
\hat{\rho}(t;\boldsymbol{r},\boldsymbol{r}^{\prime}) & = & \langle\boldsymbol{r}|\hat{\rho}(t)|\boldsymbol{r}^{\prime}\rangle=\sum_{l,m}\sum_{l^{\prime},m^{\prime}}Y^{lm*}(\hat{\boldsymbol{r}})\rho_{lm,l^{\prime}m^{\prime}}(t;r,r^{\prime})Y^{l^{\prime}m^{\prime}}(\hat{\boldsymbol{r}}^{\prime}),
\end{eqnarray}
where $Y^{lm}(\hat{\boldsymbol{r}})$ are the spherical harmonic functions
and 
\begin{eqnarray}
\rho_{lm,l^{\prime}m^{\prime}}(t;r,r^{\prime}) & = & \int{\rm d}\Omega(\hat{\boldsymbol{r}}){\rm d}\Omega(\hat{\boldsymbol{r}}^{\prime})Y^{lm}(\hat{\boldsymbol{r}})\hat{\rho}(t;\boldsymbol{r},\boldsymbol{r}^{\prime})Y^{l^{\prime}m^{\prime}*}(\hat{\boldsymbol{r}}^{\prime}).
\end{eqnarray}
Only unpolarized density operators enter the spin collision kernel
of eq.~\eqref{eq:Int-form-SReq}, so we assume only diagonal components
with $l=l^{\prime}$ and $m=m^{\prime}$ are nonzero, and define 
\begin{eqnarray}
\hat{\rho}_{l}(t;r,r^{\prime}) & = & \sum_{m}\rho_{lm,lm}=\frac{2l+1}{4\pi}\int{\rm d}\Omega(\hat{\boldsymbol{r}}){\rm d}\Omega(\hat{\boldsymbol{r}}^{\prime})P_{l}(\hat{\boldsymbol{r}}\cdot\hat{\boldsymbol{r}}^{\prime})\hat{\rho}(t;\boldsymbol{r},\boldsymbol{r}^{\prime});
\end{eqnarray}
then  
\begin{eqnarray}
_{o}\langle\boldsymbol{p}_{1}|\hat{\rho}_{o}^{\text{sin}}(t)|\boldsymbol{p}_{2}\rangle_{o} & = & \int{\rm d}^{3}\boldsymbol{r}_{1}{\rm d}^{3}\boldsymbol{r}_{2}{}_{o}\langle\boldsymbol{p}_{1}|\boldsymbol{r}_{1}\rangle\langle\boldsymbol{r}_{1}|\hat{\rho}_{o}^{\text{sin}}(t)|\boldsymbol{r}_{2}\rangle\langle\boldsymbol{r}_{2}|\boldsymbol{p}_{2}\rangle_{o}\nonumber \\
 & = & \int{\rm d}^{3}\boldsymbol{r}_{1}{\rm d}^{3}\boldsymbol{r}_{2} \Phi^*_{\boldsymbol{p}_10}(\boldsymbol{r}_1)\Phi_{\boldsymbol{p}_20}(\boldsymbol{r}_2)\frac{1}{4\pi}\rho_{l=0}^{o,\text{sin}}(t;r_{1},r_{2})\nonumber \\
 & = & 4\pi\int_{0}^{+\infty}r_{1}^{2}r_{2}^{2}{\rm d}r_{1}{\rm d}r_{2}\Phi^*_{\boldsymbol{p}_10}(r_1)\Phi_{\boldsymbol{p}_20}(r_2)\rho_{l=0}^{o,\text{sin}}(t;r_{1},r_{2}).
\label{eq:Octet-sca-number}
\end{eqnarray}
where the summation
over color indices in $\sum_{c}\langle c|\hat{\rho}^{o}|c\rangle$
is implicitly understood and the scattering-state Coulomb wavefunction is defined below.

The scattering matrix element $W_{n\lambda,\boldsymbol{p}_1,\boldsymbol{p}_2}^{k}$ in~\eqref{eq:Scattering-ME} contains
the wavefunction overlap $_{s}\langle n|\boldsymbol{p}\rangle_{o}$
and eigenenergy difference $\Delta E_{n,\boldsymbol{p}}^{so}$.
The eigenenergy of color-singlet bound state is taken from the leading-order
Coulomb interactions $V_{s}^{(0)}=-\frac{C_{{\rm F}}\alpha_{{\rm s}}}{r}$
and that of scattering state is the continuous $\frac{\boldsymbol{p}^{2}}{M}$
with $V_{o}^{(0)}=(\frac{C_{{\rm A}}}{2}-C_{{\rm F}})\frac{\alpha_{{\rm s}}}{r}$
so
\begin{eqnarray}
\Delta E_{1{\rm S},\boldsymbol{p}}^{so} & = & -\left[\frac{M^{2}C_{{\rm F}}^{2}\alpha_{{\rm s}}^{2}(a_{0}^{-1})}{4p^{2}}+1\right]\frac{p^{2}}{M},
\end{eqnarray}
with $p=\sqrt{\boldsymbol{p}^{2}}$ and $a_{0}=2/[C_{{\rm F}}M\alpha_{{\rm s}}(a_{0}^{-1})]$ being the
color-singlet Bohr radius. The scattering-state wavefunctions are
frame-dependent, but the small difference is of $\mathcal{O}(\boldsymbol{v}^{2})$
and can be thus discarded up to leading order, so 
\begin{eqnarray}
\Phi_{\boldsymbol{p}0}(\boldsymbol{r}) & = & \langle\boldsymbol{r}|\boldsymbol{p}0\rangle_{o}=\sqrt{\frac{\frac{2\pi\rho}{a_{0}p}}{e^{\frac{2\pi\rho}{a_{0}p}}-1}}e^{ipr}{}_{1}F_{1}(1+\frac{i\rho}{a_{0}p},2,-2ipr),
\end{eqnarray}
with the phase-shift factor $e^{i\delta_{0}(p)}$ being absorbed to the definition of the scattering states and therefore the overlap of our concern is
\begin{eqnarray}
 &  & \left[_{s}\langle1S|\boldsymbol{p}_{1}0\rangle_{o}\right]\left[_{o}\langle\boldsymbol{p}_{2}0|1S\rangle_{s}\right]\nonumber \\
 & = & \int_{\boldsymbol{r}_{1},\boldsymbol{r}_{2}}\phi_{1{\rm S}}^{*}(\boldsymbol{r}_{1})\Phi_{\boldsymbol{p}_{1}}(\boldsymbol{r}_{1})\Phi_{\boldsymbol{p}_{2}}^{*}(\boldsymbol{r}_{2})\phi_{1{\rm S}}(\boldsymbol{r}_{2})\nonumber \\
 & = & 2^{7}\pi^{2}\rho(\rho+1)^{2}a_{0}^{2}\left[p_{1}p_{2}\left(e^{\frac{2\pi\rho}{a_{0}p_{1}}}-1\right)\left(e^{\frac{2\pi\rho}{a_{0}p_{2}}}-1\right)\right]^{-\frac{1}{2}}\nonumber \\
 &  & \;\times\frac{1}{(1+a_{0}^{2}p_{1}^{2})^{2}}\frac{1}{(1+a_{0}^{2}p_{2}^{2})^{2}}\exp\left[\frac{2\rho}{a_{0}p_{1}}\arctan(a_{0}p_{1})+\frac{2\rho}{a_{0}p_{2}}\arctan(a_{0}p_{2})\right],
\end{eqnarray}
which is real and indicates we only need to focus on the real part
of $_{o}\langle\boldsymbol{p}_{1}|\hat{\rho}_{o}^{\text{sin}}(t)|\boldsymbol{p}_{2}\rangle_{o}$. Here $\rho=(N_{c}^{2}-1)^{-1}$. Then the anisotropic dissociation rate for $N_{c}=3$
is 
\begin{eqnarray}
\Gamma_{\text{1S}}^{s,\text{qrf}} & = & \frac{2^{8}M^{2}\alpha_{{\rm s}}(T)\alpha_{{\rm s}}^{8}(a_{0}^{-1})}{3^{6}\times5T}\int_{0}^{+\infty}{\rm d}x\frac{xH(x)e^{\frac{1}{2x}\arctan x}}{e^{\frac{\pi}{4x}}-1}\frac{e^{-H(x)}(e^{-H(x)}+1)}{\left(1-e^{-H(x)}\right)^{3}},
 \label{eq:perturbative-spin-DR}
\end{eqnarray}
where we define $H(x)=\frac{4M\alpha_{{\rm s}}^{2}(a_{0}^{-1})}{9T}(x^{2}+1)$. The environment temperature $T(t)$ evolves according to fluid
dynamics; the strong coupling constants $\alpha_{{\rm s}}(\mu)$ coming
from different scales $\{T,a_{0}^{-1}\}$ are distinguished and solved
using RunDec \cite{Chetyrkin:2000yt}, especially, $\alpha_{{\rm s}}(a_{0}^{-1})$
follows the definition of color-singlet Bohr radius. 

Lastly, we link the solution (\ref{eq:Int-form-SReq}) to the experimental observables.
The normalized spin density matrix elements at final stage $t_{{\rm F}}$ now
reduce to
\begin{eqnarray}
\delta\rho_{\lambda\lambda^{\prime}}^{n,{\text{N}}}(t_{{\rm F}},\boldsymbol{q},\boldsymbol{u}) & = & \varepsilon_{\lambda}^{i}\varepsilon_{\lambda^{\prime}}^{*j}\frac{\underset{_{\lambda_{1}}}{\sum}|\tilde{\phi}_{n,\lambda_{1}}(\boldsymbol{0})|^{2}\delta n_{ji,n\lambda_{1}}^{s}(t_{{\rm F}},\boldsymbol{q},\boldsymbol{u})}{\underset{_{\lambda_{2}}}{\sum}|\tilde{\phi}_{n,\lambda_{2}}(\boldsymbol{0})|^{2}n_{n\lambda_{2}}^{s}(t_{{\rm F}})}.
\end{eqnarray}
The lab-frame quarkonium 4-vector $q^{\mu}$ can be parameterized
in terms of transverse momentum $q_{{\rm T}}$, momentum rapidity $Y$ and
azimuthal angle $\phi_p$,
\begin{eqnarray}
q^{\mu} & = & (m_{{\rm T}}\cosh Y,q_{{\rm T}}\cos\phi_p,q_{{\rm T}}\sin\phi_p,m_{{\rm T}}\sinh Y),\label{eq:Momentum-parameter}
\end{eqnarray}
where $Y=\frac{1}{2}\log\frac{E_{p}+p_{z}}{E_{p}-p_{z}}$, $q_{{\rm T}}=\sqrt{q_{x}^{2}+q_{y}^{2}}$ and $m_{{\rm T}}=\sqrt{q_{{\rm T}}^{2}+M_{Q\bar{Q}}^{2}}$.
In general, experiments measure the $\{Y,q_{{\rm T}}\}$ spectrum in a
full azimuthal angle region $0<\phi_p<2\pi$ at ALICE and STAR, and
the centrality dependence is determined from hydrodynamic profile
$\{u^{\mu},T\}$. We recast the decay distribution eq.~\eqref{eq:Decay_distribution_massless_l}
to 
\begin{eqnarray}
\frac{{\rm d}\mathcal{R}}{q_{{\rm T}}{\rm d}q_{{\rm T}}{\rm d}Y{\rm d}\phi_p{\rm d}\cos\theta} & = & \mathcal{N}(q_{{\rm T}},Y,\phi_p)\left[1+\lambda_{\theta}(q_{{\rm T}},Y,\phi_p)\cos^{2}\theta\right],
\end{eqnarray}
where $\mathcal{N}(q_{{\rm T}},Y,\phi_p)$ is an irrelevant normalization
factor and 
\begin{eqnarray}
\lambda_{\theta}(q_{{\rm T}},y,\varphi) & = & \frac{\int_{\Sigma}{\rm d}\sigma(x)\cdot t(x)\text{Tr}\rho^{1\text{S}}_{\lambda\lambda^{\prime}}(q,x)\left[-3\delta\rho_{00}^{1\text{S,N}}(t_{{\rm F}},\boldsymbol{q},\boldsymbol{u})\right]}{\int_{\Sigma}{\rm d}\sigma(x)\cdot t(x)\text{Tr}\rho^{1\text{S}}_{\lambda\lambda^{\prime}}(q,x)\left[\frac{4}{3}+\delta\rho_{00}^{1\text{S},\text{N}}(t_{{\rm F}},\boldsymbol{q},\boldsymbol{u})\right]},\label{eq:lambda-theta-q-spectrum}
\end{eqnarray}
with $\Sigma$ the freeze-out hypersurface and $t(x)$ being its normal
vector. We may integrate over e.g. $q_{{\rm T}},\phi_p$ in both
denominator and numerator to get the rapidity dependence. From Appendix \ref{sec:Quarkonium-transport-equation}, the trace over unnormalized spin density matrix is proportional to the bound-state spin-averaged distribution $f^s_{\text{1S}}(t,\boldsymbol{q},\boldsymbol{x})$ up to the order we are interested in and is thus related to the quarkonium final-state $q_{\rm T}$ spectrum. With the same spatial homogeneity assumption used in eq.~\eqref{eq:Int-form-SReq}, this weight does not depend on the position on the freeze-out hypersurface and cancels in eq.~\eqref{eq:lambda-theta-q-spectrum} at fixed momentum.

\subsection{Freeze-out modeling and numerical results\protect\label{sec:Freeze-out-modeling}}
Evolution of both density operators and temperature depends on the
hydrodynamics, and the observable $\lambda_{\theta}$ also intimately associates with the freeze-out hypersurface. For simplicity, we will apply
the Bjorken flow profile for the evolution and characterize the freeze-out
using blast-wave model \citep{Retiere:2003kf}. We use the QTraj setup and parameters from Ref.~\cite{Brambilla:2023hkw}, which were found to reproduce the suppression pattern of bottomonium states in Pb-Pb collisions. We therefore choose the transport coefficients $\hat\kappa=4$ and $\hat\gamma=0$ and terminate the evolution at a final temperature of $T_f=190\,$MeV. While we assume the density-operator overlaps freeze out at this moment the remaining QGP still evolves according to fluid dynamics. We model the hadronic freeze-out of light degrees of freedom with the blast-wave parameters fit from experiment data such as elliptic flow and $\Lambda$ polarization \citep{Arslan:2024dwi}. 

We parameterize the spacetime coordinate $x^{\mu}=(\tau\cosh\eta,r\cos\phi_{s},r\sin\phi_{s},\tau\sinh\eta)$,
where $\phi_{s}$ is the transverse azimuthal angle in the coordinate
space, $\eta=\frac{1}{2}\log\frac{t+z}{t-z}$ the spacetime rapidity
and $\tau=\sqrt{t^{2}-z^{2}}$ the proper time. We assume the boost
invariance along the longitudinal ($z$) direction, so the longitudinal
velocity can be taken as $v^{z}=\frac{u^{z}}{u^{t}}=\frac{z}{t}$
and the freeze-out fluid velocity as a function of $\eta$, transverse
rapidity $\rho$ and velocity transverse azimuthal angle $\phi_{b}$
can be parameterized as
\begin{eqnarray}
u^{\mu}(x) & = & (\cosh\eta\cosh\rho,\sinh\rho\cos\phi_{b},\sinh\rho\sin\phi_{b},\sinh\eta\cosh\rho).\label{eq:Flow-parameter}
\end{eqnarray}
The transverse
rapidity $\rho$ is characterized by several parameters, 
\begin{eqnarray}
\rho(r,\phi_{s},\eta) & = & \tilde{r}\left[\rho_{0}+\rho_{1}(\eta)\cos\phi_{b}+\rho_{2}\cos2\phi_{b}\right],
\end{eqnarray}
where $\tan\phi_{b}=\tan\phi_{s}(R_{x}^{2}/R_{y}^{2})$ with $R_{x(y)}$
the effective radii of the elliptic source in $x(y)$ direction. We
introduce $\epsilon=R_{y}-R_{x}$ and $R=\frac{R_{x}+R_{y}}{2}$ with
$\epsilon\ll R$. Here, $\rho_{0}$ is the mean transverse rapidity,
$\rho_{1}(\eta)=\alpha_{1}\eta$ and $\rho_{2}$ characterize the
azimuthal anisotropy of the transverse rapidity. In general $\alpha_{1},\rho_{2},\epsilon R^{-1}\ll\rho_{0}$
so $\tan\phi_{b}\approx\left(1-2\frac{\epsilon}{R}\right)\tan\phi_{s}$.
The dimensionless $\tilde{r}$ is defined as 
\begin{eqnarray}
\tilde{r} & = & \sqrt{\frac{(r\cos\phi_{s})^{2}}{R_{x}^{2}}+\frac{(r\sin\phi_{s})^{2}}{R_{y}^{2}}}\approx\frac{r}{R}\left(1+\frac{1}{2}\frac{\epsilon}{R}\cos2\phi_{s}\right).
\end{eqnarray}
The freeze-out hypersurface is parameterized by 
\begin{eqnarray}
\int_{\Sigma(\tau)}{\rm d}\sigma^{\mu}(x) & = & \tau\int_{-\infty}^{+\infty}{\rm d}\eta\int_{0}^{R}r{\rm d}r\int_{0}^{2\pi}{\rm d}\phi_{s}\left(\cosh\eta,-\partial_{x}\tau,-\partial_{y}\tau,-\sinh\eta\right),\label{eq:Freezeout-hs}
\end{eqnarray}
and we will assume the freeze-out happens at a constant proper time
$\tau_{{\rm f}}$. Since the anisotropic flow is driven by collective
particle momentum flow, it would be reasonable to assume $\eta=Y$
and $\phi_{b}=\phi_{p}$ by adding $\delta(\eta-Y)\delta(\phi_{p}-\phi_{b})$
factor in the freeze-out integrands. However, this particle-flow correspondence does not hold for heavy particles, particularly bottomonium, whose center-of-mass motion is not fully thermalized, as discussed in section \ref{sec:Heavy-quarkonium-spin}. With this correspondence, $\boldsymbol{v}^2\approx 0.756$ applies near the outer edge of the hypersurface for $\rho_0=1.33$ even at $p_{\rm T}=0$, outside the assumed EFT regime $\boldsymbol{v}^2\ll1$. We therefore use this freeze-out model only for an illustrative phenomenological estimate, extrapolating the truncated collision kernel beyond its controlled domain. A more consistent calculation should follow the the physical trajectory through a hydrodynamical background as done in phenomenological studies using QTraj and restrict the relative velocities to the EFT regime.

We are interested in the polarization measured in the quarkonium rest frame in the event plane. For each event, we choose the laboratory $z$ axis along the beam direction and the $y$ axis along the event-plane normal. We use the canonical spin basis associated with rotation-free boosts between the laboratory and quarkonium-rest frames. In the quarkonium-rest frame, the spin-quantization direction is along $y$ axis with $\boldsymbol{\varepsilon}_{\rm{ep},0}=(0,1,0)$ following Ref.~\citep{Sheng:2022ffb} and then
$\lambda_{\theta}\propto\boldsymbol{\varepsilon}_{0}^{i}\boldsymbol{\varepsilon}_{0}^{*j}\delta n_{ji,{\rm 1S}}^{s}(t;\boldsymbol{v})$.
The ALICE experiment measures the polarization of quarkonia at forward
rapidity $2.5<Y<4$,  and the $q_{{\rm T}}$ distribution is 
\begin{eqnarray}
\lambda_{\theta}(q_{{\rm T}}) & = & \frac{\int\mathcal{D}[Y,\phi_{p},\eta,r,\phi_{s}]\left[-3\boldsymbol{\varepsilon}_{\text{ep},0}^{y}\boldsymbol{\varepsilon}_{\text{ep},0}^{*y}\delta n_{yy,1{\rm S}}^{s}(t_{{\rm F}},\boldsymbol{q},\boldsymbol{u})\right]}{\int\mathcal{D}[Y,\phi_{p},\eta,r,\phi_{s}]\left[\frac{4}{3}n_{{\rm 1S}}^{s}(t_{{\rm F}})+\boldsymbol{\varepsilon}_{\text{ep},0}^{y}\boldsymbol{\varepsilon}_{\text{ep},0}^{*y}\delta n_{yy,1{\rm S}}^{s}(t_{{\rm F}},\boldsymbol{q},\boldsymbol{u})\right]},\label{eq:q_T-spectrum-formulae}
\end{eqnarray}
where the integral measure is
\begin{eqnarray}
\int\mathcal{D}[Y,\phi_{p},\eta,r,\phi_{s}] & = & \int_{2.5}^{4}{\rm d}Y\int_{0}^{2\pi}{\rm d}\phi_{p}\int_{-\infty}^{+\infty}{\rm d}\eta\int_{0}^{R}r{\rm d}r\int_{0}^{2\pi}{\rm d}\phi_{s}\delta(\eta-Y)\delta(\phi_{b}-\phi_{p}).\nonumber\\
& &
\end{eqnarray}
We remark that, in general, a $q_{\rm T}$-dependent distribution weight should accompany the integrands in both numerator and denominator of eq.~\eqref{eq:q_T-spectrum-formulae}. In our simplified treatment, however, we use the same unpolarized background, evaluated at zero momentum in the quarkonium rest frame for all lab-frame momenta. The $q_{\rm T}$ dependence solely enters through $\boldsymbol{v}^i\boldsymbol{v}^j$ and the common background weight therefore cancels also after integrating the numerator and denominator over momentum. Incorporating a realistic $q_{\rm T}$ production spectrum is left for future work. 

\begin{figure}
\begin{centering}
\includegraphics[width=0.46\linewidth]{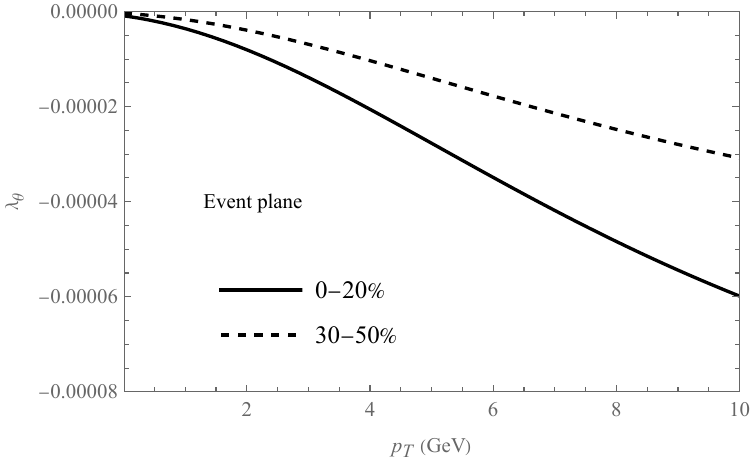}%
\hfill
\includegraphics[width=0.43\linewidth]{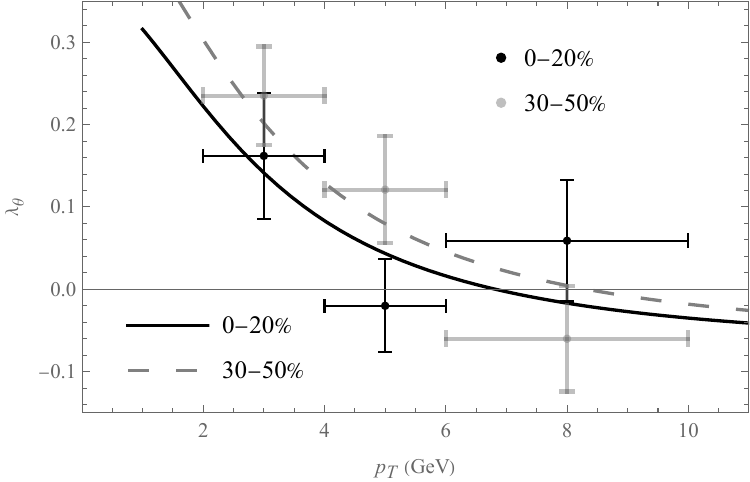}
\par\end{centering}
\caption{The $p_{\rm{T}}$ dependence of polarization parameter $\lambda_{\theta}$ with respect to the event plane for bottomonium $\Upsilon(1\rm S)$ and charmonium $J/\psi$ in Pb-Pb $\sqrt{s_{\rm NN}}=5.02\,{\rm TeV}$ at ALICE kinematic region $Y\in(2.5,4)$ and $\phi_{p}\in(0,2\pi]$ at centrality 0-20$\%$ and 30-50 $\%$. The freeze-out assumptions are discussed in section \ref{sec:Freeze-out-modeling}. The $J/\psi$ example is used as an illustrative result for the recoil mechanism by adjusting the initial condition and collision-term magnitude.\protect\label{fig:pT-spectrum}}
\end{figure}

First performing azimuthal angle integration, we find only 3 blast-wave parameters $\{\rho_{0},\rho_{2},\epsilon/R\}$
which depict the shape and size of freeze-out hypersurface, are important here
to our observable, and they take $\{1.33,0.05,0.088\}$ for $0$-20$\%$
centrality and $\{1.19,0.10,0.164\}$ for $30$-50$\%$ centrality \citep{Arslan:2024dwi}. Plugging in the solution of the spin-dependent distribution (\ref{eq:Int-form-SReq}) with $\boldsymbol{v}=\boldsymbol{v}(\boldsymbol{q},\boldsymbol{u})$
and the parametrization of the lab-frame quarkonium momentum $\boldsymbol{q}$
and fluid velocity $\boldsymbol{u}$ into eqs.~\eqref{eq:Momentum-parameter} and \eqref{eq:Flow-parameter}, we plot the $q_{{\rm T}}$ dependence of $\lambda_{\theta}$ in figure~\ref{fig:pT-spectrum}. After further integrating over $q_{{\rm T}}$ from 2 to 10 GeV without a realistic production-spectrum weight, we obtain the $q_{\rm T}$-averaged
polarization from the medium effects up to the leading fluid gradient, $\lambda_{\theta}=-2.08\times10^{-5}$ in 0-20$\%$ centrality
and $\lambda_{\theta}=-1.05\times10^{-5}$ in $30$-50$\%$ centrality. In both cases, the contribution from quarkonium dissociation dominates the polarization whereas the regeneration only gives about $10^{-3}$ portion of the total polarization parameter.

Consequently, the medium-induced polarization of $\Upsilon(1\rm{S})$ is almost zero at LHC energy in heavy-ion collisions at small $q_{\rm T}$ regime from our illustrative extrapolation. Besides, we also find the obtained $\Upsilon(1\rm S)$ polarization is negative in comparison with $J/\psi$ \cite{ALICE:2023jad}. A possible origin of such difference is a competition between dissociation and regeneration processes. Within our bottomonium calculation, the regeneration contribution from color-octet densities is much smaller than the color-singlet dissociation contribution. A sufficiently large regeneration contribution could change the sign of $\lambda_{\theta}$ for charmonia but establishing whether it occurs would require a separate calculation. Since the anisotropic dissociation rate increases for smaller heavy-quark mass, the polarization of charmonium in our framework can be larger than that of bottomonium. As a phenomenological study, we adjust the value of spin collision kernel and initial conditions and plot the $q_{\rm T}$ spectrum for charmonium the right panel of figure \ref{fig:pT-spectrum}. Parameterizing eq.~\eqref{eq:Int-form-SReq} as $\delta n_{ji}(t_{\rm{F}})=\delta n_{ji}(t_0)+\boldsymbol{v}_{i}\boldsymbol{v}_{j}C$, we choose the initial condition and normalized collision kernel $\left\{\frac{\delta n(t_0)}{n(t_{\rm F})}=-0.15,\frac{C}{n(t_{\rm F})}=48\right\}$ and $\left\{\frac{\delta n(t_0)}{n(t_{\rm F})}=-0.19,\frac{C}{n(t_{\rm F})}=64\right\}$ for 0-20$\%$ and 30-50$\%$ centrality, separately. With these adjusted parameters, the illustrative $J/\psi$ curve qualitatively reproduces the trend and magnitude of the measured polarization. A detailed, consistent treatment of regeneration and initial conditions for $J/\psi$ is left for future work.  
Other possible contributions from initial condition and feed-down effects may also influence the results. 

We also comment that the rotating-wave approximation (or $\boldsymbol{r}$-gradient expansion for the distribution function) for scattering state is not valid in general. The contribution of full color-octet density operator overlap $_{o}\langle\boldsymbol{p}_{1}|\hat{\rho}_{o}^{\text{sin}}(t)|\boldsymbol{p}_{2}\rangle_{o}$ contributes the spin collision kernel about $1.06\times10^{-7}$ while the diagonal overlap in $\boldsymbol{p}$ space gives about $7.37\times10^{-8}$ at 0-20$\%$ centrality, indicating that the diagonal approximation to the density matrix in relative-momentum space, and hence the leading-order scattering-state distribution of the $\boldsymbol{r}$ expansion, produces considerable changes to the regeneration contributions. Inclusion of off-diagonal elements retains the coherent quantum effects in the color-octet regeneration kernel. 

\subsection{Physical implications of spin anisotropic dissociation rate}

\begin{figure}
\begin{centering}
\includegraphics[scale=0.55]{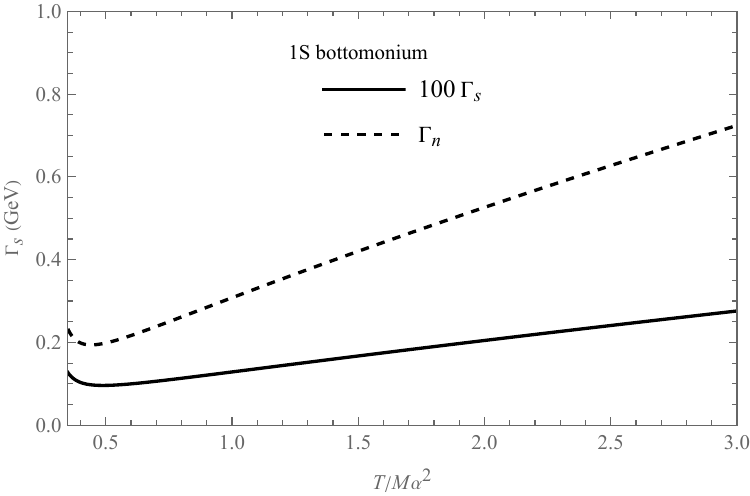}$\quad\quad$\includegraphics[scale=0.55]{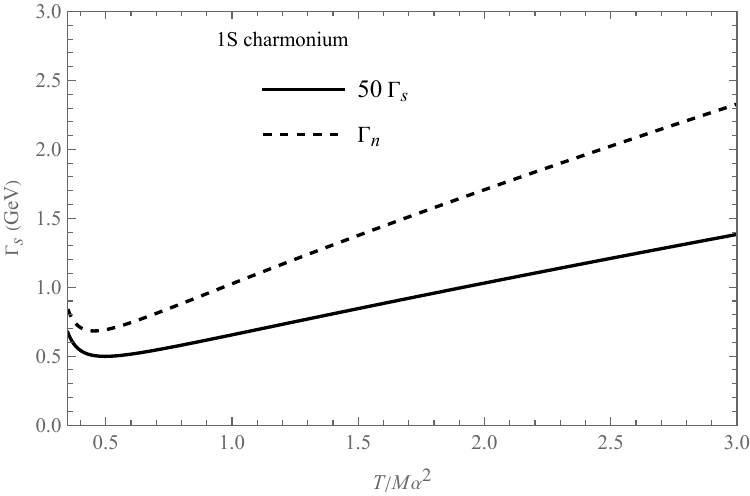}
\par\end{centering}
\caption{The heavy-quarkonium anisotropic dissociation rate $\Gamma_{s}$ and number damping
rate $\Gamma_{n}$ from perturbative calculations given in eq.~\eqref{eq:perturbative-spin-DR}
and eq.~\eqref{eq:perturbative-number-DR} for 1S bottomonium (left panel)
and charmonium (right panel) as a function of temperature.\protect\label{fig:1}}
\end{figure}

\begin{figure}
\begin{centering}
\includegraphics[scale=0.55]{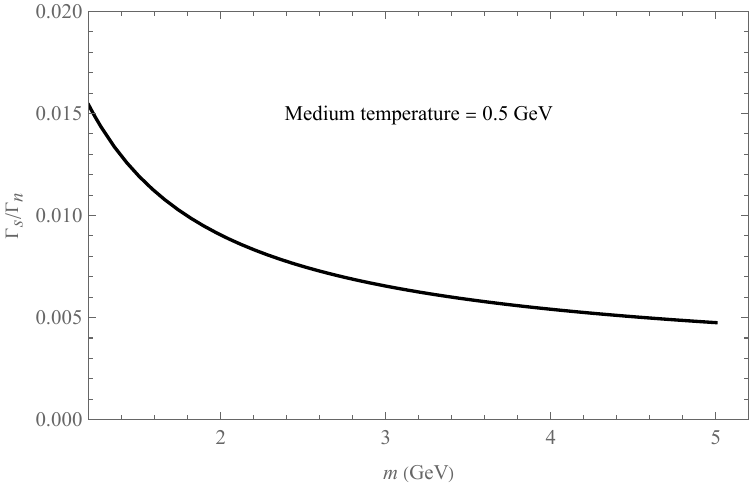}$\quad\quad$\includegraphics[scale=0.55]{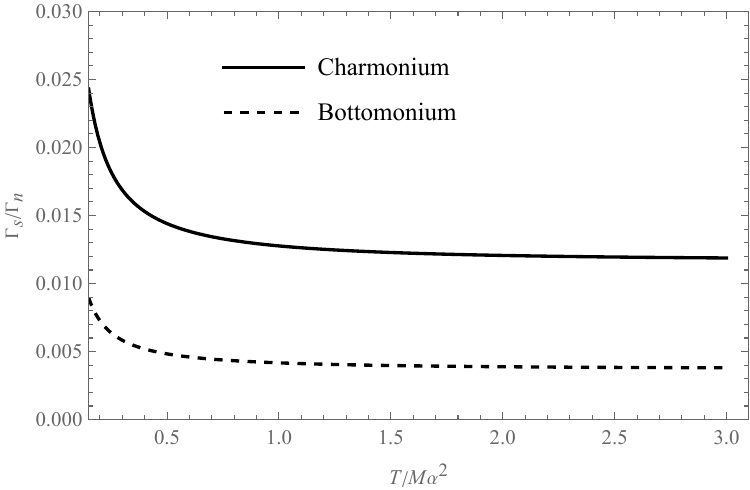}
\par\end{centering}
\caption{The ratio of heavy-quarkonium anisotropic dissociation rate and number damping
rate $\Gamma_{s}/\Gamma_{{\rm n}}$ as a function of temperature and
heavy quark mass.\protect\label{fig:2}}
\end{figure}

Lastly, we make an intuitive explanation of the magnitude of heavy-quarkonium polarization. The rate $\Gamma^{s}_{n}$ comes from Wightman gluon-field correlators $\mathcal{G}_{21}^{ji}$ as shown in eq.~\eqref{eq:CCK-qrf}. It characterizes the typical time scale of spin alignment generation of quarkonium. It originates from NRQCD spin-dependent interactions for heavy quarks, $H_{\text{int}}=c_{\text{F}}\boldsymbol{S}\cdot \frac{g\boldsymbol{B}}{M}+ic_{\text{S}}\boldsymbol{S}\cdot\frac{[\boldsymbol{D}\times,g\boldsymbol{E}]}{4M^2}$. In the weak-coupling case, we may treat heavy quarks as subsystem surrounded by gluon fields, and their spin-fluctuation correlation can be described by the same gluon-field correlators $\mathcal{G}_{21}^{ji}$ from linear response theory (see also recent works e.g.~\citep{Liang:2026jna} for related discussions). Therefore, the rate $\Gamma_{s}$ describes anisotropic generation of spin density matrix in quarkonium evolution and is thus related to the quark-antiquark spin-spin correlation encoded in that matrix. We can compare the quarkonium
spin anisotropic dissociation rate $\Gamma_{s}$ with the number damping rate $\Gamma_{n}$
under the same hydrodynamical background evolution. If the dissociation process is dominant, estimating the accumulation spin alignment over a characteristic time $\delta t\sim\Gamma_n^{-1}$ gives $\delta\rho_{00}\sim\delta n_{ji}/n\sim\boldsymbol{v}^{2}\Gamma_{s}/\Gamma_{n}$.
The rest-frame heavy quarkonium number damping rate can be similarly
derived in the quantum optical limit from a Boltzmann equation (see derivations from
self energy in Refs.~\cite{Brambilla:2011sg,Biondini:2023zcz}), e.g.
for the 1S state up to leading-recoil corrections,
\begin{eqnarray}
 \begin{split}
 \Gamma_{n}(T)& = \frac{g^{2}T_{{\rm F}}(N_{c}^{2}-1)}{N_{c}}\int_{\boldsymbol{p}}|\langle1{\rm S}|\boldsymbol{r}|\boldsymbol{p}\rangle|^{2}\frac{-(\Delta E_{1{\rm S},\boldsymbol{p}}^{so})^3}{3\pi}n_{{\rm B}}^{<}(-\Delta E_{1{\rm S},\boldsymbol{p}}^{so}) \\
 & = \frac{2^{7}17^{2}}{3^{6}}M\alpha_{{\rm s}}(T)\alpha_{{\rm s}}^{4}(a_{0}^{-1})\int_{0}^{+\infty}{\rm d}x\frac{x(x^{2}+2^{-6})e^{\frac{1}{2x}\arctan x}}{(x^{2}+1)^{3}(e^{\frac{\pi}{4x}}-1)}\frac{1}{e^{\frac{4M\alpha_{{\rm s}}^{2}(a_{0}^{-1})}{9T}(1+x^{2})}-1}.
 \end{split}
 \label{eq:perturbative-number-DR} 
\end{eqnarray}
In contrast to the anisotropic dissociation contribution $\boldsymbol{v}^i\boldsymbol{v}^j\Gamma_{s}$ that appears when we consider the relative motion between quarkonium and medium, the dissociation rate $\Gamma_n$ survives even for in-medium static quarkonium. We plot the spin anisotropic dissociation and number damping rate as a function of temperature
$T$ for 1S bottomonium with $M_{{\rm b}}\approx4.85$ GeV and 1S charmonium
with $M_{\text{c}}\approx1.27$ GeV in figure~\ref{fig:1}. We find in both cases that the anisotropic dissociation rate is much smaller than number damping
rate, both rates drop as the quark mass gets heavier, and more importantly
the anisotropic dissociation rate drops faster. The descent of both rates for
heavier quarkonium mass is easy to understand: heavier mass means larger Coulomb binding energy and suppresses the thermal dissociation rates. Although $\Gamma_s$ itself is not the complete spin relaxation rate, the heavy-quark mass dependence of $\Gamma_{s}/\Gamma_{n}$ as shown in figure~\ref{fig:2} may be compared qualitatively with the general expectation that the spin degree of freedom relaxes more slowly for heavier particles.
We also find this ratio decreases with temperature. 
For comparison, a study with four-fermion interaction in the relativistic regime finds that the typical spin relaxation time is comparable with the relaxation time of conventional hydrodynamical modes within spin hydrodynamics frameworks~\citep{Wagner:2024fhf}. As an order-of-magnitude estimation, we find
for the bottomonium, $\delta\rho_{00}\lesssim10^{-4}$ given $\Gamma_{s}/\Gamma_{n}\sim10^{-3}$,
indicating the polarization effects could be very small and are consistent with our numerical calculation based on the QTraj simulations. For the charmonium, we find at lower temperature $\Gamma_s/\Gamma_n\sim2\times10^{-2}$, and combining with $\boldsymbol{v}^2\sim0.3$, its polarization may reach $\delta\rho_{00}\sim10^{-2}$. In such case, both the regeneration contribution and the spin dynamics of individual charm quarks will be important and change the behavior of $\delta\rho_{00}$ significantly.


\section{Conclusions and outlook \protect\label{sec:Conclusions}}

In this work, we have studied the heavy-quarkonium spin alignment in a hot quark-gluon plasma (QGP) within an open quantum system framework, incorporating spin-dependent potential nonrelativistic QCD (pNRQCD) interactions. Including color-singlet dissociation and color-octet regeneration mechanisms, we retain the recoil corrections and derive a quantum master equation for the color-singlet, spin-triplet quarkonium density matrix under the energy hierarchy $M\gg Mv\gtrsim\sqrt{MT}\gg Mv^2\gtrsim T$ that may be appropriate for tightly bound quarkonia in the restricted kinematic region where the relative velocity remains small. Working in the quantum optical regime and neglecting the off-diagonal bound-state spin coherences, we perform a Wigner transformation and a semiclassical gradient expansion to obtain projected Boltzmann equations for the spin-diagonal distribution function, which determine the polarization parameter $\lambda_{\theta}$ in the quarkonium dilepton-decay channel. We also retain the off-diagonal momentum-space octet density matrix in the regeneration kernel, preserving the associated quantum-coherence effects. At the order considered, the spin-dependent source terms depend only on the unpolarized color singlet and octet density matrix elements. We evaluate the corresponding spin collision kernel perturbatively through relative order $\frac{T}{M}$ in the quarkonium rest frame.

Up to $\mathcal{O}(\rm Kn^0)$,  the leading nonzero quarkonium spin polarization arises from the relative motion $\boldsymbol{v}$ between quarkonium and medium. This mechanism couples the heavy quarkonium dynamics to the fluid velocity field, extending descriptions based primarily on the medium temperature evolution~\cite{Brambilla:2017zei,Brambilla:2022ynh}. Spin-chromomagnetic and spin-orbit interactions both contribute at the same leading nonzero order in the spin-alignment collision kernel, with the recoil effects in the Hamiltonian and spin-dipole interactions playing an essential role. These corrections generate a Moyal-product structure in the medium-rest frame collision kernel and introduce the mixed chromoelectric-chromomagnetic correlators and chromoelectric-chromoelectric correlators alongside the chromomagnetic correlators. We also identify an anisotropic dissociation coefficient $\Gamma_s$ characterizing the generation of anisotropic components of the quarkonium spin density matrix. Within our Coulombic setup, this coefficient decreases with increasing heavy-quark mass at fixed temperature.

As an illustrative phenomenological application, we study the spin alignment of $\Upsilon$(1S) in Pb-Pb collisions at $\sqrt{s_{\rm NN}}=5.02\,\rm{TeV}$ relevant to ALICE measurements. At zeroth order in fluid gradients, we reduce the transport equation to a spin-generation rate equation by treating the relative velocity as fixed and neglecting the quarkonium center-of-mass momentum evolution. For this exploratory estimate, we evaluate the analytically derived spin-dependent collision terms using background singlet and octet density matrices evolved with QTraj under a simplified temperature evolution. 
We then estimate transverse-momentum dependence of $\lambda_{\theta}$ with a blast-wave parameterization of freeze-out and quote an unweighted $q_{\rm T}$ average over $2<q_{\rm T}<10\,{\rm GeV}$, without a realistic production-spectrum weight. Within this simplified treatment, the medium-induced alignment for bottomonium is small with $|\lambda_{\theta}|\sim10^{-5}$. In the present bottomonium estimate, color-singlet dissociation dominates over color-octet regeneration. 
Momentum-space coherence effects give a non-negligible contribution relative to the diagonal contribution within the regeneration spin collision kernel (see also the recent study on position-space coherence in bottomonium evolution within the quantum Brownian regime~\cite{Brambilla:2026ivg}). We also estimate the medium-induced charmonium polarization from this anisotropic mechanism, and we can describe its $p_{\rm T}$ dependence qualitatively by adjusting specific initial conditions.

Several extensions can be made to improve the theoretical description and its phenomenological applications. The quantum master equation~\eqref{eq:QME} and its resulting Boltzmann equations~\eqref{eq:SBE-mrf} and~\eqref{eq:SBE-qrf} are derived at leading order in the coupling. Extending the framework to a strongly coupled QGP requires appropriate Wilson-line dressing of chromofield correlators \cite{Yang:2024ejk}, together with a consistent gauge-covariant treatment of the color-octet density operators and center-of-mass propagation, including the kinetic term $\frac{\boldsymbol P_{\rm cm}^2}{4M}$. Nonperturbative definitions and determinations of the corresponding spin transport coefficients would provide essential simulation inputs. The present power counting also assumes $\boldsymbol P_{\rm cm}\sim\sqrt{MT}$, corresponding to $\boldsymbol v^2\sim \frac{T}{M}\ll1$. Such thermal center-of-mass momentum scaling does not require a thermalized quarkonium distribution, but it restricts the momentum range in which the expansion is controlled. Realistic bottomonium distributions and fluid profiles can sample larger relative velocities, and thus a phenomenological extrapolation is used in this paper. Extending the description to larger center-of-mass momenta requires a consistent revision of the pNRQCD power counting. The QTraj background density matrices used in the numerical estimate are obtained in the hierarchy $T\gg Mv^2$ which differs from the quantum-optical hierarchy used in deriving the spin equation. Simulating the quantum master equations beyond $E/T$ expansion would require a different treatment from QTraj.  Within the present framework, a coupled solution of the Boltzmann equations with collision terms~\eqref{eq:qrf-pertur-disso} and~\eqref{eq:qrf-pertur-reg} is needed to incorporate momentum evolution and the spacetime-dependent hydrodynamic background. A more quantitative comparison with experiment should further use a freeze-out hypersurface from hydrodynamic simulations, include the quarkonium production-spectrum weights, and account for initial polarization and polarization transfer through feed-down decays. These developments would enable a more complete assessment of medium-induced quarkonium spin alignment and constitute important directions for future work.

\acknowledgments

We would like to thank Antonio Vairo for useful discussions. S. F. is thankful to Wen-Bo Dong for helpful discussions and providing the updated blast-wave parameter. This work is supported
in part by the DFG cluster of excellence ORIGINS funded by
the Deutsche Forschungsgemeinschaft (DFG) under Germany's Excellence Strategy-EXC-2094-390783311. This work is also supported by the European Union. Views and opinions expressed are however those of authors only and do not necessarily reflect those of the European Union or the European Research Council Executive Agency. Neither the European Union nor the granting authority can be held responsible for them. This work is supported by the advanced grant ERC-2023-ADG-Project EFT-XYZ. S. F. is supported in part by the China Scholarship Council (CSC). The work of N.B. is supported by the DFG Grant No. BR4058/5-1 ``Open quantum systems and effective field theories for hard probes of hot and/or dense medium''. The authors also acknowledge the use of Codex (OpenAI) for assistance with grammar and language improvement. 

\appendix

\section{Dilepton decay distribution in circular basis\protect\label{sec:Dilepton-decay-distribution}}

In this section, we derive the dilepton decay angular distribution
in the circular basis of $\varepsilon_{\lambda}^{\mu}$, in which
case the coefficients of decomposition of $k_{1}^{\mu}$ (\ref{eq:k1-decomposition})
can be complex. We choose 
\begin{eqnarray}
|\boldsymbol{k}_{1}|\cos\theta & = & -k_{1}\cdot\varepsilon_{0}^{*},\\
|\boldsymbol{k}_{1}|\sin\theta\frac{1}{\sqrt{2}}e^{i\phi} & = & -k_{1}\cdot\varepsilon_{1}^{*},\\
-|\boldsymbol{k}_{1}|\sin\theta\frac{1}{\sqrt{2}}e^{-i\phi} & = & -k_{1}\cdot\varepsilon_{-1}^{*},
\end{eqnarray}
and $\boldsymbol{\varepsilon}_{1}=\frac{1}{\sqrt{2}}(1,-i,0),\;\boldsymbol{\varepsilon}_{-1}=-\frac{1}{\sqrt{2}}(1,i,0)$
is a special case within such choice. Then 
\begin{eqnarray}
\frac{{\rm d}\mathcal{R}}{{\rm d}^{4}x{\rm d}^{4}q{\rm d}\Omega} & = & \frac{-Q^{2}4\alpha_{e}^{2}}{(2\pi)^{4}q^{4}}\sqrt{1-\frac{4m^{2}}{q^{2}}}\biggl\{-\frac{q^{2}}{4}{\rm Tr}\rho_{\lambda\lambda^{\prime}}+\frac{1}{2}|\boldsymbol{k}_{1}|^{2}\biggl[(1-\cos^{2}\theta){\rm Tr}\rho_{\lambda\lambda^{\prime}}\nonumber \\
 &  & \;+(3\cos^{2}\theta-1)\rho_{00}+2\sin2\theta\cos\phi\frac{{\rm Re}(\rho_{01}-\rho_{0-1})}{\sqrt{2}}+2\sin2\theta\sin\phi\frac{{\rm Im}(\rho_{01}+\rho_{0-1})}{\sqrt{2}}\nonumber \\
 &  & \;-2\sin^{2}\theta\cos2\phi{\rm Re}\rho_{1-1}+2\sin^{2}\theta\sin2\phi{\rm Im}\rho_{1-1}\biggr]\biggr\},
\end{eqnarray}
for massless leptons  we obtain the well-known formulae \cite{Faccioli:2010kd}
\begin{eqnarray}
\frac{{\rm d}\mathcal{R}}{{\rm d}^{4}x{\rm d}^{4}q{\rm d}\Omega} & = & \frac{Q^{2}\alpha_{e}^{2}}{2(2\pi)^{4}q^{2}}\left[{\rm Tr}\rho_{\lambda\lambda^{\prime}}(q)+\rho_{00}\right]\biggl\{1+\lambda_{\theta}\cos^{2}\theta+\lambda_{\theta\phi}\sin2\theta\cos\phi\nonumber \\
 &  & \;+\lambda_{\theta\phi}^{\perp}\sin2\theta\sin\phi+\lambda_{\phi}\sin^{2}\theta\cos2\phi+\lambda_{\phi}^{\perp}\sin^{2}\theta\sin2\phi\biggr\},
\end{eqnarray}
where we defined 
\begin{eqnarray}
\lambda_{\theta} & = & \frac{-3\rho_{00}+{\rm Tr}\rho_{\lambda\lambda^{\prime}}}{\rho_{00}+{\rm Tr}\rho_{\lambda\lambda^{\prime}}(q)},\\
\lambda_{\phi} & = & \frac{2{\rm Re}\rho_{1-1}(q)}{{\rm Tr}\rho_{\lambda\lambda^{\prime}}(q)+\rho_{00}},\\
\lambda_{\theta\phi} & = & -\frac{2{\rm Re}\left(\rho_{01}(q)-\rho_{0-1}(q)\right)}{\sqrt{2}({\rm Tr}\rho_{\lambda\lambda^{\prime}}(q)+\rho_{00})},\\
\lambda_{\phi}^{\perp} & = & -\frac{2{\rm Im}\rho_{1-1}(q)}{{\rm Tr}\rho_{\lambda\lambda^{\prime}}(q)+\rho_{00}},\\
\lambda_{\theta\phi}^{\perp} & = & -\frac{2{\rm Im}\left(\rho_{01}(q)+\rho_{0-1}(q)\right)}{\sqrt{2}({\rm Tr}\rho_{\lambda\lambda^{\prime}}(q)+\rho_{00})}.
\end{eqnarray}
If integrating over the azimuthal angle, 
\begin{eqnarray}
\frac{{\rm d}\mathcal{R}}{{\rm d}^{4}x{\rm d}^{4}q{\rm d}\cos\theta} & = & \frac{Q^{2}\alpha_{e}^{2}}{2(2\pi)^{3}q^{2}}{\rm Tr}\rho_{\lambda\lambda^{\prime}}(1+\rho_{00}^{{\rm N}})\left[1+\frac{1-3\rho_{00}^{{\rm N}}}{1+\rho_{00}^{{\rm N}}}\cos^{2}\theta\right].
\end{eqnarray}
The same equation can also be derived in the Cartesian basis.

\section{Spin-dependent density matrix elements from Wightman pNRQCD correlator\protect\label{sec:Spin-dependent-density-matrix}}

The 2-body fields in pNRQCD have the following mode expansions
\citep{Yao:2018nmy} for bound and scattering states separately,

\begin{eqnarray}
S_{\text{sin}}(R,\boldsymbol{r}) & = & \int_{\boldsymbol{P}}e^{i\boldsymbol{P}\cdot\boldsymbol{R}}\left(\sum_{n}e^{-iE_{n}^{\boldsymbol{P}}t}\phi_{n}(\boldsymbol{r})a_{n}^{\boldsymbol{P}}+\int_{\boldsymbol{p}}e^{-iE_{p}^{\boldsymbol{P}}t}\phi_{\boldsymbol{p}}(\boldsymbol{r})b_{\boldsymbol{p}}^{\boldsymbol{P}}\right),\label{eq:Spin_singlet_pw_decomp}\\
S_{i}(R,\boldsymbol{r}) & = & \int_{\boldsymbol{P}}\sum_{\lambda}\boldsymbol{\epsilon}_{i}^{\lambda}e^{i\boldsymbol{P}\cdot\boldsymbol{R}}\left(\sum_{n}e^{-iE_{n}^{\boldsymbol{P}}t}\phi_{n,\lambda}(\boldsymbol{r})a_{n,\lambda}^{\boldsymbol{P}}+\int_{\boldsymbol{p}}e^{-iE_{p}^{\boldsymbol{P}}t}\phi_{\boldsymbol{p},\lambda}(\boldsymbol{r})b_{\boldsymbol{p},\lambda}^{\boldsymbol{P}}\right),\label{eq:Spin_triplet_pw_decomp}
\end{eqnarray}
where $n$ is the collection of quantum numbers except for spin and
$\lambda$ is the magnetic quantum number for the spin state. The center-of-mass
momentum dependence of the interpolating field energy is explicitly
included. Such decompositions, eqs.~\eqref{eq:Spin_singlet_pw_decomp}
and \eqref{eq:Spin_triplet_pw_decomp}, are valid for $S$-wave spin-0
and spin-1 quarkonium. The first and second parts account for the
quarkonium bound state and scattering state separately. $\phi_{n(\lambda)}(\boldsymbol{r})=\langle\boldsymbol{r}|n(\lambda)\rangle_{s}$,
$\phi_{p(\lambda)}(\boldsymbol{r})=\langle\boldsymbol{r}|\boldsymbol{p}(\lambda)\rangle_{s}$
are the corresponding wavefunctions and $a_{n(\lambda)}^{\boldsymbol{P}},b_{\boldsymbol{p}(\lambda)}^{\boldsymbol{P}}$
are the annihilation operators for the interpolating two-body fields
accordingly. The creation/annihilation operators satisfy commutation
relations, e.g.
\begin{eqnarray}
[a_{n\lambda}^{\boldsymbol{p}},a_{n^{\prime}\lambda^{\prime}}^{\boldsymbol{p}^{\prime},\dagger}] & = & (2\pi)^{3}\delta^{(3)}(\boldsymbol{p}-\boldsymbol{p}^{\prime})\delta_{nn^{\prime}}\delta_{\lambda\lambda^{\prime}},\\{}
[b_{\boldsymbol{p},\lambda}^{\boldsymbol{P}},b_{\boldsymbol{p}^{\prime},\lambda^{\prime}}^{\boldsymbol{P}^{\prime},\dagger}] & = & (2\pi)^{3}\delta^{(3)}(\boldsymbol{P}-\boldsymbol{P}^{\prime})(2\pi)^{3}\delta^{(3)}(\boldsymbol{p}-\boldsymbol{p}^{\prime})\delta_{\lambda\lambda^{\prime}}.
\end{eqnarray}
where the creation operators create the states $|n\lambda;\boldsymbol{P}\rangle_{s}$
and $|\boldsymbol{p}\lambda;\boldsymbol{P}\rangle_{s}$ 
\begin{eqnarray}
|n_{1}\lambda_{1};\boldsymbol{P}_{1}\rangle_{s}=a_{n_{1},\lambda_{1}}^{\boldsymbol{P}_{1},\dagger}|0\rangle & ,\;\; & |\boldsymbol{p}_{1}\lambda_{1};\boldsymbol{P}_{1}\rangle_{s}=b_{\boldsymbol{p}_{1},\lambda_{1}}^{\boldsymbol{P}_{1},\dagger}|0\rangle,
\end{eqnarray}
and they form a complete basis for the singlet fields
\begin{eqnarray}
\boldsymbol{1} & = & \int_{\boldsymbol{P}_{1}}\left(\sum_{n_{1},\lambda_{1}}|n_{1}\lambda_{1};\boldsymbol{P}_{1}\rangle_{s}{}_{s}\langle n_{1}\lambda_{1};\boldsymbol{P}_{1}|+\int_{\boldsymbol{p}_{1}}\sum_{\lambda_{1}}|\boldsymbol{p}_{1}\lambda_{1};\boldsymbol{P}_{1}\rangle_{s}{}_{s}\langle\boldsymbol{p}_{1}\lambda_{1};\boldsymbol{P}_{1}|\right),\label{eq:Unit-operator-1}\\
\boldsymbol{1} & = & \sum_{n_{1},\lambda_{1}}|n_{1}\lambda_{1}\rangle_{s}{}_{s}\langle n_{1}\lambda_{1}|+\int_{\boldsymbol{p}_{1}}\sum_{\lambda_{1}}|\boldsymbol{p}_{1}\lambda_{1}\rangle_{s}{}_{s}\langle\boldsymbol{p}_{1}\lambda_{1}|.\label{eq:Unit-operator-2}
\end{eqnarray}
 Similar decompositions also hold for octet fields. The subscripts
$s/o$ stands for color-singlet/octet states while the spin-singlet/triplet
states are already distinguished. Up to leading recoil effects,
\begin{eqnarray}
\langle\boldsymbol{R}|\boldsymbol{P}\rangle & = & e^{i\boldsymbol{P}\cdot\boldsymbol{R}},\quad|n\lambda;\boldsymbol{P}\rangle=|n\lambda\rangle\otimes|\boldsymbol{P}\rangle,\quad|\boldsymbol{p}\lambda;\boldsymbol{P}\rangle=|\boldsymbol{p}\lambda\rangle\otimes|\boldsymbol{P}\rangle.
\end{eqnarray}

So the coordinate-space two-body color-singlet correlator in eq.~\eqref{eq:Singlet_corr_fun_2} reduces to
\begin{eqnarray}
 &  & \left\langle S_{i}^{\dagger}(R,\boldsymbol{r})S_{j}(R^{\prime},\boldsymbol{r}^{\prime})\right\rangle \nonumber \\
 & = & \int_{\boldsymbol{P},\boldsymbol{P}^{\prime}}\sum_{\lambda,\lambda^{\prime}}\boldsymbol{\epsilon}_{i}^{\lambda,*}\boldsymbol{\epsilon}_{j}^{\lambda^{\prime}}e^{i(\boldsymbol{P}^{\prime}\cdot\boldsymbol{R}^{\prime}-\boldsymbol{P}\cdot\boldsymbol{R})}\left[\sum_{n,n^{\prime}}\phi_{n,\lambda}^{*}(\boldsymbol{r})\phi_{n^{\prime},\lambda^{\prime}}(\boldsymbol{r}^{\prime}){}_{s}\langle n^{\prime}\lambda^{\prime};\boldsymbol{P}^{\prime}|\hat{\rho}_{s}^{{\rm tri}}(t^{\prime};t)|n\lambda;\boldsymbol{P}\rangle_{s}\right.\nonumber \\
 &  & \;\left.+\int_{\boldsymbol{p},\boldsymbol{p}^{\prime}}\phi_{p,\lambda}^{*}(\boldsymbol{r})\phi_{p^{\prime},\lambda^{\prime}}(\boldsymbol{r}^{\prime}){}_{s}\langle\boldsymbol{p}^{\prime}\lambda^{\prime};\boldsymbol{P}^{\prime}|\hat{\rho}_{s}^{\text{tri}}(t^{\prime};t)|\boldsymbol{p}\lambda;\boldsymbol{P}\rangle_{s}\right]\nonumber \\
 & = & \langle\boldsymbol{R}^{\prime},\boldsymbol{r}^{\prime}|\hat{\rho}_{s}^{ji}(t^{\prime};t)|\boldsymbol{R},\boldsymbol{r}\rangle\label{eq:Singlet_12_correlator-1}
\end{eqnarray}
where in the first step we used
\begin{eqnarray}
{\rm Tr}\left(\hat{\rho}_{s}^{{\rm tri}}(t^{\prime};t)a_{n,\lambda}^{\boldsymbol{P},\dagger}a_{n^{\prime},\lambda^{\prime}}^{\boldsymbol{P}^{\prime}}\right) & = & \int_{\boldsymbol{P}_{1}}\sum_{n_{1},\lambda_{1}}\langle n_{1}\lambda_{1};\boldsymbol{P}_{1}|\hat{\rho}_{s}^{{\rm tri}}(t^{\prime};t)a_{n,\lambda}^{\boldsymbol{P},\dagger}a_{n^{\prime},\lambda^{\prime}}^{\boldsymbol{P}^{\prime}}a_{n_{1},\lambda_{1}}^{\boldsymbol{P}_{1},\dagger}|0\rangle\nonumber \\
 & = & _{s}\langle n^{\prime}\lambda^{\prime};\boldsymbol{P}^{\prime}|\hat{\rho}_{s}^{{\rm tri}}(t^{\prime};t)|n\lambda;\boldsymbol{P}\rangle_{s},
\end{eqnarray}
with $h_{s}^{\text{tri}}|n_{1}\lambda_{1};\boldsymbol{P}_{1}\rangle_{s}=E_{n_{1}}^{\boldsymbol{P}_{1}}|n_{1};\boldsymbol{P}_{1}\rangle_{s}$ where the spin magnetic quantum number is degenerate for a given eigenenergy $E_n^{\boldsymbol{P}}$.
We defined the spin-triplet color-singlet density operator
\begin{eqnarray}
\hat{\rho}_{s}^{{\rm tri}}(t^{\prime};t) & = & e^{-ih_{s}^{{\rm tri}}(t^{\prime}-t_0)}\hat{\rho}(t_{0})e^{ih_{s}^{{\rm tri}}(t-t_0)}.
\end{eqnarray}
This definition guarantees the tree-level density operator $\hat{\rho}_{s,0}^{{\rm tri}}(t^{\prime};t)$
in the absence of system-environment interaction evolves as 
\begin{eqnarray}
\hat{\rho}_{s,0}^{{\rm tri}}(t^{\prime};t) & = & e^{-ih_{s}^{{\rm tri}}(t^{\prime}-t_{0})}\hat{\rho}_{s,0}^{{\rm tri}}(t_{0};t_{0})e^{ih_{s}^{{\rm tri}}(t-t_{0})}.\label{eq:Spin-tri-color-s-operator}
\end{eqnarray}
In the second step we defined the spin-dependent color-singlet density
operator, 
\begin{eqnarray}
\hat{\rho}_{s}^{ji}(t^{\prime};t) & = & \sum_{\lambda,\lambda^{\prime}}\boldsymbol{\epsilon}_{i}^{\lambda,*}\boldsymbol{\epsilon}_{j}^{\lambda^{\prime}}\left[\sum_{n,n^{\prime}}|n^{\prime}\lambda^{\prime}\rangle{}_{s}{}_{s}\langle n^{\prime}\lambda^{\prime}|\hat{\rho}_{s}^{{\rm tri}}(t^{\prime};t)|n\lambda\rangle_{s}{}_{s}\langle n\lambda|\right.\nonumber \\
 &  & \;\left.+\int_{\boldsymbol{p},\boldsymbol{p}^{\prime}}|\boldsymbol{p}^{\prime}\lambda^{\prime}\rangle_{s}{}_{s}\langle\boldsymbol{p}^{\prime}\lambda^{\prime}|\hat{\rho}_{s}^{\text{tri}}(t^{\prime};t)|\boldsymbol{p}\lambda\rangle_{s}{}_{s}\langle\boldsymbol{p}\lambda|\right],\label{eq:Spin-dependent-DO}
\end{eqnarray}
which satisfies the following identity for spin-diagonal matrix elements 
\begin{eqnarray}
{}_s\langle n^{\prime}(\boldsymbol{p^{\prime}})\lambda|\hat{\rho}_{s}^{kk}(t^{\prime};t) |n(\boldsymbol{p})\lambda\rangle_s& = & {}_s\langle n^{\prime}(\boldsymbol{p^{\prime}})\lambda|\hat{\rho}_{s}^{{\rm tri}}(t^{\prime};t)|n(\boldsymbol{p})\lambda\rangle_s,
\end{eqnarray}
and its tree-level evolution preserve the properties of operators
in the Heisenberg picture following eq.~\eqref{eq:Spin-tri-color-s-operator}
\begin{eqnarray}
\hat{\rho}_{s,0}^{ji}(t^{\prime};t) & = & e^{-ih_{s}^{{\rm tri}}(t^{\prime}-t_{0})}\hat{\rho}_{s,0}^{ji}(t_{0};t_{0})e^{ih_{s}^{{\rm tri}}(t-t_{0})}.
\end{eqnarray}

The spin-dependent states
differ from their spin-independent counterparts up to a higher-order
correction: the states can be solved from a Lippmann-Schwinger equation, 
\begin{eqnarray}
|n\lambda\rangle & = & |n\rangle_{0}+\frac{1}{E-h_{0}}\delta V|n\lambda\rangle,
\end{eqnarray}
where $|n\rangle_{0}$ is the eigenstate of Hamiltonian $h_{0}=\frac{\boldsymbol{p}^{2}}{M}+\frac{\boldsymbol{P}^{2}}{4M}+V_{\text{Coulomb}}$
and $\delta V$ is the spin-dependent potential, and 
\begin{eqnarray}
\frac{1}{E-h_{0}}\delta V & \sim & \mathcal{O}\left(\frac{1}{M^{2}}\right)
\end{eqnarray}
in the weakly-coupled pNRQCD. Therefore up to leading order, the orbital wavefunction does not depend on the spin projection. For an initially unpolarized case, the system evolution generated by the leading-order Hamiltonian satisfies,
\begin{eqnarray}
\hat{\rho}_{s}^{ij}(t^{\prime};t) & \approx & \frac{1}{d-1}\left(\sum_{\lambda}\boldsymbol{\epsilon}_{i}^{\lambda,*}\boldsymbol{\epsilon}_{j}^{\lambda}\right)\sum_{n,n^{\prime}}|n^{\prime}\rangle{}_{s,0}{}_{s,0}\langle n^{\prime}|\hat{\rho}_{s}^{{\rm tri}}(t^{\prime};t)|n\rangle_{s,0}{}_{s,0}\langle n|\nonumber \\
 & = & \frac{\delta_{ij}}{d-1}\hat{\rho}_{s}^{{\rm tri}}(t^{\prime};t)+\mathcal{O}(M^{-2}).\label{eq:LO-Spin-dep-DO}
\end{eqnarray}
and thus
\begin{eqnarray}
\langle S_{i}^{\dagger}(R,\boldsymbol{r})S_{j}(R^{\prime},\boldsymbol{r}^{\prime})\rangle & \approx & \frac{\delta_{ij}}{d-1}{}_{s}\langle\boldsymbol{R}^{\prime},\boldsymbol{r}^{\prime}|\hat{\rho}_{s}^{\text{tri}}(t^{\prime};t)|\boldsymbol{R},\boldsymbol{r}\rangle_{s}.\label{eq:Leaing_Spin-t-color-s}
\end{eqnarray}
Eq.~\eqref{eq:Leaing_Spin-t-color-s} will be used as the zeroth-order background evolution in the collision kernel whereas the spin-dependent system-medium interaction generates the traceless correlator. The diagonal color-octet density operator is related to ``12'' Wightman function in a nontrivial way and discussed in the main text.

\section{Calculation of collision kernel}\label{Sec:Calculation-of-collision-kernel}

We list the calculation details of the Wigner transform of $-\frac{1}{d-1}{}_{s}\langle n\lambda;\boldsymbol{P}^{\prime}|\Sigma_{s,ji}(t)\hat{\rho}_{s}^{{\rm tri}}(t)|n\lambda;\boldsymbol{P}\rangle_{s}+{\rm H.c.}$
and $_{s}\langle n\lambda;\boldsymbol{P}^{\prime}|\Xi_{s,ji}\left(\rho_{o}^{{\rm sin}}(t);t\right)|n\lambda;\boldsymbol{P}\rangle_{s}$.
They give dissociation and regeneration collision terms of the spin
Boltzmann equation (\ref{eq:SBE-mrf}), respectively.

We first calculate the dissociation sector in detail.

\begin{eqnarray}
& &\Delta_{\lambda_2\lambda_3}^{ji}\mathcal{C}_{ji,n\lambda}^{-,{\rm mrf}}(t,\boldsymbol{x},\frac{\boldsymbol{P}^{\prime}+\boldsymbol{P}}{2}) \nonumber\\& = & \Delta_{\lambda_2\lambda_3}^{ji}\frac{1}{d-1}\int_{\boldsymbol{P}^{\prime}-\boldsymbol{P}}e^{i(\boldsymbol{P}^{\prime}-\boldsymbol{P})\cdot\boldsymbol{x}}{}_{s}\langle n\lambda;\boldsymbol{P}^{\prime}|\hat{\Sigma}_{s,ji}(t)\hat{\rho}_{s}^{{\rm tri}}(t)|n\lambda;\boldsymbol{P}\rangle_{s}+{\rm H.c.}\nonumber \\
 & = & \Delta_{\lambda_2\lambda_3}^{ji}\frac{1}{d-1}\int_{\boldsymbol{P}^{\prime}-\boldsymbol{P}}e^{i(\boldsymbol{P}^{\prime}-\boldsymbol{P})\cdot\boldsymbol{x}}\sum_{n_{1}\lambda_{1}}\int_{\boldsymbol{P}_{1}}{}_{s}\langle n\lambda;\boldsymbol{P}^{\prime}|\hat{\Sigma}_{s,ji}(t)|n_{1}\lambda_{1};\boldsymbol{P}_{1}\rangle_{s}\nonumber \\
 &  & \quad\times\delta_{n_{1}n}\;{}_{s}\langle n_{1}\lambda_{1};\boldsymbol{P}_{1}|\hat{\rho}_{s}^{{\rm tri}}(t)|n\lambda;\boldsymbol{P}\rangle_{s}+{\rm H.c.}
\end{eqnarray}
where we plug in the completeness relation (\ref{eq:Unit-operator-3}).
According to rotating-wave approximation, we drop the overlap $_{s}\langle\boldsymbol{p}_{1}\lambda_{1};\boldsymbol{P}_{1}|\hat{\rho}_{s}^{{\rm tri}}(t)|n\lambda;\boldsymbol{P}\rangle_{s}$
and estimate the $_{s}\langle n_{1}\lambda_{1};\boldsymbol{P}_{1}|\hat{\rho}_{s}^{kk}(t)|n\lambda;\boldsymbol{P}\rangle_{s}$
at its the diagonal limit in $n$ space. 
Ergo only the following overlap of self-energy
operator matters,
\begin{eqnarray}
 &  & \Delta_{\lambda_2\lambda_3}^{ji}{}_{s}\langle n\lambda;\boldsymbol{P}^{\prime}|\hat{\Sigma}_{s,ji}(t)|n\lambda_1;\boldsymbol{P}_{1}\rangle_{s}\nonumber \\
 & = & \delta_{\lambda\lambda_1}\Delta_{\lambda_2\lambda_3}^{ji}\frac{g^{2}T_{{\rm F}}}{M^{2}N_{c}}\int_{t_{0}}^{t}{\rm d}t_{2}\theta(t-t_{2})e^{i\left(\Delta E_{n,p_{2}}^{so}+\frac{\boldsymbol{P}_{1}^{2}-\boldsymbol{P}_{2}^{2}}{4M}\right)(t-t_{2})}{}_{s}\langle n;\boldsymbol{P}^{\prime}|\nonumber \\
 &  & \quad\left\langle \mathcal{T}\left(\boldsymbol{B}^{a,j}(\hat{\boldsymbol{R}},t)+\epsilon^{jmn}\frac{\boldsymbol{P}^{\prime,m}\boldsymbol{E}^{a,n}(\hat{\boldsymbol{R}},t)+\boldsymbol{E}^{a,n}(\hat{\boldsymbol{R}},t)\boldsymbol{P}_{2}^{m}}{8M}\right)\right.\int_{\boldsymbol{p}_{2},\boldsymbol{P}_{2}}|\boldsymbol{p}_{2},\boldsymbol{P}_{2}\rangle_{o}{}_{o}\langle\boldsymbol{p}_{2},\boldsymbol{P}_{2}|\nonumber \\
 &  & \;\left.\left(\boldsymbol{B}^{a,i}(\hat{\boldsymbol{R}},t_{2})+\epsilon^{ipq}\frac{\boldsymbol{P}_{2}^{p}\boldsymbol{E}^{a,q}(\hat{\boldsymbol{R}},t_{2})+\boldsymbol{E}^{a,q}(\hat{\boldsymbol{R}},t_{2})\boldsymbol{P}_{1}^{p}}{8M}\right)\right\rangle |n;\boldsymbol{P}_{1}\rangle_{s},
\end{eqnarray}
where we insert the other unit operator in eq.~\eqref{eq:Unit-operator-3}
adjacent to $e^{-ih_{o}^{{\rm sin}}(t-t_{2})}$, pick up the scattering-state
sector, and use
\begin{eqnarray}
h_{o}^{{\rm sin}}|\boldsymbol{p};\boldsymbol{P}\rangle_{o}=\left(\frac{\boldsymbol{p}^{2}}{M}+\frac{\boldsymbol{P}^{2}}{4M}\right)|\boldsymbol{p};\boldsymbol{P}\rangle_{o} & ,\; & h_{s}^{{\rm tri}}|n\lambda;\boldsymbol{P}_{1}\rangle_{s}=\left(E_{n}^{s}+\frac{\boldsymbol{P}_{1}^{2}}{4M}\right)|n\lambda;\boldsymbol{P}_{1}\rangle_{s},
\end{eqnarray}
to eliminate all the evolution operators. Here we also use that the spin-dependent potential enters in higher order so $|n,\lambda\rangle=|n\rangle\otimes|\lambda\rangle$ and gives $\langle\lambda|\lambda^{\prime}\rangle=\delta_{\lambda\lambda^{\prime}}$ without imposing rotating-wave approximation~\eqref{eq:RW-approx-fs} for these $\lambda$ states. The center-of-mass momentum
operators reduce to c-numbers due to $\hat{\boldsymbol{P}}|\boldsymbol{P}\rangle=\boldsymbol{P}|\boldsymbol{P}\rangle$.
The energy difference is $\Delta E_{n,p_{2}}^{so}=E_{n}^{s}-E_{p_{2}}^{o}=E_{n}^{s}-\frac{\boldsymbol{p}_{2}^{2}}{M}$.
To turn the gauge-field operators to c-fields, we further plug
the $r$-space unit operator of eq.~\eqref{eq:Unit-operator-r} in between them,
and make use of the definition of wavefunctions in eqs.~\eqref{eq:Singlet-WF}
and \eqref{eq:Octet-WF},
\begin{eqnarray}
 &  & \Delta_{\lambda_2\lambda_3}^{ji}{}_{s}\langle n\lambda;\boldsymbol{P}^{\prime}|\hat{\Sigma}_{s,ji}(t)|n\lambda_1;\boldsymbol{P}_{1}\rangle_{s}\nonumber \\
 & = & \delta_{\lambda\lambda_1}\Delta_{\lambda_2\lambda_3}^{ji}\frac{g^{2}T_{{\rm F}}}{M^{2}N_{c}}\int_{\boldsymbol{R}_{2},\boldsymbol{R}_{1}}\int_{\boldsymbol{p}_{2},\boldsymbol{P}_{2}}\left|_{o}\langle\boldsymbol{p}_{2}|n\rangle_{s}\right|^{2}\int_{t_{0}}^{t}{\rm d}t_{2}e^{i\left(\Delta E_{n,p_{2}}^{so}+\frac{\boldsymbol{P}_{1}^{2}-\boldsymbol{P}_{2}^{2}}{4M}\right)(t-t_{2})}\nonumber \\
 &  & \;\times e^{i(\boldsymbol{P}_{2}-\boldsymbol{P}^{\prime})\cdot\boldsymbol{R}_{1}}e^{i(\boldsymbol{P}_{1}-\boldsymbol{P}_{2})\cdot\boldsymbol{R}_{2}}\left\langle \mathcal{T}\left(\boldsymbol{B}_{{\rm rf}}^{a,j}(\boldsymbol{R}_{1},t)+\epsilon^{jmn}\frac{\boldsymbol{P}^{\prime,m}+\boldsymbol{P}_{2}^{m}}{8M}\boldsymbol{E}_{{\rm rf}}^{a,n}(\boldsymbol{R}_{1},t)\right)\right.\nonumber \\
 &  & \;\left.\left(\boldsymbol{B}_{{\rm rf}}^{a,i}(\boldsymbol{R}_{2},t_{2})+\epsilon^{ipq}\frac{\boldsymbol{P}_{2}^{p}+\boldsymbol{P}_{1}^{p}}{8M}\boldsymbol{E}_{{\rm rf}}^{a,q}(\boldsymbol{R}_{2},t_{2})\right)\right\rangle .\label{eq:SE-temp-1}
\end{eqnarray}
Here we label the c-fields with ``rf'' to indicate they are defined
in the medium-rest frame, and thus the gluon distribution functions
there are $f_{{\rm B}}(p_{0})=1\left/\left(e^{\beta p_{0}}-1\right)\right.$.
The wavefunction overlap $_{o}\langle\boldsymbol{p}_{2}|n\rangle_{s}$
is a shorthand of their inner product
\begin{eqnarray}
_{o}\langle\boldsymbol{p}_{2}|n\rangle_{s} & \equiv & \int_{\boldsymbol{r}}\Phi_{\boldsymbol{p}_{2}}^{*}(\boldsymbol{r})\phi_{n}(\boldsymbol{r}).
\end{eqnarray}
 We now introduce the momentum-space gauge-field correlator (\ref{eq:p-space-GF-correlator}),
\begin{eqnarray}
\tilde{G}_{BB,\alpha\beta}^{ji}(q) & = & \int_{R_{1}-R_{2}}e^{iq\cdot(R_{1}-R_{2})}\langle T_{\alpha\beta}\boldsymbol{B}_{{\rm rf}}^{a,j}(R_{1})\boldsymbol{B}_{{\rm rf}}^{a,i}(R_{2})\rangle,
\end{eqnarray}
and local equilibrium of gluons are implicitly assumed. Then integrating
over $\boldsymbol{R}_{1},\boldsymbol{R}_{2}$, eq.~\eqref{eq:SE-temp-1}
now reduces to
\begin{eqnarray}
& &\Delta_{\lambda_2\lambda_3}^{ji}{}_{s}\langle n\lambda;\boldsymbol{P}^{\prime}|\hat{\Sigma}_{s,ji}(t)|n\lambda_1;\boldsymbol{P}_{1}\rangle_{s} \nonumber\\& = & \delta_{\lambda\lambda_1}\Delta_{\lambda_2\lambda_3}^{ji}\frac{g^{2}T_{{\rm F}}}{M^{2}N_{c}}(2\pi)^{3}\delta^{(3)}(\boldsymbol{P}_{1}-\boldsymbol{P}^{\prime})\int_{q}\int_{\boldsymbol{p}_{2}}\left|_{o}\langle\boldsymbol{p}_{2}|n\rangle_{s}\right|^{2}\nonumber \\
 &  & \;\times\int_{t_{0}}^{t}{\rm d}t_{2}e^{i\left(\Delta E_{n,p_{2}}^{so}+\frac{\boldsymbol{P}_{1}^{2}-(\boldsymbol{P}_{1}-\boldsymbol{q})^{2}}{4M}-q_{0}\right)(t-t_{2})}\tilde{\mathcal{G}}_{11}^{ji}(q,P^{\prime}),
\end{eqnarray}
where $\tilde{\mathcal{G}}_{11}^{ji}(q,P)$ is defined in eq.~\eqref{eq:Gauge-field-correlator}.

Then the dissociation collision kernel can be evaluated with the definition
of spin-averaged distribution function
\begin{eqnarray}
 &  & \Delta_{\lambda_2\lambda_3}^{ji}\mathcal{C}_{ji,n\lambda}^{-,{\rm mrf}}(t,\boldsymbol{x},\boldsymbol{P}_{{\rm cm}})\nonumber \\
 & = & \Delta_{\lambda_2\lambda_3}^{ji}\frac{g^{2}T_{{\rm F}}}{M^{2}(d-1)N_{c}}\int_{\boldsymbol{P}_{{\rm rel}}}\int_{\boldsymbol{p}_{2}}\left|_{o}\langle\boldsymbol{p}_{2}|n\rangle_{s}\right|^{2}\int_{q}\int_{t_{0}}^{t}{\rm d}t_{2}e^{i\left(\Delta E_{n,p_{2}}^{so}+\frac{2\boldsymbol{P}_{{\rm cm}}\cdot\boldsymbol{q}-\boldsymbol{q}^{2}}{4M}-q_{0}\right)(t-t_{2})}\nonumber \\
 &  & \;\times e^{i\frac{\boldsymbol{P}_{{\rm rel}}\cdot\boldsymbol{q}}{4M}(t-t_{2})}\tilde{\mathcal{G}}_{11}^{ji}(q,P_{{\rm cm}}+\frac{P_{{\rm rel}}}{2})\int_{\boldsymbol{x}_{1}}e^{i\boldsymbol{P}_{{\rm rel}}\cdot(\boldsymbol{x}-\boldsymbol{x}_{1})}f_{n\lambda}^{s}(t,\boldsymbol{x}_{1},\boldsymbol{P}_{{\rm cm}})+{\rm H.c.}\label{eq:SE-temp-2}
\end{eqnarray}
where we denote $\boldsymbol{P}_{{\rm rel}}=\boldsymbol{P}^{\prime}-\boldsymbol{P}$
and $\boldsymbol{P}_{{\rm cm}}=\frac{\boldsymbol{P}^{\prime}+\boldsymbol{P}}{2}$.
The integration over $\boldsymbol{P}_{{\rm rel}}$ is involving because
the $e^{i\frac{\boldsymbol{P}_{{\rm rel}}\cdot\boldsymbol{q}}{4M}(t-t_{2})}$
factor effectively shifts $\boldsymbol{x}\to\boldsymbol{x}+\frac{\boldsymbol{q}}{4M}(t-t_{2})$.
We can alternatively use the identity to turn it as a Moyal-star product,
\begin{eqnarray}
 &  & e^{i\left(\Delta E_{n,p_{2}}^{so}+\frac{2\boldsymbol{P}_{{\rm cm}}\cdot\boldsymbol{q}-\boldsymbol{q}^{2}}{4M}-q_{0}\right)(t-t_{2})}\left[i\frac{\boldsymbol{P}_{{\rm rel}}\cdot\boldsymbol{q}}{4M}(t-t_{2})\right]e^{i\boldsymbol{P}_{{\rm rel}}\cdot\boldsymbol{x}}\nonumber \\
 & = & e^{i\left(\Delta E_{n,p_{2}}^{so}+\frac{2\boldsymbol{P}_{{\rm cm}}\cdot\boldsymbol{q}-\boldsymbol{q}^{2}}{4M}-q_{0}\right)(t-t_{2})}\left[\frac{-i}{2}\overleftarrow{\boldsymbol{\nabla}}_{\boldsymbol{P}_{{\rm cm}}}\cdot\boldsymbol{\nabla}_{\boldsymbol{x}}\right]e^{i\boldsymbol{P}_{{\rm rel}}\cdot\boldsymbol{x}}.
\end{eqnarray}
Similarly replacing $P_{{\rm cm}}+\frac{P_{{\rm rel}}}{2}\to P_{{\rm cm}}+\frac{-i\nabla_{x}}{2}$,
we can integrate over $\boldsymbol{P}_{{\rm rel}}$ and $\boldsymbol{x}_{1}$,
\begin{eqnarray}
 &  & \Delta_{\lambda_2\lambda_3}^{ji}\mathcal{C}_{ji,n\lambda}^{-,{\rm mrf}}(t,\boldsymbol{x},\boldsymbol{P}_{{\rm cm}})\nonumber \\
 & = & \Delta_{\lambda_2\lambda_3}^{ji}\frac{g^{2}T_{{\rm F}}}{M^{2}(d-1)N_{c}}\int_{q}\int_{\boldsymbol{p}_{2}}\left|_{o}\langle\boldsymbol{p}_{2}|n\rangle_{s}\right|^{2}\int_{t_{0}}^{t}{\rm d}t_{2}e^{i\left(\Delta E_{n,p_{2}}^{so}+\frac{2\boldsymbol{P}_{{\rm cm}}\cdot\boldsymbol{q}-\boldsymbol{q}^{2}}{4M}-q_{0}\right)(t-t_{2})}\nonumber \\
 &  & \;\times\exp\left(-\frac{i}{2}\overleftarrow{\boldsymbol{\nabla}}_{\boldsymbol{P}_{{\rm cm}}}\cdot\boldsymbol{\nabla}_{\boldsymbol{x}}\right)\tilde{\mathcal{G}}_{11}^{ji}(q,P_{{\rm cm}}-i\frac{\nabla_{\boldsymbol{x}}}{2})f_{n\lambda}^{s}(t,\boldsymbol{x},\boldsymbol{P}_{{\rm cm}})+{\rm H.c.}\nonumber \\
 & = & \Delta_{\lambda_2\lambda_3}^{ji}\frac{g^{2}T_{{\rm F}}}{M^{2}(d-1)N_{c}}\int_{q}\int_{\boldsymbol{p}_{2}}\left|_{o}\langle\boldsymbol{p}_{2}|n\rangle_{s}\right|^{2}\frac{i}{\Delta E_{n,p_{2}}^{so}+\frac{2\boldsymbol{P}_{{\rm cm}}\cdot\boldsymbol{q}-\boldsymbol{q}^{2}}{4M}-q_{0}+i\epsilon}\nonumber \\
 &  & \;\times\exp\left(-\frac{i}{2}\overleftarrow{\boldsymbol{\nabla}}_{\boldsymbol{P}_{{\rm cm}}}\cdot\boldsymbol{\nabla}_{\boldsymbol{x}}\right)\tilde{\mathcal{G}}_{11}^{ji}(q,P_{{\rm cm}}-i\frac{\nabla_{\boldsymbol{x}}}{2})f_{n\lambda}^{s}(t,\boldsymbol{x},\boldsymbol{P}_{{\rm cm}})+{\rm H.c.}
\end{eqnarray}
where in the last step we set $t_{0}\to-\infty$ in the Markovian
limit, thereby neglecting the initial-time memory effects. 

Finally, we make use of the hierarchy of the energy scales. We note
that 
\begin{eqnarray}
\boldsymbol{\nabla}_{\boldsymbol{x}}\sim q\sim T & ,\; & \boldsymbol{\nabla}_{\boldsymbol{P}_{{\rm cm}}}\sim(\sqrt{MT})^{-1},
\end{eqnarray}
so $\overleftarrow{\boldsymbol{\nabla}}_{\boldsymbol{P}_{{\rm cm}}}\cdot\boldsymbol{\nabla}_{\boldsymbol{x}}\sim\sqrt{\frac{T}{M}}$
is a small quantity. We wish to keep the equation with the leading spin-dependent collision kernel with recoil corrections up to relative order
$\frac{T}{M}$, and thus the correlator of interest is,
\begin{eqnarray}
\tilde{\mathcal{G}}_{11}^{ji}(q,P_{{\rm cm}}-\frac{i\boldsymbol{\nabla}_{\boldsymbol{x}}}{2}) & \simeq & \tilde{\mathcal{G}}_{11}^{ji}(q,P_{{\rm cm}})+\left(\epsilon^{imn}\tilde{G}_{BE,11}^{jn}(q)+\epsilon^{jmn}\tilde{G}_{EB,11}^{ni}(q)\right)\frac{-i\boldsymbol{\nabla}_{\boldsymbol{x}}^{m}}{8M},
\end{eqnarray}
and thus 
\begin{eqnarray}
 &  & \Delta_{\lambda_2\lambda_3}^{ji}\mathcal{C}_{ji,n\lambda}^{-,{\rm mrf}}(t,\boldsymbol{x},\boldsymbol{P}_{{\rm cm}})\nonumber \\
 & \simeq & \Delta_{\lambda_2\lambda_3}^{ji}\frac{g^{2}T_{{\rm F}}}{M^{2}(d-1)N_{c}}2{\rm Re}\int_{\boldsymbol{p}_{2}}\left|_{o}\langle\boldsymbol{p}_{2}|n\rangle_{s}\right|^{2}\nonumber\\& &\times\int_{q}\frac{i}{\Delta E_{n,p_{2}}^{so}+\frac{2\boldsymbol{P}_{{\rm cm}}\cdot\boldsymbol{q}-\boldsymbol{q}^{2}}{4M}-q_{0}+i\epsilon}\exp\left(-\frac{i}{2}\overleftarrow{\boldsymbol{\nabla}}_{\boldsymbol{P}_{{\rm cm}}}\cdot\boldsymbol{\nabla}_{\boldsymbol{x}}\right)\nonumber \\
 &  & \;\times\left[\tilde{\mathcal{G}}_{11}^{ji}(q,P_{{\rm cm}})+\left(\epsilon^{imn}\tilde{G}_{BE,11}^{jn}(q)+\epsilon^{jmn}\tilde{G}_{EB,11}^{ni}(q)\right)\frac{-i\boldsymbol{\nabla}_{\boldsymbol{x}}^{m}}{8M}\right]f_{n\lambda}^{s}(t,\boldsymbol{x},\boldsymbol{P}_{{\rm cm}}).
\end{eqnarray}

Similar calculation can be applied to $_{s}\langle n\lambda;\boldsymbol{P}^{\prime}|\Xi_{s,ji}\left(\rho_{o}^{{\rm sin}}(t);t\right)|n\lambda;\boldsymbol{P}\rangle_{s}$
and gives
\begin{eqnarray}
 &  & \Delta_{\lambda_2\lambda_3}^{ji}\mathcal{C}_{ji,n\lambda}^{+,{\rm mrf}}(t,\boldsymbol{x},\boldsymbol{P}_{{\rm cm}})\nonumber \\
 & = & \Delta_{\lambda_2\lambda_3}^{ji}\int_{\boldsymbol{P}^{\prime}-\boldsymbol{P}}e^{i(\boldsymbol{P}^{\prime}-\boldsymbol{P})\cdot\boldsymbol{x}}\langle n,\lambda;\boldsymbol{P}^{\prime}|\Xi_{s,ji}\left(\rho_{o}^{{\rm sin}}(t);t\right)|n,\lambda;\boldsymbol{P}\rangle\nonumber \\
 & = & \Delta_{\lambda_2\lambda_3}^{ji}\frac{g^{2}T_{{\rm F}}}{M^{2}(N_{c}^{2}-1)N_{c}}2{\rm Re}\int_{q}\int_{\boldsymbol{p}_{1},\boldsymbol{p}_{2}}\left[_{s}\langle n|\boldsymbol{p}_{1}\rangle_{o}\right]\left[_{o}\langle\boldsymbol{p}_{2}|n\rangle_{s}\right]\frac{i}{\Delta E_{n,p_{2}}^{so}+\frac{2\boldsymbol{P}_{{\rm cm}}\cdot\boldsymbol{q}-\boldsymbol{q}^{2}}{4M}-q_{0}+i\epsilon}\nonumber \\
 &  & \;\times\exp\left(\frac{i}{2}\overleftarrow{\boldsymbol{\nabla}}_{\boldsymbol{P}_{{\rm cm}}}\cdot\boldsymbol{\nabla}_{\boldsymbol{x}}\right)\left[\tilde{\mathcal{G}}_{22}^{ji}(q,P_{{\rm cm}})+\left(\epsilon^{imn}\tilde{G}_{BE,22}^{jn}(q)-\epsilon^{jmn}\tilde{G}_{EB,22}^{ni}(q)\right)\frac{i\boldsymbol{\nabla}_{\boldsymbol{x}}^{m}}{8M}\right]\nonumber \\
 &  & \;\times\int_{\boldsymbol{r}}e^{-i(\boldsymbol{p}_{1}-\boldsymbol{p}_{2})\cdot\boldsymbol{r}}f_{{\rm sin}}^{o}(t,\boldsymbol{x},\boldsymbol{P}_{{\rm cm}}-\boldsymbol{q};\boldsymbol{r},\frac{\boldsymbol{p}_{1}+\boldsymbol{p}_{2}}{2})
\end{eqnarray}
Note that we do not take the diagonal limit of $\langle\boldsymbol{p}_{1}|\hat{\rho}_{{\rm sin}}^{o}(t)|\boldsymbol{p}_{2}\rangle$
by setting $\boldsymbol{p}_{1}=\boldsymbol{p}_{2}$ due to the same reason describing in section \ref{subsec:Boltzmann-equation-of}. This is a
critical difference of this work from the treatment of scattering-state
distribution function in \citep{Yao:2020eqy,Yang:2024ejk}.

The quarkonium-rest frame counterpart can be derived by simply setting
$P_{{\rm cm}}=0$ and replacing $\tilde{\mathcal{G}}^{ji}\to\mathcal{G}^{ji}$
in eq.~\eqref{eq:SE-temp-2}.
\begin{eqnarray}
 &  & \Delta_{\lambda_2\lambda_3}^{ji}\mathcal{C}_{ji,n\lambda}^{-,{\rm qrf}}(t,\boldsymbol{x})\nonumber \\
 & = & \Delta_{\lambda_2\lambda_3}^{ji}\frac{g^{2}T_{{\rm F}}}{M^{2}(d-1)N_{c}}\int_{\boldsymbol{P}_{{\rm rel}}}\int_{\boldsymbol{p}_{2}}\left|_{o}\langle\boldsymbol{p}_{2}|n\rangle_{s}\right|^{2}\int_{q}\int_{t_{0}}^{t}{\rm d}t_{2}e^{i\left(\Delta E_{n,p_{2}}^{so}-\frac{\boldsymbol{q}^{2}}{4M}-q_{0}\right)(t-t_{2})}\nonumber \\
 &  & \;\times e^{i\frac{\boldsymbol{P}_{{\rm rel}}\cdot\boldsymbol{q}}{4M}(t-t_{2})}\mathcal{G}_{11}^{ji}(q,\frac{P_{{\rm rel}}}{2})\int_{\boldsymbol{x}_{1}}e^{i\boldsymbol{P}_{{\rm rel}}\cdot(\boldsymbol{x}-\boldsymbol{x}_{1})}f_{n\lambda}^{s}(t,\boldsymbol{x}_{1})+{\rm H.c.}
\end{eqnarray}
and the only difference step from now on is the treatment of $e^{i\frac{\boldsymbol{P}_{{\rm rel}}\cdot\boldsymbol{q}}{4M}(t-t_{2})}$.
We use 
\begin{eqnarray}
e^{-i\frac{\boldsymbol{q}^{2}}{4M}(t-t_{2})}\boldsymbol{P}_{{\rm rel}}\cdot\frac{-i\boldsymbol{q}}{4M}(t-t_{2})e^{i\boldsymbol{P}_{{\rm rel}}\cdot\boldsymbol{x}} & = & e^{i(-\frac{\boldsymbol{q}^{2}}{4M})(t-t_{2})}\frac{-i}{2}\overleftarrow{\boldsymbol{\nabla}}_{\boldsymbol{q}}\cdot\overrightarrow{\boldsymbol{\nabla}}_{\boldsymbol{x}}e^{i\boldsymbol{P}_{{\rm rel}}\cdot\boldsymbol{x}},
\end{eqnarray}
and get up to leading-order $\hbar$-expansion
\begin{eqnarray}
 &  & \Delta_{\lambda_2\lambda_3}^{ji}\mathcal{C}_{ji,n\lambda}^{-,{\rm qrf}}(t,\boldsymbol{x})\nonumber \\
 & = & \Delta_{\lambda_2\lambda_3}^{ji}\frac{g^{2}T_{{\rm F}}}{M^{2}(d-1)N_{c}}2{\rm Re}\int_{\boldsymbol{p}_{2}}\left|_{o}\langle\boldsymbol{p}_{2}|n\rangle_{s}\right|^{2}\int_{q}\frac{i}{\Delta E_{n,p_{2}}^{so}-\frac{\boldsymbol{q}^{2}}{4M}-q_{0}+i\epsilon}\left(1+\frac{i}{2}\overleftarrow{\boldsymbol{\nabla}}_{\boldsymbol{q}}\cdot\boldsymbol{\nabla}_{\boldsymbol{x}}\right)\nonumber \\
 &  & \;\times\left[\mathcal{G}_{11}^{ji}(q)+\left(\epsilon^{imn}G_{BE,11}^{jn}(q)+\epsilon^{jmn}G_{EB,11}^{ni}(q)\right)\frac{-i\boldsymbol{\nabla}_{\boldsymbol{x}}^{m}}{8M}\right]f_{n\lambda}^{s}(t,\boldsymbol{x}).
\end{eqnarray}
The Moyal product $\exp\left(\frac{i}{2}\overleftarrow{\boldsymbol{\nabla}}_{\boldsymbol{q}}\cdot\boldsymbol{\nabla}_{\boldsymbol{x}}\right)$ structure does not apply here because $\overleftarrow{\boldsymbol{\nabla_q}}$ only acts on $\frac{\boldsymbol{q}^2}{4M}$ term. Similar derivation of $\mathcal{C}_{ji,n\lambda}^{+,{\rm qrf}}(t,\boldsymbol{x})$
will not be repeated here, we list it in the main text (\ref{eq:CK-regen-2}).

\section{Quarkonium number transport equation with recoil effects\protect\label{sec:Quarkonium-transport-equation}}

The spin Boltzmann equation is coupled with the quarkonium number
transport equation, including the spin-1 and spin-0 sectors. In this
section, for completeness, we first derive the quantum master
equation for such density operators with center-of-mass corrections
and then derive the corresponding Boltzmann equations.

Total differential rate is defined by integrating out the azimuthal-angle
dependence for the dilepton distribution \eqref{eq:Decay_distribution_massless_l},
\begin{eqnarray}
\frac{{\rm d}\mathcal{R}}{{\rm d}^{4}X{\rm d}^{4}q} & = & \int_{-1}^{+1}{\rm d}\cos\theta\frac{{\rm d}\mathcal{R}}{{\rm d}^{4}X{\rm d}^{4}q{\rm d}\cos\theta}=\frac{2\pi Q^{2}e^{4}}{3(2\pi)^6q^{2}}J_{kk}(q,X).
\end{eqnarray}
and only depends on $\hat{\rho}_{s}^{\text{tri}}(t^{\prime};t)$.
For a state characterized by a set of quantum number $\{n\}$, the
nuclear modification factor compared to $pp$ collision reads
\begin{eqnarray}
\mathcal{R}^{n}(q) & = & \frac{\int_{\boldsymbol{R}}|\tilde{\phi}_{n}(\boldsymbol{0})|^{2}f_{n}^{s}(t_{{\rm F}},\boldsymbol{R},\boldsymbol{q})}{\int_{\boldsymbol{R}}|\tilde{\phi}_{n}(\boldsymbol{0})|^{2}f_{n}^{s}(t_{0},\boldsymbol{R},\boldsymbol{q})}=\frac{\langle n;\boldsymbol{q}|\hat{\rho}_{s}^{{\rm tri}}(t_{{\rm F}})|n;\boldsymbol{q}\rangle}{\langle n;\boldsymbol{q}|\hat{\rho}_{s}^{{\rm tri}}(t_{0})|n;\boldsymbol{q}\rangle}.
\end{eqnarray}
where we neglect the polarization dependence of the wavefunctions
and distribution up to the leading order and reproduce eq.~(17) of
\citep{Brambilla:2017zei}. The previous OQS works like \citep{Brambilla:2017zei,Brambilla:2020qwo,Brambilla:2022ynh}
did not include the center-of-mass motion in the theoretical framework,
they obtain the $p_{{\rm T}}$ dependence in Ref.~\citep{Brambilla:2022ynh}
for example by adding a weight $p_{{\rm T}}\left/\left(p_{{\rm T}}^{2}+\langle M\rangle^{2}\right)^{2}\right.$distribution
in the 3-momentum integral. In this
section, we derive a self-consistent quantum master equation and Boltzmann
equation by the minimal extension of chromoelectric dipole interactions
with leading center-of-mass corrections. We focus on the spin-1 sector
for simplicity, and the spin-0 equation is of exact same form except
for different initial conditions.

Up to leading-order multipole expansion in pNRQCD, the tree-level
interactions for spin-triplet fields are written as, from eq.~\eqref{eq:NLO-pNRQCD-L}
\begin{eqnarray}
\mathcal{L}_{\text{int}}^{\text{st}} & = & \sqrt{\frac{T_{{\rm F}}}{N_{c}}}\int_{\boldsymbol{r}}\biggl[S_{m}^{\dagger}\boldsymbol{r}^{i}\left(g\boldsymbol{E}^{i,a}+\frac{\epsilon^{ijk}}{4M}\{P_{R}^{j},g\boldsymbol{B}^{k,a}\}\right)O_{m}^{a}\nonumber \\
 &  & \;+O_{m}^{\dagger,a}\boldsymbol{r}^{i}\left(g\boldsymbol{E}^{i,a}+\frac{\epsilon^{ijk}}{4M}\{P_{R}^{j},g\boldsymbol{B}^{k,a}\}\right)S_{m}\biggr]\nonumber \\
 &  & +\frac{d^{abc}}{4T_{{\rm F}}}\int_{\boldsymbol{r}}O_{m}^{\dagger,a}\boldsymbol{r}^{i}\left(g\boldsymbol{E}^{i,c}+\frac{\epsilon^{ijk}}{4M}\{P_{R}^{j},g\boldsymbol{B}^{k,c}\}\right)O_{m}^{b},
\end{eqnarray}
similar vertices for the spin-singlet sectors are obtained by replacing
$S_{m}\to S_{{\rm sin}}$. Here we keep only the leading recoil
corrections to the interactions and drop the interaction of same order
e.g. $\frac{\alpha_{{\rm s}}}{Mr}\boldsymbol{r}\cdot g\boldsymbol{E}$. 

From the Kadanoff-Baym equations for color-singlet and color-octet correlators,
\begin{eqnarray}
 &  & \langle\boldsymbol{R}^{\prime},\boldsymbol{r}^{\prime}|\hat{\rho}_{s}^{{\rm tri}}(t^{\prime};t)|\boldsymbol{R},\boldsymbol{r}\rangle-\langle\boldsymbol{R}^{\prime},\boldsymbol{r}^{\prime}|\hat{\rho}_{s,0}^{{\rm tri}}(t^{\prime};t)|\boldsymbol{R},\boldsymbol{r}\rangle\nonumber \\
 & = & i^{2}\frac{g^{2}T_{{\rm F}}}{N_{c}}\int_{{\rm C}}{\rm d}t_{1}{\rm d}t_{2}\int_{\boldsymbol{R}_{1},\boldsymbol{R}_{2}}\int_{\boldsymbol{r}_{1},\boldsymbol{r}_{2}}\langle T_{1\alpha}S_{i}(R^{\prime},\boldsymbol{r}^{\prime})S_{m_{1}}^{\dagger}(R_{1},\boldsymbol{r}_{1})\rangle^{0}\nonumber \\
 &  & \quad\times\langle T_{\alpha\beta}\left(\boldsymbol{E}^{i_{1},a_{1}}(R_{1})+\frac{\epsilon^{i_{1}j_{1}k_{1}}}{4M}\{P_{R_{1}}^{j_{1}},\boldsymbol{B}^{k_{1},a_{1}}(R_{1})\}\right)\nonumber \\
 &  & \quad\times\boldsymbol{r}_{1}^{i_{1}}\langle T_{\alpha\beta}O_{m_{1}}^{a_{1}}(R_{1},\boldsymbol{r}_{1})O_{m_{2}}^{\dagger,a_{2}}(R_{2},\boldsymbol{r}_{2})\rangle^{0}\boldsymbol{r}_{2}^{i_{2}}\nonumber \\
 &  & \quad\times\left(\boldsymbol{E}^{i_{2},a_{2}}(R_{2})+\frac{\epsilon^{i_{2}j_{2}k_{2}}}{4M}\{P_{R_{2}}^{j_{2}},\boldsymbol{B}^{k_{2},a_{2}}(R_{2})\}\right)\rangle\langle T_{\beta2}S_{m_{2}}(R_{2},\boldsymbol{r}_{2})S_{i}^{\dagger}(R,\boldsymbol{r})\rangle^{0}
\end{eqnarray}
and
\begin{eqnarray}
 &  & \langle\boldsymbol{R}^{\prime},\boldsymbol{r}^{\prime}|\hat{\rho}_{o}^{{\rm tri}}(t^{\prime};t)|\boldsymbol{R},\boldsymbol{r}\rangle-\langle\boldsymbol{R}^{\prime},\boldsymbol{r}^{\prime}|\hat{\rho}_{o,0}^{{\rm tri}}(t^{\prime};t)|\boldsymbol{R},\boldsymbol{r}\rangle\nonumber \\
 & = & i^{2}g^{2}\int_{{\rm C}}{\rm d}t_{1}{\rm d}t_{2}\int_{\boldsymbol{R}_{1},\boldsymbol{R}_{2}}\int_{\boldsymbol{r}_{1},\boldsymbol{r}_{2}}\langle T_{1\alpha}O_{i}^{a}(R^{\prime},\boldsymbol{r}^{\prime})O_{m_{1}}^{\dagger,a_{1}}(R_{1},\boldsymbol{r}_{1})\rangle^{0}\nonumber \\
 &  & \times\langle T_{\alpha\beta}\left(\boldsymbol{E}^{i_{1},c_{1}}(\boldsymbol{R}_{1},t_{1})+\frac{\epsilon^{i_{1}j_{1}k_{1}}}{4M}\{P_{R_{1}}^{j_{1}},\boldsymbol{B}^{k_{1},c_{1}}(\boldsymbol{R}_{1},t_{1})\}\right)\nonumber \\
 &  & \times\boldsymbol{r}_{1}^{i_{1}}\left[\frac{T_{{\rm F}}}{N_{c}}\delta^{a_{1}c_{1}}\langle T_{\alpha\beta}S_{m_{1}}(R_{1},\boldsymbol{r}_{1})S_{m_{2}}^{\dagger}(R_{2},\boldsymbol{r}_{2})\rangle^{0}\delta^{b_{2}c_{2}}\right.\nonumber \\
 &  & \quad\left.+\left(\frac{d^{a_{1}b_{1}c_{1}}}{4T_{{\rm F}}}\frac{d^{a_{2}b_{2}c_{2}}}{4T_{{\rm F}}}\right)\langle T_{\alpha\beta}O_{m_{1}}^{b_{1}}(R_{1},\boldsymbol{r}_{1})O_{m_{2}}^{\dagger,a_{2}}(R_{2},\boldsymbol{r}_{2})\rangle^{0}\right]\boldsymbol{r}_{2}^{i_{2}}\nonumber \\
 &  & \times\left(\boldsymbol{E}^{i_{2},c_{2}}(\boldsymbol{R}_{2},t_{2})+\frac{\epsilon^{i_{2}j_{2}k_{2}}}{4M}\{P_{R_{2}}^{j_{2}},\boldsymbol{B}^{k_{2},c_{2}}(\boldsymbol{R}_{2},t_{2})\}\right)\rangle\langle T_{\beta2}O_{m_{2}}^{b_{2}}(R_{2},\boldsymbol{r}_{2})O_{i}^{\dagger,a}(R,\boldsymbol{r})\rangle^{0}.\nonumber \\
 & &
\end{eqnarray}
Similar to discussions in the main text, we focus on the leading-coupling case and drop the cumbersome Wilson-line dressing in the chromofields, and thus the 12 propagators are related with density operator by 
\begin{eqnarray}
\langle T_{12}S_{i}(R^{\prime},\boldsymbol{r}^{\prime})S_{i}^{\dagger}(R,\boldsymbol{r})\rangle & = & \langle\boldsymbol{R}^{\prime},\boldsymbol{r}^{\prime}|\hat{\rho}_{s}^{{\rm tri}}(t^{\prime};t)|\boldsymbol{R},\boldsymbol{r}\rangle,\\
\langle T_{12}O_{i}^{a}(R^{\prime},\boldsymbol{r}^{\prime})O_{i}^{\dagger,a}(R,\boldsymbol{r})\rangle & = & \langle\boldsymbol{R}^{\prime},\boldsymbol{r}^{\prime}|\hat{\rho}_{o}^{{\rm tri}}(t^{\prime};t)|\boldsymbol{R},\boldsymbol{r}\rangle.
\end{eqnarray}
We derive the quantum master equations following the same strategy
listed in section \ref{sec:Evolution-of-heavy-quarkonium},
\begin{eqnarray}
\frac{{\rm d}}{{\rm d}t}\hat{\rho}_{s}^{{\rm tri}}(t) & = & -i[h_{s}^{{\rm tri}},\hat{\rho}_{s}^{{\rm tri}}(t)]-\hat{\Sigma}_{s}(t)\hat{\rho}_{s}^{{\rm tri}}(t)-\hat{\rho}_{s}^{{\rm tri}}(t)\hat{\Sigma}_{s}^{\dagger}(t)+\hat{\Xi}_{so}\left(\hat{\rho}_{o}^{{\rm tri}}(t);t\right)\label{eq:QME-SA-s}\\
\frac{{\rm d}}{{\rm d}t}\hat{\rho}_{o}^{{\rm tri}}(t) & = & -i[h_{o}^{{\rm tri}},\hat{\rho}_{o}^{{\rm tri}}(t)]-\hat{\Sigma}_{o}(t)\hat{\rho}_{o}^{{\rm tri}}(t)-\hat{\rho}_{o}^{{\rm tri}}(t)\hat{\Sigma}_{o}^{\dagger}(t)\nonumber \\
 &  & \quad+\hat{\Xi}_{os}\left(\hat{\rho}_{s}^{{\rm tri}}(t);t\right)+\hat{\Xi}_{oo}\left(\hat{\rho}_{o}^{{\rm tri}}(t);t\right).\label{eq:QME-SA-o}
\end{eqnarray}
where the self-energy and transition operators for color-singlet field
are
\begin{eqnarray}
\hat{\Sigma}_{s}(t_{1}) & = & \frac{g^{2}T_{{\rm F}}}{N_{c}}\int_{t_{0}}^{t_{1}}{\rm d}t_{2}\langle \mathcal{T}\Biggl(\boldsymbol{E}^{i_{1},a_{1}}(\hat{\boldsymbol{R}},t_{1})\nonumber \\
 &  & \;+\frac{\epsilon^{i_{1}j_{1}k_{1}}}{4M}\{\hat{P}^{j_{1}},\boldsymbol{B}^{k_{1},a_{1}}(\hat{\boldsymbol{R}},t_{1})\}\Biggr)\hat{\boldsymbol{r}}^{i_{1}}e^{-ih_{o}^{{\rm tri}}(t_{1}-t_{2})}\hat{\boldsymbol{r}}^{i_{2}}\nonumber \\
 &  & \left(\boldsymbol{E}^{i_{2},a_{1}}(\hat{\boldsymbol{R}},t_{2})+\frac{\epsilon^{i_{2}j_{2}k_{2}}}{4M}\{\hat{P}^{j_{2}},\boldsymbol{B}^{k_{2},a_{1}}(\hat{\boldsymbol{R}},t_{2})\}\right)\rangle e^{ih_{s}^{{\rm tri}}(t_{1}-t_{2})},\\
\hat{\Xi}_{so}\left(\hat{\rho}_{o}^{{\rm tri}}(t_{1});t_{1}\right) & = & \frac{g^{2}T_{{\rm F}}}{N_{c}(N_{c}^{2}-1)}\int_{t_{0}}^{t_{1}}{\rm d}t_{2}\langle \bar{\mathcal{T}}\Biggl(\boldsymbol{E}^{i_{1},a_{1}}(\hat{\boldsymbol{R}},t_{1})\nonumber \\
 &  & \;+\frac{\epsilon^{i_{1}j_{1}k_{1}}}{4M}\{\hat{P}^{j_{1}},\boldsymbol{B}^{k_{1},a_{1}}(\hat{\boldsymbol{R}},t_{1})\}\Biggr)\hat{\boldsymbol{r}}^{i_{1}}\hat{\rho}_{o}^{{\rm tri}}(t_{1})e^{ih_{o}^{{\rm tri}}(t_{2}-t_{1})}\hat{\boldsymbol{r}}^{i_{2}}\nonumber \\
 &  & \left(\boldsymbol{E}^{i_{2},a_{1}}(\hat{\boldsymbol{R}},t_{2})+\frac{\epsilon^{i_{2}j_{2}k_{2}}}{4M}\{\hat{P}^{j_{2}},\boldsymbol{B}^{k_{2},a_{1}}(\hat{\boldsymbol{R}},t_{2})\}\right)\rangle e^{ih_{s}^{{\rm tri}}(t_{1}-t_{2})}+{\rm H.c.}\nonumber \\
 & &
\end{eqnarray}
and for color-octet field are
\begin{eqnarray}
\hat{\Sigma}_{o}(t_{1}) & = & g^{2}\frac{T_{{\rm F}}}{N_{c}(N_{c}^{2}-1)}\int_{t_{0}}^{t_{1}}{\rm d}t_{2}\langle \mathcal{T}\Biggl(\boldsymbol{E}^{i_{1},c_{1}}(\hat{\boldsymbol{R}},t_{1})+\frac{\epsilon^{i_{1}j_{1}k_{1}}}{4M}\{\hat{P}^{j_{1}},\boldsymbol{B}^{k_{1},c_{1}}(\hat{\boldsymbol{R}},t_{1})\}\Biggr)\nonumber \\
 &  & \;\hat{\boldsymbol{r}}^{i_{1}}\left[e^{-ih_{s}^{{\rm tri}}(t_{1}-t_{2})}+\frac{N_{c}^{2}-4}{16T_{{\rm F}}^{3}}e^{-ih_{o}^{{\rm tri}}(t_{1}-t_{2})}\right]\hat{\boldsymbol{r}}^{i_{2}}\nonumber \\
 &  & \;\left(\boldsymbol{E}^{i_{2},c_{1}}(\hat{\boldsymbol{R}},t_{2})+\frac{\epsilon^{i_{2}j_{2}k_{2}}}{4M}\{\hat{P}^{j_{2}},\boldsymbol{B}^{k_{2},c_{1}}(\hat{\boldsymbol{R}},t_{2})\}\right)\rangle e^{-ih_{o}^{{\rm tri}}(t_{2}-t_{1})}\\
\hat{\Xi}_{os}\left(\hat{\rho}_{s}^{{\rm tri}}(t_{1});t_{1}\right) & = & g^{2}\frac{T_{{\rm F}}}{N_{c}}\int_{t_{0}}^{t_{1}}{\rm d}t_{2}\langle \bar{\mathcal{T}}\Biggl(\boldsymbol{E}^{i_{1},c_{1}}(\hat{\boldsymbol{R}},t_{1})\nonumber \\
 &  & \;+\frac{\epsilon^{i_{1}j_{1}k_{1}}}{4M}\{\hat{P}^{j_{1}},\boldsymbol{B}^{k_{1},c_{1}}(\hat{\boldsymbol{R}},t_{1})\}\Biggr)\boldsymbol{r}^{i_{1}}\hat{\rho}_{s}^{{\rm tri}}(t_{1})e^{ih_{s}^{{\rm tri}}(t_{2}-t_{1})}\boldsymbol{r}^{i_{2}}\nonumber \\
 &  & \left(\boldsymbol{E}^{i_{2},c_{1}}(\hat{\boldsymbol{R}},t_{2})+\frac{\epsilon^{i_{2}j_{2}k_{2}}}{4M}\{\hat{P}^{j_{2}},\boldsymbol{B}^{k_{2},c_{1}}(\hat{\boldsymbol{R}},t_{2})\}\right)\rangle e^{ih_{o}^{{\rm tri}}(t_{1}-t_{2})}+{\rm H.c.}\\
\hat{\Xi}_{oo}\left(\hat{\rho}_{o}^{{\rm tri}}(t_{1});t_{1}\right) & = & g^{2}\frac{N_{c}^{2}-4}{16T_{{\rm F}}^{2}(N_{c}^{2}-1)N_{c}}\int_{t_{0}}^{t_{1}}{\rm d}t_{2}\langle \bar{\mathcal{T}}\Biggl(\boldsymbol{E}^{i_{1},c_{1}}(\hat{\boldsymbol{R}},t_{1})\nonumber \\
 &  & \;+\frac{\epsilon^{i_{1}j_{1}k_{1}}}{4M}\{\hat{P}^{j_{1}},\boldsymbol{B}^{k_{1},c_{1}}(\hat{\boldsymbol{R}},t_{1})\}\Biggr)\boldsymbol{r}^{i_{1}}\hat{\rho}_{o}^{{\rm tri}}(t_{1})e^{ih_{o}^{{\rm tri}}(t_{2}-t_{1})}\boldsymbol{r}^{i_{2}}\nonumber \\
 &  & \left(\boldsymbol{E}^{i_{2},c_{1}}(\hat{\boldsymbol{R}},t_{2})+\frac{\epsilon^{i_{2}j_{2}k_{2}}}{4M}\{\hat{P}^{j_{2}},\boldsymbol{B}^{k_{2},c_{1}}(\hat{\boldsymbol{R}},t_{2})\}\right)\rangle e^{ih_{o}^{{\rm tri}}(t_{1}-t_{2})}+{\rm H.c.}
\end{eqnarray}

In QGP, the color-singlet two-body fields are bound while the color-octet
quark-antiquark fields are in scattering state, therefore, we deduce
the Boltzmann equations by sandwiching eq.~\eqref{eq:QME-SA-s} with
$_{s}\langle n;\boldsymbol{P}^{\prime}|$ and $|n;\boldsymbol{P}\rangle_{s}$
and eq.~\eqref{eq:QME-SA-o} with $_{o}\langle\boldsymbol{p}^{\prime};\boldsymbol{P}^{\prime}|$
and $|\boldsymbol{p};\boldsymbol{P}\rangle_{o}$, and then Wigner transforming
the equations, we just present the final equations here:

\begin{eqnarray}
\left(\frac{{\rm d}}{{\rm d}t}+\frac{\boldsymbol{P}_{{\rm cm}}}{2M}\cdot\boldsymbol{\nabla}_{\boldsymbol{R}}\right)f_{{\rm tri},n}^{s}(t,\boldsymbol{R},\boldsymbol{P}_{{\rm cm}}) & = & -\mathcal{C}_{s,n}^{-,{\rm mrf}}[f_{{\rm tri},n}^{s}]+\mathcal{C}_{s,n}^{+,{\rm mrf}}[f_{{\rm tri}}^{o}],\nonumber\\
& & \label{eq:NBE-CS}\\
\left(\frac{{\rm d}}{{\rm d}t}+\frac{2\boldsymbol{p}_{{\rm cm}}}{M}\cdot\boldsymbol{\nabla}_{\boldsymbol{r}}+\frac{\boldsymbol{P}_{{\rm cm}}}{2M}\cdot\boldsymbol{\nabla}_{\boldsymbol{R}}\right)f_{{\rm tri}}^{o}(t,\boldsymbol{r},\boldsymbol{p}_{{\rm cm}};\boldsymbol{R},\boldsymbol{P}_{{\rm cm}}) & = & -\mathcal{C}_{o}^{-,{\rm mrf}}[f_{{\rm tri}}^{o}]+\mathcal{C}_{s,n}^{+,{\rm mrf}}[f_{{\rm tri},n}^{s},f_{{\rm tri}}^{o}],\nonumber\\
& & \label{eq:NBE-CO}
\end{eqnarray}
where up to leading order, the distribution functions have no dependence
on polarization so we drop the spin quantum number $\lambda$. The
dissociation and regeneration collision kernels for spin-singlet color-singlet
and color-octet sectors with leading recoil effects are,
\begin{eqnarray}
\mathcal{C}_{s,n}^{-,{\rm mrf}}[f_{{\rm tri},n}^{s}] & = & \frac{g^{2}T_{{\rm F}}}{N_{c}}2{\rm Re}\int_{q}\int_{\boldsymbol{p}_{2}}\langle n|\hat{\boldsymbol{r}}^{i_{1}}|\boldsymbol{p}_{2}\rangle\langle\boldsymbol{p}_{2}|\hat{\boldsymbol{r}}^{i_{2}}|n\rangle\frac{i}{\Delta E_{np_{2}}^{so}+\frac{2\boldsymbol{P}_{{\rm cm}}\cdot\boldsymbol{q}-\boldsymbol{q}^{2}}{4M}-q_{0}+i\epsilon}\nonumber \\
 &  & \;\times\exp\left(-\frac{i}{2}\overleftarrow{\boldsymbol{\nabla}}_{\boldsymbol{P}_{{\rm cm}}}\cdot\boldsymbol{\nabla}_{\boldsymbol{R}}\right)\tilde{\mathfrak{G}}_{11}^{i_{1}i_{2}}(q,P_{{\rm cm}}-\frac{i\nabla_{R}}{2})f_{{\rm tri},n}^{s}(t,\boldsymbol{R},\boldsymbol{P}_{{\rm cm}})\\
\mathcal{C}_{s,n}^{+,{\rm mrf}}[f_{{\rm tri}}^{o}] & = & 2{\rm Re}\frac{g^{2}T_{{\rm F}}}{N_{c}(N_{c}^{2}-1)}\int_{q}\int_{\boldsymbol{p}_{1},\boldsymbol{p}_{2}}\langle n|\hat{\boldsymbol{r}}^{i_{1}}|\boldsymbol{p}_{1}\rangle\langle\boldsymbol{p}_{2}|\hat{\boldsymbol{r}}^{i_{2}}|n\rangle\nonumber \\
 &  & \;\times\frac{i}{\Delta E_{np_{2}}^{so}+\frac{2\boldsymbol{P}_{{\rm cm}}\cdot\boldsymbol{q}-\boldsymbol{q}^{2}}{4M}-q_{0}+i\epsilon}\exp\left(\frac{i}{2}\overleftarrow{\boldsymbol{\nabla}}_{\boldsymbol{P}_{{\rm cm}}}\cdot\boldsymbol{\nabla}_{\boldsymbol{R}}\right)\nonumber \\
 &  & \;\times\left[\tilde{\mathfrak{G}}_{22}^{i_{1}i_{2}}(q,P_{{\rm cm}})+\frac{i\nabla_{R}^{j_{1}}}{4M}\left(\epsilon^{i_{2}j_{1}k_{1}}\tilde{G}_{EB,22}^{i_{1}k_{1}}(q)-\epsilon^{i_{1}j_{1}k_{1}}\tilde{G}_{BE,22}^{k_{1}i_{2}}(q)\right)\right]\nonumber \\
 &  & \;\times\int_{\boldsymbol{r}}e^{-i(\boldsymbol{p}_{1}-\boldsymbol{p}_{2})\cdot\boldsymbol{r}}f_{{\rm tri}}^{o}(t,\boldsymbol{R},\boldsymbol{P}_{{\rm cm}}-\boldsymbol{q};\boldsymbol{r},\frac{\boldsymbol{p}_{1}+\boldsymbol{p}_{2}}{2}).
\end{eqnarray}
and 
\begin{eqnarray}
 &  & \mathcal{C}_{o}^{-,{\rm mrf}}[f_{{\rm tri}}^{o}]\nonumber \\
 & = & \frac{g^{2}T_{{\rm F}}}{N_{c}(N_{c}^{2}-1)}2{\rm Re}\int_{\boldsymbol{p}^{\prime}-\boldsymbol{p}}\int_{\boldsymbol{p}_{1}}e^{i(\boldsymbol{p}^{\prime}-\boldsymbol{p})\cdot\boldsymbol{r}}\int_{q}\nonumber \\
 &  & \;\times\left[\sum_{n}\langle\boldsymbol{p}^{\prime}|\hat{\boldsymbol{r}}^{i_{1}}|n\rangle\langle n|\hat{\boldsymbol{r}}^{i_{2}}|\boldsymbol{p}_{1}\rangle\frac{i}{\Delta E_{p_{1}n}^{os}+\frac{2\boldsymbol{P}_{{\rm cm}}\cdot\boldsymbol{q}-\boldsymbol{q}^{2}}{4M}-q_{0}+i\epsilon}\right.\nonumber \\
 &  & \quad\left.+\frac{N_{c}^{2}-4}{16T_{{\rm F}}^{3}}\int_{\boldsymbol{p}_{2}}\langle\boldsymbol{p}^{\prime}|\hat{\boldsymbol{r}}^{i_{1}}|\boldsymbol{p}_{2}\rangle\langle\boldsymbol{p}_{2}|\hat{\boldsymbol{r}}^{i_{2}}|\boldsymbol{p}_{1}\rangle\frac{i}{\Delta E_{p_{1}p_{2}}^{oo}+\frac{2\boldsymbol{P}_{{\rm cm}}\cdot\boldsymbol{q}-\boldsymbol{q}^{2}}{4M}-q_{0}+i\epsilon}\right]\nonumber \\
 &  & \quad\exp\left(-\frac{i}{2}\overleftarrow{\boldsymbol{\nabla}}_{\boldsymbol{P}_{{\rm cm}}}\cdot\boldsymbol{\nabla}_{\boldsymbol{R}}\right)\tilde{\mathfrak{G}}_{11}^{i_{1}i_{2}}(q,P_{{\rm cm}}-\frac{i\nabla_{R}}{2})\nonumber \\
 &  & \;\times\int_{\boldsymbol{r}_{1}}e^{-i(\boldsymbol{p}_{1}-\boldsymbol{p})\cdot\boldsymbol{r}_{1}}f_{{\rm tri}}^{o}(t,\boldsymbol{R},\boldsymbol{P}_{{\rm cm}};\boldsymbol{r}_{1},\frac{\boldsymbol{p}_{1}+\boldsymbol{p}}{2})\\
\nonumber \\ &  & \mathcal{C}_{s,n}^{+,{\rm mrf}}[f_{{\rm tri},n}^{s},f_{{\rm tri}}^{o}]\nonumber \\
 & = & \frac{g^{2}T_{{\rm F}}}{N_{c}}2{\rm Re}\int_{q}\int_{\boldsymbol{p}^{\prime}-\boldsymbol{p}}e^{i(\boldsymbol{p}^{\prime}-\boldsymbol{p})\cdot\boldsymbol{r}}\sum_{n}\langle\boldsymbol{p}^{\prime}|\hat{\boldsymbol{r}}^{i_{1}}|n\rangle\langle n|\hat{\boldsymbol{r}}^{i_{2}}|\boldsymbol{p}\rangle\nonumber \\
 &  & \;\frac{i}{\Delta E_{pn}^{os}+\frac{2\boldsymbol{P}_{{\rm cm}}\cdot\boldsymbol{q}-\boldsymbol{q}^{2}}{4M}-q_{0}+i\epsilon}\exp\left(\frac{i}{2}\overleftarrow{\boldsymbol{\nabla}}_{\boldsymbol{P}_{{\rm cm}}}\cdot\boldsymbol{\nabla}_{\boldsymbol{R}}\right)\Biggl[\tilde{\mathfrak{G}}_{22}^{i_{1}i_{2}}(q,P_{{\rm cm}})\nonumber \\
 &  & \;+\left(\epsilon^{i_{2}j_{1}k_{1}}\tilde{G}_{EB,22}^{i_{1}k_{1}}(q)-\epsilon^{i_{1}j_{1}k_{1}}\tilde{G}_{BE,22}^{k_{1}i_{2}}(q)\right)\frac{i\nabla_{R}^{j_{1}}}{4M}\Biggr]f_{{\rm tri},n}^{s}(t,\boldsymbol{R},\boldsymbol{P}_{{\rm cm}}-\boldsymbol{q})\nonumber \\
 &  & +\frac{g^{2}(N_{c}^{2}-4)}{16T_{{\rm F}}^{2}(N_{c}^{2}-1)N_{c}}2{\rm Re}\int_{\boldsymbol{p}^{\prime}-\boldsymbol{p}}e^{i(\boldsymbol{p}^{\prime}-\boldsymbol{p})\cdot\boldsymbol{r}}\int_{\boldsymbol{p}_{1},\boldsymbol{p}_{2}}\langle\boldsymbol{p}^{\prime}|\hat{\boldsymbol{r}}^{i_{1}}|\boldsymbol{p}_{2}\rangle\langle\boldsymbol{p}_{1}|\hat{\boldsymbol{r}}^{i_{2}}|\boldsymbol{p}\rangle\nonumber \\
 &  & \quad\int_{q}\frac{i}{\Delta E_{pp_{1}}^{oo}+\frac{2\boldsymbol{P}_{{\rm cm}}\cdot\boldsymbol{q}-\boldsymbol{q}^{2}}{4M}-q_{0}+i\epsilon}\exp\left(\frac{i}{2}\overleftarrow{\boldsymbol{\nabla}}_{\boldsymbol{P}_{{\rm cm}}}\cdot\boldsymbol{\nabla}_{\boldsymbol{R}}\right)\nonumber \\
 &  & \;\times\left[\tilde{\mathfrak{G}}_{22}^{i_{1}i_{2}}(q,P_{{\rm cm}})+\left(\epsilon^{i_{2}j_{1}k_{1}}\tilde{G}_{EB,22}^{i_{1}k_{1}}(q)-\epsilon^{i_{1}j_{1}k_{1}}\tilde{G}_{BE,22}^{k_{1}i_{2}}(q)\right)\frac{i\nabla_{R}^{j_{1}}}{4M}\right]\nonumber \\
 &  & \quad\times\int_{\boldsymbol{r}_{1}}e^{-i(\boldsymbol{p}_{2}-\boldsymbol{p}_{1})\cdot\boldsymbol{r}_{1}}f_{{\rm tri}}^{o}(t,\boldsymbol{R},\boldsymbol{P}_{{\rm cm}}-\boldsymbol{q};\boldsymbol{r}_{1},\frac{\boldsymbol{p}_{2}+\boldsymbol{p}_{1}}{2}).
\end{eqnarray}
where we define $\Delta E_{pn}^{os}=E_{p}^{o}-E_{n}^{s}$ and $\Delta E_{pp^{\prime}}^{oo}=E_{p}^{o}-E_{p^{\prime}}^{o}$.
Here we defined the medium-rest frame chromo electric-electric correlator
with recoil corrections 
\begin{eqnarray}
\tilde{\mathfrak{G}}_{11}^{i_{1}i_{2}}(q,P^{\prime}) & \equiv & \tilde{G}_{EE,11}^{i_{1}i_{2}}(q)+\epsilon^{i_{2}j_{2}k_{2}}\frac{2P^{\prime,j_{2}}-q^{j_{2}}}{4M}\tilde{G}_{EB,11}^{i_{1}k_{2}}(q)+\epsilon^{i_{1}j_{1}k_{1}}\frac{2P^{\prime,j_{1}}-q^{j_{1}}}{4M}\tilde{G}_{BE,11}^{k_{1}i_{2}}(q)\nonumber \\
 &  & +\epsilon^{i_{1}j_{1}k_{1}}\epsilon^{i_{2}j_{2}k_{2}}\frac{2P^{\prime,j_{1}}-q^{j_{1}}}{4M}\frac{2P^{\prime,j_{2}}-q^{j_{2}}}{4M}\tilde{G}_{BB,11}^{k_{1}k_{2}}(q).
\end{eqnarray}
We keep the leading relative $\mathcal{O}(\frac{T}{M})$ corrections
in the collision term. The color-octet collision term receives the
contributions from both bound-state dissociation and scattering-state
excitation. The Boltzmann equations (\ref{eq:NBE-CS}) and (\ref{eq:NBE-CO})
for the color-singlet and color-octet distribution functions are closed
and the coupled equation array can be solved for given initial conditions. 

To make a comparison with the existing literature \citep{Yao:2018nmy,Yao:2020eqy,Akamatsu:2020ypb},
we list the Boltzmann equation in the classical limit by dropping
the $R$ gradient in the collision kernels, and in terms of Wightman
gluon correlators, 
\begin{eqnarray}
 &  & \left(\frac{{\rm d}}{{\rm d}t}+\frac{\boldsymbol{P}_{{\rm cm}}}{2M}\cdot\boldsymbol{\nabla}_{\boldsymbol{R}}\right)f_{{\rm tri},n}^{s}(t,\boldsymbol{R},\boldsymbol{P}_{{\rm cm}})\nonumber \\
 & = & {\rm Re}\frac{g^{2}T_{{\rm F}}}{N_{c}}\int_{q}\int_{\boldsymbol{p}_{1},\boldsymbol{p}_{2}}\langle n|\hat{\boldsymbol{r}}^{i_{1}}|\boldsymbol{p}_{1}\rangle\langle\boldsymbol{p}_{2}|\hat{\boldsymbol{r}}^{i_{2}}|n\rangle2\pi\delta\left(\Delta E_{np_{2}}^{so}+\frac{2\boldsymbol{P}_{{\rm cm}}\cdot\boldsymbol{q}-\boldsymbol{q}^{2}}{4M}-q_{0}\right)\nonumber \\
 &  & \;\times\Biggl[\tilde{\mathfrak{G}}_{12}^{i_{1}i_{2}}(q,P_{{\rm cm}})\int_{\boldsymbol{r}}e^{-i(\boldsymbol{p}_{1}-\boldsymbol{p}_{2})\cdot\boldsymbol{r}}\frac{f_{{\rm tri}}^{o}(t,\boldsymbol{R},\boldsymbol{P}_{{\rm cm}}-\boldsymbol{q};\boldsymbol{r},\frac{\boldsymbol{p}_{1}+\boldsymbol{p}_{2}}{2})}{N_{c}^{2}-1}\nonumber \\
 &  & \quad\;-\delta_{\boldsymbol{p}_{1}\boldsymbol{p}_{2}}\tilde{\mathfrak{G}}_{21}^{i_{1}i_{2}}(q,P_{{\rm cm}})f_{{\rm tri},n}^{s}(t,\boldsymbol{R},\boldsymbol{P}_{{\rm cm}})\Biggr]\label{eq:NBE-CS-CL}
\end{eqnarray}
and 
\begin{eqnarray}
 &  & \left(\frac{{\rm d}}{{\rm d}t}+\frac{2\boldsymbol{p}_{{\rm cm}}}{M}\cdot\boldsymbol{\nabla}_{\boldsymbol{r}}+\frac{\boldsymbol{P}_{{\rm cm}}}{2M}\cdot\boldsymbol{\nabla}_{\boldsymbol{R}}\right)f_{{\rm tri}}^{o}(t,\boldsymbol{r},\boldsymbol{p}_{{\rm cm}};\boldsymbol{R},\boldsymbol{P}_{{\rm cm}})\nonumber \\
 & = & \frac{g^{2}T_{{\rm F}}}{N_{c}}{\rm Re}\int_{q}\int_{\boldsymbol{p}^{\prime}-\boldsymbol{p}}e^{i(\boldsymbol{p}^{\prime}-\boldsymbol{p})\cdot\boldsymbol{r}}\nonumber \\
 &  & \;\times\int_{\boldsymbol{p}_{1}}\sum_{n}\langle\boldsymbol{p}^{\prime}|\hat{\boldsymbol{r}}^{i_{1}}|n\rangle\langle n|\hat{\boldsymbol{r}}^{i_{2}}|\boldsymbol{p}_{1}\rangle2\pi\delta\left(\Delta E_{p_{1}n}^{os}+\frac{2\boldsymbol{P}_{{\rm cm}}\cdot\boldsymbol{q}-\boldsymbol{q}^{2}}{4M}-q_{0}\right)\nonumber \\
 &  & \quad\Biggl[\tilde{\mathfrak{G}}_{12}^{i_{1}i_{2}}(q,P_{{\rm cm}})\delta_{\boldsymbol{p}_{1}\boldsymbol{p}}f_{{\rm tri},n}^{s}(t,\boldsymbol{R},\boldsymbol{P}_{{\rm cm}}-\boldsymbol{q})\nonumber \\
 &  & \quad-\tilde{\mathfrak{G}}_{21}^{i_{1}i_{2}}(q,P_{{\rm cm}})\int_{\boldsymbol{r}_{1}}e^{-i(\boldsymbol{p}_{1}-\boldsymbol{p})\cdot\boldsymbol{r}_{1}}\frac{f_{{\rm tri}}^{o}(t,\boldsymbol{R},\boldsymbol{P}_{{\rm cm}};\boldsymbol{r}_{1},\frac{\boldsymbol{p}_{1}+\boldsymbol{p}}{2})}{N_{c}^{2}-1}\Biggr]\nonumber \\
 &  & +\frac{g^{2}(N_{c}^{2}-4)}{16T_{{\rm F}}^{2}N_{c}(N_{c}^{2}-1)}{\rm Re}\int_{q}\int_{\boldsymbol{p}^{\prime}-\boldsymbol{p}}e^{i(\boldsymbol{p}^{\prime}-\boldsymbol{p})\cdot\boldsymbol{r}}\nonumber \\
 &  & \times\int_{\boldsymbol{p}_{1},\boldsymbol{p}_{2}}\int_{\boldsymbol{p}_{3},\boldsymbol{p}_{4}}\langle\boldsymbol{p}^{\prime}|\hat{\boldsymbol{r}}^{i_{1}}|\boldsymbol{p}_{3}\rangle\langle\boldsymbol{p}_{2}|\hat{\boldsymbol{r}}^{i_{2}}|\boldsymbol{p}_{4}\rangle2\pi\delta\left(\Delta E_{p_{4}p_{2}}^{oo}+\frac{2\boldsymbol{P}_{{\rm cm}}\cdot\boldsymbol{q}-\boldsymbol{q}^{2}}{4M}-q_{0}\right)\nonumber \\
 &  & \quad\Biggl[\delta_{\boldsymbol{p}\boldsymbol{p}_{4}}\delta_{\boldsymbol{p}_{1}\boldsymbol{p}_{3}}\tilde{\mathfrak{G}}_{12}^{i_{1}i_{2}}(q,P_{{\rm cm}})\int_{\boldsymbol{r}_{1}}e^{-i(\boldsymbol{p}_{1}-\boldsymbol{p}_{2})\cdot\boldsymbol{r}_{1}}f_{{\rm tri}}^{o}(t,\boldsymbol{R},\boldsymbol{P}_{{\rm cm}}-\boldsymbol{q};\boldsymbol{r}_{1},\frac{\boldsymbol{p}_{1}+\boldsymbol{p}_{2}}{2})\nonumber \\
 &  & \quad-\delta_{\boldsymbol{p}_{2}\boldsymbol{p}_{3}}\delta_{\boldsymbol{p}_{1}\boldsymbol{p}_{4}}\tilde{\mathfrak{G}}_{21}^{i_{1}i_{2}}(q,P_{{\rm cm}})\int_{\boldsymbol{r}_{1}}e^{-i(\boldsymbol{p}_{1}-\boldsymbol{p})\cdot\boldsymbol{r}_{1}}f_{{\rm tri}}^{o}(t,\boldsymbol{R},\boldsymbol{P}_{{\rm cm}};\boldsymbol{r}_{1},\frac{\boldsymbol{p}_{1}+\boldsymbol{p}}{2})\Biggr]\label{eq:NBE-CO-CL}
\end{eqnarray}
where we introduce the notation $\delta_{\boldsymbol{p}\boldsymbol{p}^{\prime}}=(2\pi)^3\delta^{(3)}(\boldsymbol{p}-\boldsymbol{p}^{\prime})$.
If we follow the $\boldsymbol{r}$ expansion up to leading order and
drop all the $P_{{\rm cm}}$ dependence in the gluon correlators,
eq.~\eqref{eq:NBE-CS-CL} can recover the color-singlet equation derived
in e.g. Ref.~\citep{Yao:2018nmy} and eqs.(4.28), (4.29) of Ref.~\citep{Yao:2020eqy}.
The octet equation (\ref{eq:NBE-CO-CL}) derived
in this paper incorporates both recoil effects and
octet-octet transition.

\section{Perturbative collision term in medium rest frame \protect\label{sec:Perturbative-collision-term}}

For completeness, we present the spin Boltzmann equation in the medium-rest frame (\ref{eq:SBE-mrf}) with perturbative collision
term:
\begin{eqnarray}
 &  & \left(\frac{{\rm d}}{{\rm d}t}+\frac{\boldsymbol{P}_{{\rm cm}}}{2M}\cdot\boldsymbol{\nabla}_{\boldsymbol{R}}\right)\left[\Delta_{\lambda_2\lambda_3}^{ji}\delta f_{ji,n\lambda}^{s}(t,\boldsymbol{R},\boldsymbol{P}_{{\rm cm}})\right]\nonumber \\
 & = & \Delta_{\lambda_2\lambda_3}^{ji}\Gamma_{n}^{{s,\rm mrf}}\frac{P_{{\rm cm}}^{i}P_{{\rm cm}}^{j}}{4M^{2}}\frac{f_{n\lambda}^{s}(t,\boldsymbol{R},\boldsymbol{P}_{{\rm cm}})}{d-1}-\Delta_{\lambda_2\lambda_3}^{ji}\int_{\boldsymbol{p}_{1},\boldsymbol{p}_{2}}\Biggl[\mathcal{W}_{n,\boldsymbol{p}_{1},\boldsymbol{p}_{2}}^{\text{mrf},3}\frac{P_{{\rm cm}}^{i}P_{{\rm cm}}^{j}}{4M^{2}}\nonumber \\
 &  & \quad-\mathcal{W}_{n,\boldsymbol{p}_{1},\boldsymbol{p}_{2}}^{\text{mrf},4}\left(\frac{P_{{\rm cm}}^{i}}{2M}\boldsymbol{\nabla}_{\boldsymbol{P}_{{\rm cm}}}^{j}+\frac{P_{{\rm cm}}^{j}}{2M}\boldsymbol{\nabla}_{\boldsymbol{P}_{{\rm cm}}}^{i}\right)+\mathcal{W}_{n,\boldsymbol{p}_{1},\boldsymbol{p}_{2}}^{\text{mrf},5}\boldsymbol{\nabla}_{\boldsymbol{P}_{{\rm cm}}}^{i}\boldsymbol{\nabla}_{\boldsymbol{P}_{{\rm cm}}}^{j}\Biggr]\nonumber \\
 &  & \;\times n_{{\rm B}}^{<}(\Delta E_{n,\boldsymbol{p}_{1}}^{so})\int_{\boldsymbol{r}}e^{-i(\boldsymbol{p}_{1}-\boldsymbol{p}_{2})\cdot\boldsymbol{r}}\frac{f_{{\rm sin}}^{o}(t,\boldsymbol{R},\boldsymbol{P}_{{\rm cm}};\boldsymbol{r},\frac{\boldsymbol{p}_{1}+\boldsymbol{p}_{2}}{2})}{N_{c}^{2}-1}.
\end{eqnarray}
The definitions for $\Gamma_{n}^{s,\text{mrf}}$ and $\mathcal{W}_{n,\boldsymbol{p}_1,\boldsymbol{p}_2}^{\text{mrf},\{3,4,5\}}$
are as follows. 

\begin{eqnarray}
\Gamma_{n}^{{s, \rm mrf}} & = & \int_{\boldsymbol{p}_{1}}n_{{\rm B}}^{>}(\Delta E_{n,\boldsymbol{p}_{1}}^{so})\left[\frac{17}{4}W_{n,\boldsymbol{p}_{1},\boldsymbol{p}_{1}}^{3}-\frac{3}{2}W_{n,\boldsymbol{p}_{1},\boldsymbol{p}_{1}}^{4}\beta n_{{\rm B}}^{<}(\Delta E_{n,\boldsymbol{p}_{1}}^{so})\right.\nonumber \\
 &  & \quad\left.+\frac{1}{10}W_{n,\boldsymbol{p}_{1},\boldsymbol{p}_{1}}^{5}\beta^{2}n_{{\rm B}}^{<}(\Delta E_{n,\boldsymbol{p}_{1}}^{so})\left(1+2n_{{\rm B}}^{<}(\Delta E_{n,\boldsymbol{p}_{1}}^{so})\right)\right],\\
\mathcal{W}_{n,\boldsymbol{p}_{1},\boldsymbol{p}_{2}}^{\text{mrf},3} & = & \frac{17}{4}W_{n,\boldsymbol{p}_{1},\boldsymbol{p}_{2}}^{3}-\frac{3}{2}\beta W_{n,\boldsymbol{p}_{1},\boldsymbol{p}_{2}}^{4}n_{{\rm B}}^{>}(\Delta E_{n,\boldsymbol{p}_{1}}^{so})\nonumber \\
 &  & \qquad+\frac{1}{10}W_{n,\boldsymbol{p}_{1},\boldsymbol{p}_{2}}^{5}\beta^{2}n_{{\rm B}}^{>}(\Delta E_{n,\boldsymbol{p}_{1}}^{so})\left(1+2n_{{\rm B}}^{<}(\Delta E_{n,\boldsymbol{p}_{1}}^{so})\right),\\
\mathcal{W}_{n,\boldsymbol{p}_{1},\boldsymbol{p}_{2}}^{\text{mrf},4} & = & \frac{3}{4}W_{n,\boldsymbol{p}_{1},\boldsymbol{p}_{2}}^{4}-\frac{1}{10}\beta W_{n,\boldsymbol{p}_{1},\boldsymbol{p}_{2}}^{5}\left(1+n_{{\rm B}}^{<}(\Delta E_{n,\boldsymbol{p}_{1}}^{so})\right),\\
\mathcal{W}_{n,\boldsymbol{p}_{1},\boldsymbol{p}_{2}}^{\text{mrf},5} & = & \frac{1}{10}W_{n,\boldsymbol{p}_{1},\boldsymbol{p}_{2}}^{5}.
\end{eqnarray}
We
note that these scattering amplitudes are different in the two different
frames.

\bibliography{EFTs}
\bibliographystyle{JHEP}

\end{document}